\documentclass[aps,prd,notitlepage,superscriptaddress,nofootinbib,showkeys]{revtex4-1}

\usepackage{graphicx}
\usepackage{amsmath}
\usepackage{amsfonts}
\usepackage{placeins}
\usepackage{mathtools}
\usepackage{todonotes}
\usepackage{physics}
\usepackage{amssymb}
\usepackage{bm}
\usepackage{csquotes}
\usepackage{epstopdf}
\usepackage{float}

\usepackage{subfigure}
\usepackage{capt-of}
\allowdisplaybreaks
\usepackage{comment}
\usepackage{hyperref}

\hypersetup{
     colorlinks   = true,
     citecolor    = red,
     linkcolor    = blue,
     urlcolor     = blue,
}
\definecolor{deeppink}{rgb}{0.9, 0.17, 0.31}

\def\mH{\mathcal{H}}
\def\HI{H_{\rm inf}}
\def\Hkpv{H_{\kpv}}
\def\ke{k_{\rm end}}
\def\kpv{k_{*}}

\def\kre{k_{\rm re}}

\def\wre{w_{\rm re}}
\def\Tre{T_{\rm re}}
\def\Mp{M_{\rm P}}
\def\Mpl{M_{\rm pl}}

\def\ee{\eta_{\rm end}}
\def\ere{\eta_{\rm re}}
\def\ae{a_{\rm end}}

\def\mR{\mathcal{R}}
\def\mP{\mathcal{P}}
\def\mPt{\mathcal{P}_{\rm T}}
\def\mPts{\mathcal{P}_{\rm T}^{\rm sec}}
\def\mPtv{\mathcal{P}_{\rm T}^{\rm vac}}
\def\mPb{\mathcal{P}_{\rm B}}
\def\mPe{\mathcal{P}_{\rm E}}
\def\mPc{\mathcal{P}_{\zeta}}
\def\mPcs{\mathcal{P}_{\zeta}^{\rm ind}}
\def\mPcv{\mathcal{P}_{\zeta}^{\rm vac}}

\def\zetai{\zeta^{\rm ind}}
\def\zetav{\zeta^{\rm vac}}

\def\vq{\mathbf{q}}
\def\vk{\mathbf{k}}

\def\l{\left}
\def\r{\right}

\def\are{a_{\rm re}}
\def\Hre{H_{\rm re}}

\def\bJ{\mathrm{J}}
\def\bY{\mathrm{Y}}

\def\nb{n_{\rm B}}

\def\ns{n_{\rm s}}
\def\As{A_{\rm s}}

\def\gsre{g^{*}_{\rm re}}

\def\Mpc{\mathrm{Mpc}}
\def\GeV{\mathrm{GeV}}

\def\rhob{\rho_{\rm B}}
\def\rhoe{\rho_{\rm E}}
\def\rhoem{\rho_{\rm EM}}
\def\drho{\dot{\rho}}
\def\rhop{\rho_\phi}
\def\dphi{\dot{\phi}}

\def\ddbphi{\ddot{\bphi}}
\def\dbphi{\dot{\bphi}}
\def\bphi{\bar{\phi}}

\def\phik{\phi_{\vk}}
\def\bphikpv{\bphi_{\kpv}}
\def\dbphikpv{\dot{\bphi}_{\kpv}}

\def\mFnb{\mathcal{F}_{\rm nb}}

\def\vx{\mathbf{x}}

\def\mGk{\mathcal{G}_k}
\def\d{\mathrm{d}}
\def\nn{\nonumber}
\def\xe{x_{\rm end}}
\def\mI{\mathcal{I}}

\def\umin{u_{\rm min}}
\def\umax{u_{\rm max}}
\def\mGki{\mGk^{\rm inf}}

\def\mSem{\mathcal{S}_{\rm EM}}
\def\uk{u_{\vk}^{\lambda}}
\def\mVk{\mathcal{V}_{\vk}^{\lambda}}
\def\ddmVk{\ddot{\mathcal{V}}_{\vk}^{\lambda}}

\def\Mpbh{\mathrm{M}_{\text{PBH}}}
\def\th{t_{\rm f}}
\def\rhopbh{\rho_{\text{pbh}}}
\def\rhot{\rho_{\text{total}}}
\def\rM{\rm{M}}
\def\Meq{\rm{M}_{\rm eq}}
\def\gseq{g_{*,\rm eq}}
\def\gsk{g_{*,k}}
\def\keq{k_{\rm eq}}
\def\Msolar{\rm{M}_{\odot}}

\def\tev{t_{\rm evp}}

\def\gst{g_{\rm s, *}}

\def\hkl{h^{\lambda}_{\vk}}
\def\mcB{\mathcal{C}_{\rm B}}
\def\Fn{\mathcal{F}_n}

\def\xre{x_{\rm re}}

\def\Png{\mathrm{P}_{\rm NG}}

\def\eerd{\eta_{\rm eRD}}
\def\tmPb{\tilde{\mP}_{\rm B}}
\def\xerd{x_{\rm eRD}}
\def\mGkerd{\mathcal{G}_{k}^{\rm eRD}}
\def\tmGkerd{\tilde{\mathcal{G}}_{k}^{\rm eRD}}
\def\tmGkrd{\tilde{\mathcal{G}}_{k}^{\rm RD}}
\def\mGkmd{\mathcal{G}_k^{\rm eMD}}
\def\tmGkmd{\tilde{\mathcal{G}}_k^{\rm eMD}}
\def\mPtsmd{\mathcal{P}^{\rm sec}_{\rm T,eMD}}
\def\mPtsrd{\mathcal{P}^{\rm sec}_{\rm T,RD}}
\def\mPti{\mathcal{P}_{\rm T,inf}^{\rm sec}}
\def\mPts{\mathcal{P}_{\rm T}^{\rm sec}}

\def\mIemd{\mathcal{I}_{\rm eMD}}

\def\ogwp{\Omega_{\rm gw}^{\rm pri}}
\def\omegara{\Omega_{\rm ra}}
\def\fre{f_{\rm re}}
\def\fpbh{f_{\rm pbh}}
\def\fe{f_{\rm end}}
\def\ogws{\Omega_{\rm gw}^{\rm sec}}
\def\npbh{n_{\rm PBH}}
\def\npbhi{n_{\rm PBH}^i}
\def\Qgw{Q_{\rm garv}}
\def\ogwevp{\Omega_{\rm gw,0}^{\rm evp}}
\def\gm{\mathrm{gm}}
\def\tf{t_{\rm f}}
\def\tev{t_{\rm ev}}

\def\ogwh{\Omega_{\rm gw }h^2}
\def\TBH{T_{\rm BH}}

\def\kc{k_c}
\def\gc{\alpha_{\rm c}}

\def\mPbi{\mathcal{P}_{\rm B}^{\rm inf}}
\def\mPei{\mathcal{P}_{\rm E}^{\rm inf}}

\def\rmAk{\mathrm{A}_{k}}
\def\rmBk{\mathrm{B}_k}

\graphicspath{{./figs/}}
\begin{document}

 \title{
 The Magnetic Origin of Primordial Black Holes:\\
Ultralight PBHs and Secondary GWs}
 \author{Subhasis Maiti}
\email{E-mail: subhashish@iitg.ac.in}
\affiliation{Department of Physics, Indian Institute of Technology, Guwahati, 
Assam, India}
\author{Debaprasad Maity}
\email{E-mail: debu@iitg.ac.in}
\affiliation{Department of Physics, Indian Institute of Technology, Guwahati, 
Assam, India}
\begin{abstract}
Ultralight primordial black holes (PBHs) provide a compelling window into early-Universe cosmology. 
Following our earlier work, we explore a mechanism for the formation of ultralight PBHs sourced by primordial inflationary magnetic fields, without invoking an ultra-slow-roll phase of inflation. We propose a magnetogenesis model in which large curvature perturbations are induced at small scales, leading to the efficient production of ultralight PBHs across a broad mass spectrum.
We analyze the phenomenological implications of these ultralight PBHs for early-Universe cosmology, particularly during reheating. 
We compute the resulting stochastic gravitational wave (GW) background generated by both the electromagnetic spectrum and evaporating PBHs, 
which exhibits distinctive features tied to the underlying magnetogenesis model parameters.
Our results demonstrate that inflationary magnetic fields can serve as a viable and testable origin for ultralight PBHs, opening new avenues for probing the interplay between inflation, magnetogenesis, PBHs, and primordial gravitational waves.
    
\end{abstract}
 
\maketitle
\section{Introduction}

Magnetic fields are ubiquitous throughout the Universe. Any conventional magnetohydrodynamic mechanism requires an appropriate seed magnetic field to explain the observed cosmic magnetic fields. One of the most well-known mechanisms for generating such seed fields is inflationary magnetogenesis, in which the conformal invariance of the electromagnetic sector is broken through non-conformal couplings of the form $f(\phi, R) F_{\mu\nu}F^{\mu\nu}$ \cite{Ratra:1991bn, PhysRevD.37.2743, PhysRevD.52.6694, PhysRevD.70.063502, Bamba:2008ja, Tripathy:2021sfb,Ng:2014lyb, Maiti:2025awl, Maiti:2025cbi} or $f(\phi, R) F_{\mu\nu}\tilde{F}^{\mu\nu}$ \cite{Campanelli:2008kh,Jain:2012jy,Caprini:2014mja,Sharma:2018kgs,Bamba:2021wyx, Ragavendra:2026fgs}. In conventional inflationary magnetogenesis models, the coupling function $f(\phi,R)$ is chosen such that the generated magnetic field attains sufficient present-day strength while simultaneously avoiding the strong coupling and backreaction problems \cite{Maiti:2025awl, Maiti:2025cbi}. 

In addition, axion-like fields naturally couple to gauge fields through the interaction term $\phi F_{\mu\nu}\tilde{F}^{\mu\nu}$, thereby breaking conformal invariance and leading to significant amplification of primordial magnetic fields. Several magnetogenesis scenarios based on such axion--photon couplings have been extensively studied in the literature~\cite{Corba:2024tfz, Peloso:2022ovc, Caprini:2017vnn, Figueroa:2023oxc, Barbon:2025wjl, Sharma:2024nfu, Klose:2022knn, Brandenburg:2024awd}, where $\phi$ is treated as an axion-like field. These models can lead to substantial production of helical magnetic fields and may also source secondary gravitational waves (GWs)~\cite{Barbon:2025wjl, Adshead:2019aac}. Furthermore, such amplified gauge fields can induce large curvature perturbations at small scales, potentially leading to the formation of ultralight primordial black holes (PBHs)~\cite{Bugaev:2013fya, Unal:2023srk, Erfani:2015rqv, Franciolini:2026cps, Barbon:2025wjl}.

Once generated, primordial magnetic fields undergo subsequent evolution and can leave observable imprints on a variety of cosmological probes, including the cosmic microwave background (CMB) \cite{Durrer:2013pga, Subramanian:2015lua, Planck:2015zrl, Paoletti:2008ck, Zucca:2016iur, Jedamzik:2020krr}, large-scale structure \cite{Sethi:2008eq, Kristiansen:2008tx}, and cosmic reionization \cite{Chluba:2015lpa}. Most studies of inflationary magnetogenesis have primarily focused on magnetic fields at large scales, where their amplitudes are strongly constrained by CMB B-mode polarization measurements \cite{Paoletti:2008ck, Planck:2015zrl, Zucca:2016iur, paoletti2022constraints, BICEP2:2017lpa} and recent blazar observations \cite{neronov2010evidence,Essey:2010nd,PhysRevD.98.083518}. Combining these observations suggests that the present-day magnetic field strength on scales of $\sim 1~{\rm Mpc}$ lies within the range $\sim 10^{-9}$--$10^{-19}$ Gauss~\cite{Planck:2015zrl, Paoletti:2008ck,PhysRevLett.116.191302}.

In contrast, the evolution of magnetic fields at small scales is considerably more complicated, as nonlinear plasma effects become important and depend sensitively on the microscopic properties of the surrounding medium. Consequently, inflationary magnetic fields at small scales remain comparatively weakly constrained. While most previous studies have concentrated on generating large-scale magnetic fields consistent with current observations, their possible impact on cosmological perturbations at small scales remains relatively unexplored.

The absence of strong observational constraints on small-scale magnetic fields motivates the search for scenarios capable of simultaneously generating observationally viable large-scale magnetic fields together with enhanced curvature perturbations at small scales, potentially leading to the formation of primordial black holes (PBHs) in the early Universe~\cite{Bugaev:2013fya, Unal:2023srk, Erfani:2015rqv, Franciolini:2026cps}.

The study of PBHs has attracted significant attention in recent years due to several compelling motivations. First, PBHs can constitute a viable dark matter candidate. Depending on their mass spectrum, they may account for a substantial fraction, or even the entirety, of the present-day dark matter abundance. Several mass windows, particularly in the asteroid-to-lunar mass range, remain weakly constrained by current observations and therefore continue to allow this possibility. Second, the evaporation of light PBHs through Hawking radiation has been invoked in a variety of scenarios addressing baryogenesis, dark radiation, and the origin of several cosmological backgrounds~\cite{Majumdar:1995yr, Baumann:2007yr, Hook:2014mla, Smyth:2021lkn, Datta:2020bht, Boudon:2020qpo, Morrison:2018xla, PhysRevD.59.041301, Bernal:2022pue, Schmitz:2023pfy, Borah:2024lml, Barman:2022pdo, RiajulHaque:2023cqe, Hamada:2016jnq, DeLuca:2022bjs, DeLuca:2021oer}. Third, PBHs formed in the early Universe can temporarily dominate the cosmic energy density and subsequently reheat the Universe through their evaporation, thereby providing suitable initial conditions for successful Big Bang Nucleosynthesis (BBN) \cite{Baumann:2007yr, RiajulHaque:2023cqe}. Moreover, heavier PBHs may act as seeds for the formation of supermassive black holes (SMBHs) observed at high redshifts \cite{Kawasaki:2012kn, Ziparo:2024nwh}, and may also play an important role in galaxy formation at later cosmological epochs \cite{Liu:2022bvr,Carr:2018rid}.

From a theoretical perspective, PBH formation is extremely sensitive to the amplitude and shape of the primordial curvature power spectrum. While CMB and large-scale structure observations tightly constrain the curvature spectrum on large scales, small-scale fluctuations remain largely unconstrained, thereby allowing the possibility of significant enhancement capable of triggering PBH formation~\cite{SDSS:2003eyi, Khatri:2013dha, Planck:2018jri, DES:2021wwk}. Consequently, inflationary scenarios involving an ultra-slow-roll (USR) phase have received considerable attention as a mechanism for PBH production. In such scenarios, the inflaton potential develops a plateau, an inflection point, or a localized bump~\cite{Garcia-Bellido:2017mdw, Motohashi:2017kbs, Byrnes:2018txb, Ballesteros:2018wlw, Raveendran:2022dtb, Ragavendra:2020sop, Garcia-Bellido:1996mdl, PhysRevD.103.083510, Solbi:2021wbo, Figueroa:2021zah, Frolovsky:2022qpg}, causing a temporary slowdown of the inflaton field. This transient phase amplifies superhorizon curvature perturbations and produces a sharp enhancement in the curvature power spectrum~\cite{Garcia-Bellido:2017mdw, Motohashi:2017kbs, Byrnes:2018txb, Ballesteros:2018wlw, Raveendran:2022dtb, Ragavendra:2020sop, Garcia-Bellido:1996mdl, PhysRevD.103.083510, Solbi:2021wbo, Figueroa:2021zah, Frolovsky:2022qpg}.

Other proposed mechanisms for PBH formation include bubble collisions during first-order phase transitions~\cite{PhysRevD.109.123030, PhysRevD.26.2681, Rubin:2001yw, Ai:2024cka} and the collapse of cosmic string loops~\cite{Hawking:1987bn, Polnarev:1988dh, PhysRevD.45.3447, Balaji:2025tun, Balaji:2024rvo, Balaji:2022rsy}. In addition, primordial magnetic fields can not only source curvature perturbations but may also directly collapse to form PBHs~\cite{Maiti:2025ijr,Kushwaha:2024zhd, Franciolini:2026cps}.

Besides the conventional enhancement of curvature perturbations through an ultra-slow-roll phase~\cite{Khlopov:1980mg, Bullock:1996at, Garcia-Bellido:1996mdl, Yokoyama:1998pt, Kawasaki:1997ju, Garcia-Bellido:2017mdw, PhysRevD.103.083510, Bhaumik:2020dor, Solbi:2021wbo, Figueroa:2021zah, Frolovsky:2022qpg, Bugaev:2013fya, Unal:2023srk, Erfani:2015rqv, Ozsoy:2023ryl}, the enhancement of the curvature power spectrum through secondary sources provides another interesting possibility that remains comparatively unexplored~\cite{Saga_2020,Kushwaha:2024zhd, Maiti:2025ijr}. In our earlier work~\cite{Maiti:2025ijr}, we demonstrated such a possibility within an inflationary magnetogenesis framework where the magnetic field acts as a secondary source of curvature perturbations. Adopting a well-studied saw-tooth-type gauge coupling function $f(\phi,R)$, we showed that the curvature power spectrum can be sufficiently enhanced to trigger the production of PBHs with masses in the range $\sim 10^{16}$--$10^{17}$ g, potentially constituting a viable dark matter candidate. Such a unified framework opens the possibility of constructing more realistic early-Universe scenarios that can be tested or constrained through multiple cosmological observations.

In the present work, we extend this investigation by considering the production of ultra-light PBHs and their associated gravitational wave signatures. Specifically, we adopt a well-studied helical inflationary magnetogenesis model~\cite{Sharma:2018kgs, Subramanian:2015lua, Tripathy:2021sfb} and demonstrate that it admits a viable parameter space capable of simultaneously generating large-scale magnetic fields free from strong coupling and backreaction problems, together with enhanced small-scale curvature perturbations that trigger the formation of ultra-light PBHs. We further investigate the resulting stochastic gravitational wave background generated by both the magnetic field and PBH evaporation, and discuss their potentially observable signatures in future experiments.

This paper is organized as follows. In Sec.~II, we discuss the inflationary dynamics within the $\alpha$-Attractor-E model and analyze the evolution of the first slow-roll parameter. In Sec.~III, we study the inflationary magnetogenesis scenario with a modified coupling function that generates non-helical magnetic fields at large scales and helical magnetic fields at small scales relevant for PBH formation. We also discuss the strong coupling and backreaction problems together with the dependence of the present-day magnetic field on the magnetogenesis parameters and reheating dynamics. In Sec.~IV, we investigate the generation of the curvature power spectrum from both vacuum inflaton fluctuations and perturbations sourced by the blue-tilted magnetic field. In Sec.~V, we analyze the conditions for PBH formation and compute the resulting PBH abundance. In Sec.~VI, we study the secondary gravitational wave background induced by the magnetic field and the associated curvature perturbations. In Sec.~VII, we present the main phenomenological results and discuss the allowed parameter space compatible with PBH formation and present-day magnetic field observations. Finally, in Sec.~VIII, we summarize our main conclusions and discuss their implications for inflationary magnetogenesis, PBH cosmology, and early-Universe physics.

\section{Inflation in brief }
Inflation is a phase where our universe expands exponentially, which not only solves the classical problems of Big-Bang cosmology, such as the horizon problem, flatness, and monopole problems, but also generates the primordial density perturbation that seeds the cosmic microwave background (CMB) anisotropies and the large-scale structure formation of the universe~\cite{Guth:1980zm, Linde:1981mu, Albrecht:1982wi, Starobinsky:1980te, PhysRevD.50.7222}. This accelerated expansion is typically driven by a scalar field known as the inflaton, whose dynamics in a Friedmann-Lemaitre-Robertson-Walker (FLRW) background, $ds^2 = -dt^2 + a(t)^2(dx^2 +dy^2 +dz^2)$, is~\cite{PhysRevD.50.7222, Cook:2015vqa}
\begin{align}
    \ddbphi +3H\dbphi+V'(\bphi)=0,\label{eq:phi_back}
\end{align}
where $\bphi$ is the background field which only depends on time. The Hubble parameter $H = \dot{a}/a$. Alongside the relevant Friedman equation governing the dynamics of the  cosmological scale factor during inflation is
\begin{align}
    H^2=\frac{1}{3\Mp^2}\l( \frac{1}{2}\dbphi^2+V(\phi)\r).\label{eq:hubble_inf}
\end{align}
Inflation occurs when the potential energy dominates over the kinetic energy, i.e., $V(\phi)>>\dbphi^2$, leading to an approximately constant Hubble parameter and an exponential growth of the scale factor $a(t)$. In the slow-roll approximation, we can always drop $\ddbphi$ there from Eq.\eqref{eq:phi_back}, and write it as $\dbphi\simeq-V_{,\bphi}(\phi)/3H$, and 
%as we have mention that during inflation the entire energy budget is came from the inflaton-potential, so we can write 
Eq.\eqref{eq:hubble_inf} as $H^2\simeq V(\phi)/2\Mp^2$. Utilizing these two approximations, we can write the inflaton field velocity as $\dbphi^2\simeq 2\epsilon\Mp^2H^2$, where $\epsilon$ is the 1'st slow-roll parameter defined as~,
 $\epsilon =({\Mp^2}/{2})\l({V'
    }/{V}\r)^2$~\cite{PhysRevD.28.679}.
\begin{figure}[t]
\centering
\includegraphics[scale=0.424]{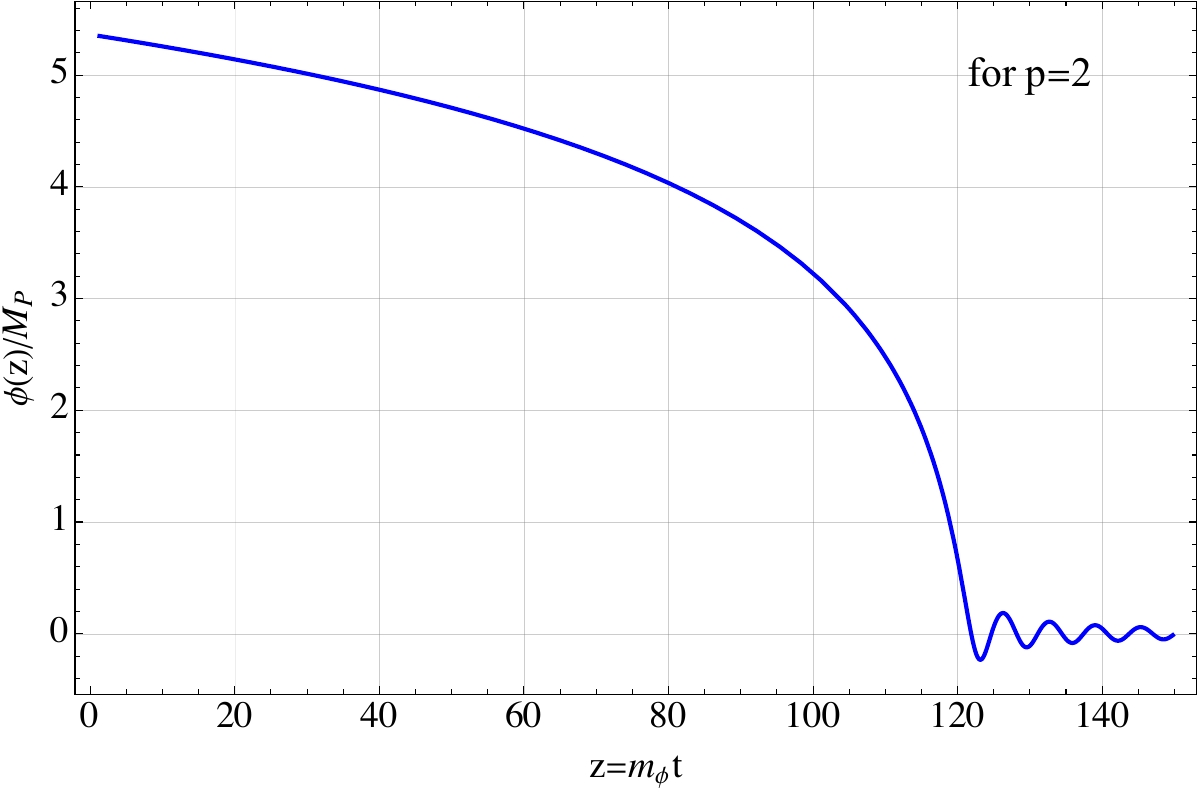}
\includegraphics[scale=0.45]{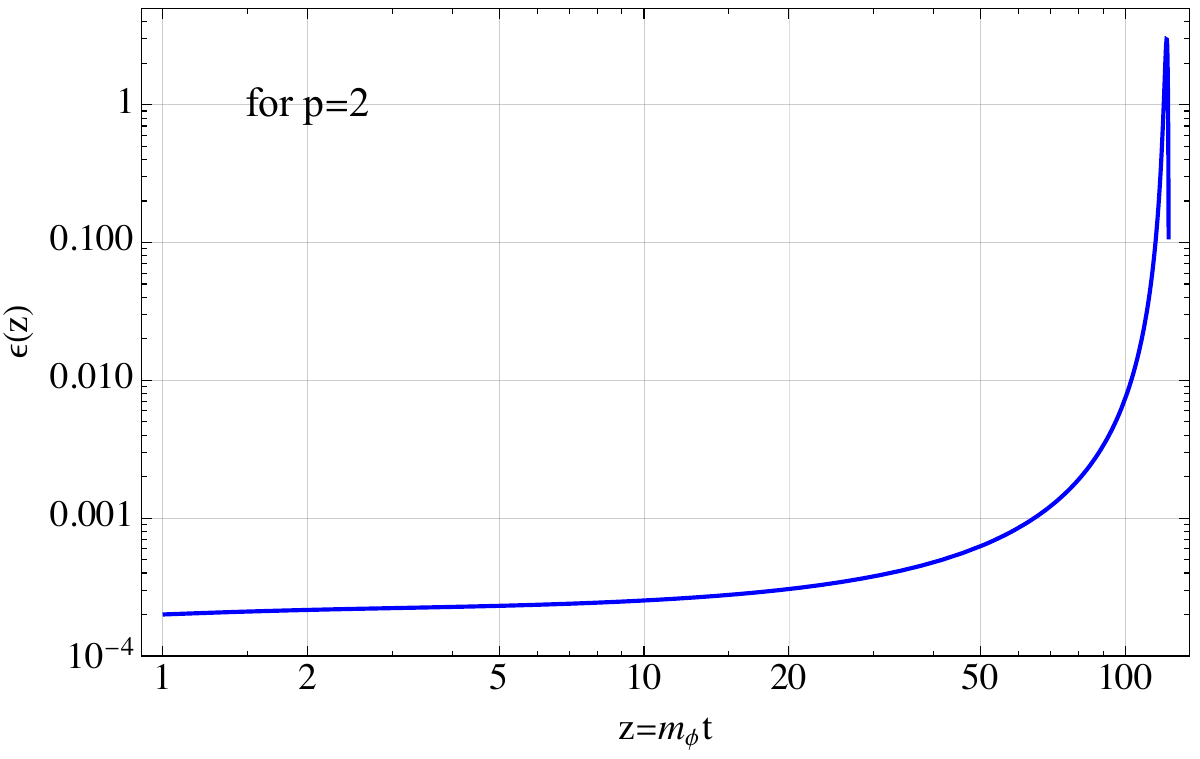}
\caption{In the above figures, we illustrate the time evolution of the inflaton field for the \(\alpha\)-attractor model with $p=2$. In the right panel, we show how the first slow-roll parameter $\epsilon$ evolves with time $(z=m_\phi t)$ during inflation.
}
\label{fig:phi_evo}
\end{figure}
The slow-roll parameter plays a central role in inflationary dynamics, as it determines the duration of the inflationary phase of the universe. In general, inflation persists as long as $\epsilon \ll 1$, and it terminates once $\epsilon \simeq 1$. 

Beyond the homogeneous evolution of the inflation field and background, quantum fluctuations of both become crucial in the context of inflationary density perturbations that will source the large-scale structure formation. These fluctuations initially generated on sub-horizon scales are stretched to super-horizon scales by the rapid expansion. The EoM for the inflaton perturbation $\phi(\vx,t)=\bphi(t)+\delta\phi(\vx,t)$ satisfies 
%the Klein-Gordan equation,
%for $\delta\phi(\vx,t)$ in a spatially flat FLRW background is
\begin{align}
    \ddot{\delta\phi}(\vx,t)+3H\dot{\delta\phi}(\vx,t)-\frac{1}{a^2}\nabla^2\delta\phi(\vx,t)+V''(\bphi)\delta\phi(\vx,t)=0
\end{align}
where $V''(\bphi)=\partial^2V(\bphi)/\partial\bphi^2$. To exactly know the behaviour of the fluctuation field, we need to dynamically solve the coupled equations, but for a simple de-sitter approximation, in Fourier space, the solution of the inflation field fluctuations is well known 
\begin{align}\label{eq:sol_delta_phik}
    \delta\phik(\eta)\simeq \frac{\HI}{\sqrt{2k^3}}(1-ik\eta)e^{-ik\eta}
\end{align}
where in de Sitter inflationary background $\HI$ is the constant Hubble throughout the inflationary evolution. Here, $\delta\phik(\eta)$ is the Fourier mode of the inflaton fluctuation $\delta \phi(x,\eta)$, where $\eta$ is the conformal time $\d\eta =dt/a(t)$.

Now, for the time being, to solve the dynamical equations, we consider a simple, well-known inflationary model, known as the $\alpha-$Attractor E-model with the potential \cite{Kallosh:2013yoa}
\begin{align}\label{eq:potential_V}
    V(\phi)=\Lambda^4\l(1- \exp\l[-\sqrt{\frac{2}{3\alpha}}\frac{\phi}{\Mp}\r]\r)^{2p} .
\end{align}
Here $\Mp=1/\sqrt{8\pi\rm{G}}\simeq 2.435\times 10^{18}\,\GeV$ is the reduced \textit{Plank} mass. With the appropriate initial conditions at the beginning of inflation, Eq.\eqref{eq:phi_back} can be solved %t the biginning of inflation, we need to set the initial condition for $\bphi(\eta=\eta_i)$ and the initial velocity for the inflation field i.e. $\dbphi(\eta=\eta_i)$ 
in terms of model parameters $p$ and $\alpha$ along with the scalar spectral index $\ns$ as
\begin{align}\label{phik-f}
    \bphikpv
    =\sqrt{\frac{3\alpha}{2}}\Mp \ln \l[ 1+\frac{4p+2\sqrt{4p^2+6p\alpha(1+p)(1-\ns)}}{3\alpha(1-\ns)} \r]
\end{align}
and the initial velocity of the background inflation field is $\dbphikpv\simeq \sqrt{2\epsilon_{\kpv}}\Hkpv\Mp$,
where $\Hkpv=\pi\Mp(r_{\kpv}\As/2)^{1/2}$ is the value of the Hubble constant when the pivot scale left the horizon during inflation. Here $r_{\kpv}$ is the tensor-to-scalar ratio produced due to inflation field fluctuations at the pivot scale $\kpv=0.05\,\Mpc^{-1}$. If all curvature fluctuations originate from fluctuations in the inflaton field, then, to get the scalar amplitude of the curvature perturbation $A_s\simeq 2.1\times 10^{-9}$~\cite{Planck:2018jri} for the inflationary Hubble constant $\HI\simeq 10^{-5}$, the predicted slow-roll parameter $\epsilon_{\kpv}\simeq 2\times 10^{-4}$. We can set the initial parameters of the models and solve the EoM of the inflation field.

In Fig.\ref{fig:phi_evo}, we show the evolution of the inflaton field during inflation for a representative inflationary scenario, namely the $\alpha$–attractor E-model defined in Eq.\eqref{eq:potential_V}. For illustration, we fix the parameter to $p=2$, which leads to an effective post-inflationary equation of state $w_\phi = 1/3$, which we consider through out the present paper. The right panel of Fig.\ref{fig:phi_evo} displays the corresponding evolution of the first slow-roll parameter $\epsilon$, as a function of time $z = m_\phi t$ for the same choice $p=2$. Here $m^2_\phi = {V''(\bar{\phi})}$ is the effective mass of the inflaton field. Inflation ends when $\epsilon \simeq 1$, after which it oscillates around unity, as can be seen in the Fig.\ref{fig:phi_evo}. 
%The behavior of $\epsilon$ plays a central role in determining the spectrum of curvature perturbations induced by the magnetic field. 
Although we have illustrated this using a specific $\alpha$–attractor potential, most standard slow-roll inflationary models predict a qualitatively similar evolution of the first slow-roll parameter: $\epsilon$ remains nearly constant at the beginning of the inflation, then rapidly grows to unity at the end of the inflationary era. However, if one considers an inflationary model in which the evolution of $\epsilon$ departs significantly from this generic behavior, the resulting induced curvature spectrum can be modified accordingly.

\section{Inflationary Magnetic Fields (IMFs)}

The observation of large-scale magnetic fields in cosmic voids and the intergalactic medium (IGM) strongly suggests its primordial origin, potentially tracing back to the inflationary epoch. However, within the framework of standard Maxwell theory in a conformally flat FLRW background, electromagnetic fields do not experience amplification due to conformal invariance. Consequently, the generation of significant magnetic fields during inflation requires the introduction of an explicit coupling that breaks the conformal invariance of the gauge sector.

A wide range of models addressing this issue have been proposed in the literature~\cite{Guth:1980zm, Linde:1981mu, Albrecht:1982wi, Starobinsky:1980te, Kandus:2010nw, Durrer:2013pga, Ferreira:2013sqa, Subramanian:2015lua, Kobayashi:2014sga, Haque:2020bip, Tripathy:2021sfb, Li:2022yqb, Adshead:2016iae, Maiti:2025cbi, PhysRevD.94.043523, Sharma:2017eps, Papanikolaou:2024cwr, Hortua:2014wna, Maiti:2025rkn, Maiti:2025awl, Martin:2007ue}, demonstrating that substantially large-scale magnetic fields can indeed be generated during inflation through suitable modifications of the gauge sector. In this work, we mainly focus on the helical magnetic field produced during inflation. The action for the gauge field is given by
\begin{align}
    \mSem=-\int d^4x \sqrt{-g}\,\frac{I^2(\phi,R)}{4}\l[ F_{\mu\nu}F^{\mu\nu}+\gamma F_{\mu\nu}\tilde{F}^{\mu\nu}\r].
\end{align}
Here, $F_{\mu\nu}=\partial_\mu A_{\nu}-\partial_{\nu}A_{\mu}$ denotes the field strength tensor associated with the gauge field $A_{\mu}$, while $\tilde{F}^{\mu\nu}=\frac{1}{2}\epsilon^{\mu\nu\alpha\beta}F_{\alpha\beta}$ represents its dual, with $\epsilon^{\mu\nu\alpha\beta}=\frac{1}{\sqrt{-g}}\eta^{\mu\nu\alpha\beta}$. The parity-violating term proportional to $\gamma$ can lead to the generation of helical magnetic fields, which are of particular interest due to their enhanced stability and potential observational signatures.

Keeping in mind the well-known backreaction and strong-coupling problems, we adopt a sawtooth-type magnetogenesis scenario during inflation~\cite{Ferreira:2013sqa, Cecchini:2023bqu}. Furthermore, observational constraints, particularly the non-detection of primordial tensor perturbations at cosmic microwave background (CMB) scales, impose stringent bounds on the tensor-to-scalar ratio which yields tight constraints on the possible large scale magnetic field production. Furthermore, we aim to produce ultralight PBHs 
%During a de Sitter inflationary phase, a blue-tilted magnetic spectrum leads to both primary and secondary tensor power spectra that are nearly scale-invariant on superhorizon scales. However, when the secondary tensor modes are sourced by electromagnetic fields, their amplitude is directly determined by the strength of the generated gauge fields~\cite{Teuscher:2025xke, Teuscher:2025jhq}.
potentially contributing to the reheating of the Universe without requiring an ultra-slow-roll phase during inflation. Taking into account these important effects, we consider the following phenomenological functional form for the coupling function:
\begin{align}\label{eq:coupling_function}
    I(\eta)=\l\{ 
    \begin{matrix}
        \mathrm{C} & \text{for}~ \eta < \eta_c\\
        (a(\eta)/\ae)^n & ~~~~~~~~~~~\text{for}~\eta_c<\eta<\ee\\
        1 & ~~\text{for}~\eta>\ee
    \end{matrix}
    \r.
\end{align}
Here $\ae=-1/\HI\ee$ is the scale factor defined at the end of inflation, where $\ee$ is the conformal time defined at the end of inflation. While the type of coupling $I(\eta)$ we assumed changes the canonical kinetic term of the gauge field, it effectively renormalizes the coupling of all the electrically charged particles through $e_{\text{phy}}=e/I(\eta)$, where `$e$' is the bare charge. Therefore, $I(\eta)<<1$, at the beginning, the effective gauge coupling became large, i.e., $e_{\text{phy}}>>1$. This means that interaction involving charged particles becomes non-perturbative, and the theory enters into a strong-coupling regime during inflation. 
%As a result, any calculations based on perturbative analysis become unreliable, undermining the predictions of the model. To avoid such an issue, we work in the regime $n<0$. 
Therefore, to solve the strong coupling problem the constant $C=(a(\eta_c)/a(\ee))^{n}>>1$. However, for phenomenological purpose we keep our discussion for both possibilities of $n$. In a de Sitter inflationary background, the secondary tensor power spectrum is approximately scale-invariant, with its amplitude directly determined by the strength of the generated electric and magnetic fields. To avoid excessive production of tensor modes at Cosmic Microwave Background (CMB) scales, we impose the condition $\eta_c\gg\eta_{k_*}$, where $\eta_{k_*}$ corresponds to the conformal time at which the pivot scale exits the horizon during inflation. This choice effectively suppresses unwanted tensor contributions on CMB scales while allowing significant production at smaller scales.

Assuming spatial flatness, the vector potential can be decomposed into irreducible components as 
$A_\mu=\left(A_0,\partial_i S + A_i\right)$, subject to the transverse condition 
$\delta^{ij}\partial_i A_j=0$. In Fourier space, the gauge field can be expanded as
\begin{align}
    A_i(\vx,\eta)=\sum_{\lambda=\pm}\int\frac{d^3\vk}{(2\pi)^{3/2}}\epsilon_i^{\lambda}(\vk)e^{i \vk\cdot\vx}\uk(\eta),
\end{align}
with the reality condition $u_{-\vk}^{\lambda} = u_{\vk}^{\lambda *}$. Here, $\epsilon_i^{\lambda}(\vk)$ denotes the polarization vector corresponding to the polarization state $\lambda=\pm$, satisfying $\epsilon_i^{\lambda}(\vk)\vk^i=0$ and $\delta^{ij}\epsilon_i^{\lambda}(\vk)\epsilon_j^{\lambda'}(\vk)=\delta^{\lambda\lambda'}$.

In a flat FLRW background, the equation of motion for the mode function $\uk$ is given by~\cite{Campanelli:2008kh, Jain:2012jy, Caprini:2014mja, Sharma:2018kgs, Bamba:2021wyx}
\begin{align}\label{eq:uk_gen}
    {\uk}''+2\frac{I'(\eta)}{I(\eta)}{\uk}'+\l(2\gamma\lambda k\frac{I'(\eta)}{I(\eta)}+k^2\r)\uk=0.
\end{align}

The coupling function $I(\eta)$ evolves differently across conformal time. For $\eta>\eta_c$, it remains constant ($I'=0$), and no excitation of the gauge field occurs. In contrast, for $\eta<\eta_c$, its time dependence sources the gauge field, leading to mode excitation.

For convenience, we define the rescaled variable $\mVk = I(\eta)\uk$. In terms of this variable, Eq.~\eqref{eq:uk_gen} becomes
\begin{align}\label{eq:mVk_inf}
    {\mVk}''+\l( k^2+2\lambda\gamma k\frac{I'}{I}-\frac{I''}{I}\r)\mVk=0.
\end{align}

\paragraph{\underline{Sub-horizon regime:}}
When modes are deep inside the horizon, the condition
$k^2 \gg \lambda\gamma k \frac{I'}{I},\, I''/I$
holds, and subleading terms can be neglected. The equation reduces to
\begin{align}
    {\mVk}''+k^2\mVk=0.
\end{align}
In the asymptotic limit $|-k\eta|\rightarrow \infty$, the Bunch--Davies vacuum solution is
${\mVk}_{,\rm BD}(k)=\frac{e^{-ik\eta}}{\sqrt{2k}}$.

Even after horizon crossing, the excitation of modes depends on the characteristic time $\eta_c$. Modes exiting the horizon at $\eta<\eta_c$ remain in the vacuum state, whereas for $\eta>\eta_c$, both newly exiting modes and those already outside the horizon become excited due to the time-dependent coupling.
Modes exiting before $\eta_c$ remain non-helical, while those exiting after $\eta_c$ acquire helicity for $\gamma\neq 0$. 

To analyze the behaviour of the mode function, we define a dimensionless variable $x=-k\eta$, and rewrite the above Eq.\eqref{eq:mVk_inf} in terms of this new variable, we get 
\begin{align}\label{eq:mVk_inf_s}
    \frac{\partial^2\mVk}{\partial x^2}+\l(1-2\lambda\gamma\frac{n}{x}-\frac{n(n+1)}{x^2}\r)\mVk=0.
\end{align}
The late-time activation of the coupling enhances all super-horizon modes. To analyze this, we divide modes into two regimes: $k<\kc$ and $\kc<k<\ke$. Modes with $k<\kc$ exit the horizon before $\eta_c$ and are later affected by the coupling, while modes in $\kc<k<\ke$ are excited near horizon crossing.

\paragraph{\underline{solution for the super horizon modes $k<\kc$}:}
In the super-horizon limit ($x\ll 1$) for those modes which are already outside the horizon, we can simplify the above Eq.\eqref{eq:mVk_inf_s} as
\begin{align}
  \frac{\partial^2\mVk}{\partial x^2}-\frac{n(n+1)}{x^2}\mVk=0.  
\end{align}
In this regime, both polarization states are equally excited, implying non-helical behavior. Therefore, we can write the solution of the above equation as
\begin{align}
    \mVk(k<\kc,\eta>\eta_c)=\rmAk \eta^{n+1}+\rmBk \eta^{-n}.
\end{align}
For $n<0$, the term $\rmAk \eta^{n+1}$ grows as $\eta\to 0$, while $\rmBk \eta^{-n}$ represents the decaying mode. And for $n>0$ it is reversed.

Imposing continuity at $\eta=\eta_c$, we obtain
\begin{align}
    \rmAk(k<\kc,\eta_c)\simeq \frac{1}{\sqrt{2k}} \frac{n-ik\eta_c}{2n+1}\eta_c^{-(n+1)}, \quad
    \rmBk(k<\kc,\eta_c)\simeq \frac{1}{\sqrt{2k}} \frac{n+1+ik\eta_c}{2n+1}\eta_c^n.
\end{align}
Oscillatory factors $e^{-ik\eta}$ are neglected since these modes are already super-horizon ($-k\eta\ll 1$) and effectively frozen.

\begin{figure}[h]
\centering
\includegraphics[scale=0.45]{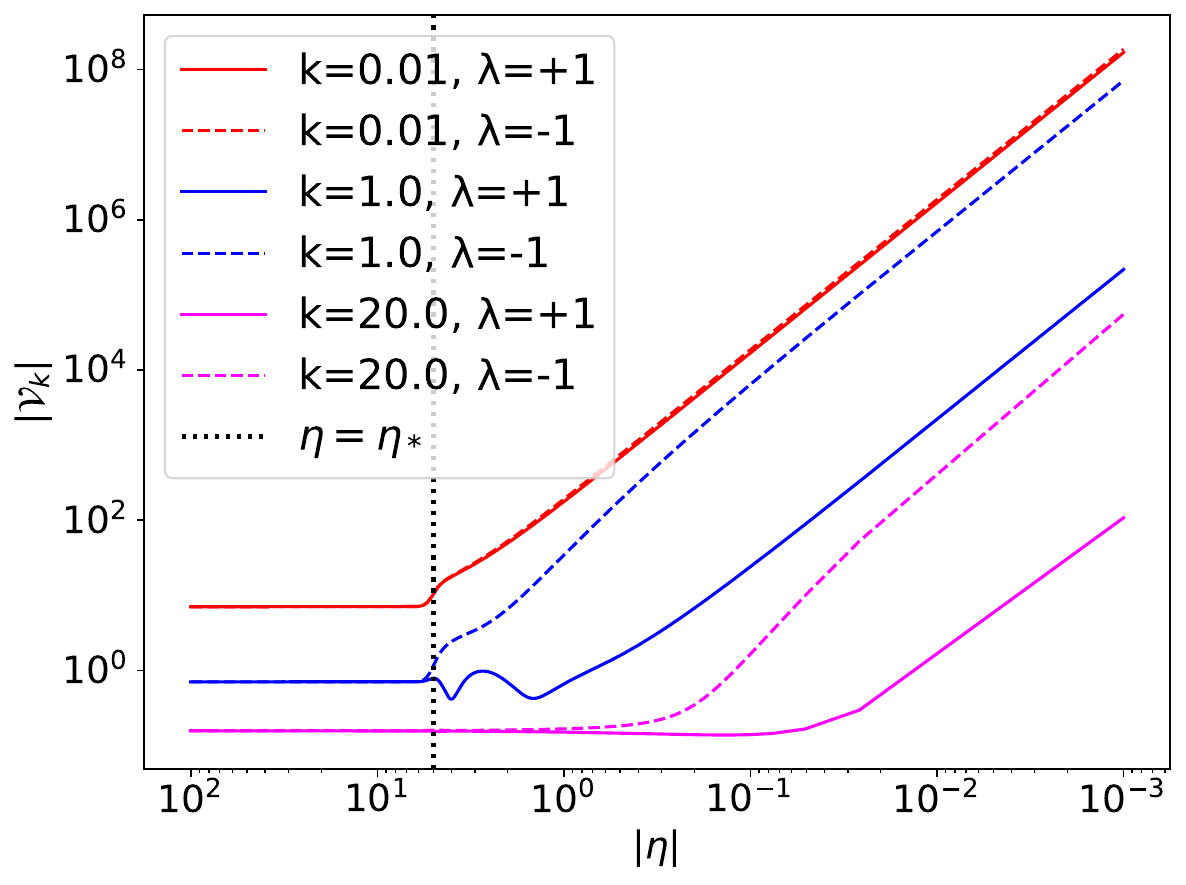}
\caption{We plot the time evolution of the mode function $\mVk$ as a function of the normalized conformal time $|\eta|$. The normalization is chosen such that the mode with $k=1.0$ exits the horizon at $|\eta|=1.0$. We consider three representative wavenumbers, $k=0.01$, $k=1.0$, and $k=20$. The mode with $k=0.01$ is already outside the horizon when the coupling function becomes active, whereas the modes with $k=1.0$ and $k=20$ exit the horizon after the coupling is activated.
The solid and dashed lines represent the two helicity modes, corresponding to $\lambda=+$ and $\lambda=-$, respectively. The vertical black dashed line indicates the normalized conformal time at which the coupling becomes active.}
\label{fig:gauge_field}
\end{figure}

Using these results, the magnetic and electric energy densities per logarithmic interval are
\begin{align}\label{eq:mpb_kc}
    \mPbi(k<\kc,\ee)\simeq \frac{\HI^4}{4\pi^2}\frac{n^2+k^2\eta_c^2}{(1-2|n|)^2}\l(\frac{\kc}{\ke}\r)^{2(1-|n|)}\l(\frac{k}{\ke}\r)^{4},\\
    \mPei(k<\kc,\ee)\simeq \frac{\HI^4}{4\pi^2}(n^2+k^2\eta_c^2)\l(\frac{\kc}{\ke}\r)^{2(1-|n|)}\l(\frac{k}{\ke}\r)^{2}.
\end{align}

These expressions show that long-wavelength modes are enhanced even after horizon exit due to the late-time activation of the coupling. 
Here we have found that the magnetic field spectrum goes as $\mPb(k<\kc,\ee)\propto k^{4}$ whereas the electric field spectrum goes as $\mPe(k<\kc,\ee)\propto k^{2}$. This implies that modes that are outside the horizon when the coupling function becomes active are always blue-tilted.  

We address the issue of strong coupling and backreaction by restricting our analysis to the parameter range $-2.2<n<0$. Within this regime, the generated magnetic fields are always blue-tilted. Therefore, to explain large-scale magnetic fields, a non-trivial evolution during the reheating phase is required, as discussed in \cite{Maiti:2025cbi}. In the absence of post-inflationary enhancement, it is not possible to generate a sufficiently large amplitude of magnetic fields on large scales consistent with present-day observational constraints.

%A shorter activation period of the coupling function reduces the enhancement, not only if we consider a non-standard reheating scenario, but it will also impact the present-day magnetic field strength~\cite{Maiti:2025awl}.

\paragraph{Solution for $\kc<k<\ke$:}
For these modes, the equation of motion is given by Eq.~\eqref{eq:uk_gen}. Defining $\mVk=I(\eta)\uk$ and introducing the dimensionless variable $z=2ik\eta$ along with $\kappa=in\gamma\lambda$, the equation can be recast as~\cite{Sharma:2018kgs}
\begin{align}
    \ddmVk(z)+\left[\frac{1}{z^2}\left(\frac{1}{4}-\left(n+\frac{1}{2}\right)^2\right)+\frac{\kappa}{z}-\frac{1}{4}\right]\mVk(z)=0 .
\end{align}
Here, the dot denotes differentiation with respect to $z$, i.e., $\ddmVk=\partial^2\mVk/\partial z^2$.

To fix the initial condition, we assume that at the onset of inflation all relevant modes are deeply subhorizon and reside in the Bunch--Davies vacuum. Accordingly, in the asymptotic limit $|-k\eta|\rightarrow \infty$, the modified mode function behaves as ${\mVk}_{,\rm BD}(k)=e^{-ik\eta}/\sqrt{2k}$. Imposing this condition, the solution for the mode function is obtained as
\begin{align}\label{sol_mode}
    \mVk(\eta)=\frac{1}{\sqrt{2k}}e^{i\pi\kappa}W_{\kappa,\left(n+{1}/{2}\right)}(2ik\eta) .
\end{align}
where $W_{\kappa,n+1/2}(z)$ denotes the \text{Whittaker} function.

By the end of inflation, the modes of interest lie well outside the horizon. Using the above solution, the electric and magnetic power spectra in the super-horizon limit $|-k\ee|<<1$ (see~\cite{Tripathy:2021sfb} for details) are given by
\begin{subequations}\label{eq:mPbe_inf}
   \begin{align}
    \mPe(k>\kc,\ee) &=\frac{\partial\rhob(k,\ee)}{\partial\ln(k)}=\HI^4\,\mcB(n,\gamma)\l(\frac{k}{\ke}\r)^{\nb}\\
    \mPb(k>\kc,\ee) &=\frac{\partial\rhoe(k,\ee)}{\partial\ln(k)}=\HI^4\, \mcB(n,\gamma)\l(\frac{k}{\ke}\r)^{\nb}\l\{\gamma^2+\frac{(1-2|n|\gamma^2)^2}{(1-2|n|)^2}\l(\frac{k}{\ke}\r)^2\r\}
\end{align} 
\end{subequations}
where $\kc$ corresponds to the comoving wave number that exits the horizon at the characteristic time $\eta_c$. These expressions are consistent with the expected power-law for the power-law type behaviour of the coupling function~\cite{Tripathy:2021sfb, Subramanian:2015lua}.
Here, the magnetic spectral index is defined as $\nb=4-2|n|$, and the function $\mcB(n,\gamma)$ is given by
\begin{align}\label{eq:fn}
    \mcB(n,\gamma)=\frac{1}{4\pi^2}\frac{\cosh(n\pi\gamma)}{(1+\gamma^2)^2}\frac{\Gamma^2(2|n|)}{2^{2|n|-2}\Gamma(|n|-i\gamma|n|)}
\end{align}

For a non-vanishing helical coupling ($\gamma\neq 0$), both the electric and magnetic fields exhibit the same spectral scaling, differing only in amplitude on super-horizon scales $k<\ke$. In particular, their ratio is approximately given by $\mPb(k,\ee)/\mPe(k,\ee)\simeq \gamma^2$. For $\gamma>1$, the magnetic field dominates, although both fields remain of comparable order. In the limit $\gamma=0$, the spectrum reduces to the well-known non-helical case~\cite{Subramanian:2015lua, Tripathy:2021sfb, Maiti:2025cbi}.

In this magnetogenesis model, large-scale magnetic fields can be generated. However, once the strong coupling issue is properly addressed, the resulting magnetic field spectrum becomes strongly blue-tilted. Consequently, explaining the observed large-scale magnetic fields requires additional dynamics, such as mechanisms that convert electrical energy density into magnetic energy density during reheating in the limit of vanishing electrical conductivity, as discussed in \cite{Kobayashi:2019uqs, Maiti:2025cbi, Haque:2020bip}. 

On the other hand, if the strong coupling issue is ignored, couplings with $n>1/2$ can be considered. In such cases, a nearly scale-invariant magnetic field spectrum can be generated at CMB scales for $n\sim 2$, which can naturally account for the observed large-scale magnetic field strength~\cite{Tripathy:2021sfb, Ragavendra:2026fgs, Maiti:2025cbi}.

\paragraph{\underline{Constraint due to Backreaction}}

It is important to note that the modes that exit the horizon before the coupling function becomes active, i.e., $k<\kc$, always exhibit a blue-tilted spectrum irrespective of the sign of the parameter $n$. For case $n>-2.0$, the modes generated after the characteristic time $\eta_c$, namely those satisfying $k>\kc$, obey blue tilted spectrum, with the maximum electromagnetic (EM) energy density stored near $k\simeq\ke$.

In contrast, when $n<-2.0$, the modes generated after $\eta_c$ ($k>\kc$) acquire a red-tilted spectrum, whereas the modes satisfying $k<\kc$ remain blue tilted. In this case, the dominant contribution to the magnetic energy density is concentrated around the characteristic scale $\kc$.

To ensure the validity of perturbation theory, we restrict our analysis to the regime where the total energy density of the generated EM field always remains subdominant compared to the inflaton energy density, namely $\rhoem<\rho_c=3\HI^2\Mp^2$. The total EM energy density at the end of inflation can then be written as
\begin{align}
    \rhoem(\ee)=\int_{\kpv}^{\ke}\d \ln(k)\{\mPb(k,\ee)+\mPe(k,\ee)\}
    \simeq \HI^4
    \l\{
    \begin{matrix}
    \frac{(n^2+1)}{2}\l(\frac{\kc}{\ke}\r)^{\nb} & \nb<0\\
       \frac{ \mcB(\nb)}{\nb} \l(1+\gamma^2+\frac{(1-2|n|\gamma^2)^2}{(1-2|n|)^2}\r) & \nb\geq 0
    \end{matrix}
    \r.
\end{align}

At present, the largest observable wavelength corresponds to the CMB pivot scale, and therefore we set the infrared cutoff at $\kpv$. On the other hand, in the regime $-2\leq n<2$, both the magnetic and electric spectra are blue tilted. To regulate the corresponding ultraviolet divergence, we introduce a UV cutoff scale $k_{\rm UV}\simeq\ke$, where $\ke$ denotes the largest mode that exited the horizon before the end of inflation.

Modes with $k>\ke$ remain inside the horizon throughout inflation and therefore stay in the Bunch-Davies vacuum state. Moreover, within the framework of UV renormalization, the energy density associated with subhorizon modes decays exponentially at large $k$ values (see~\cite{Kolb:2023ydq}).

\begin{figure}[h]
\centering
\includegraphics[scale=0.45]{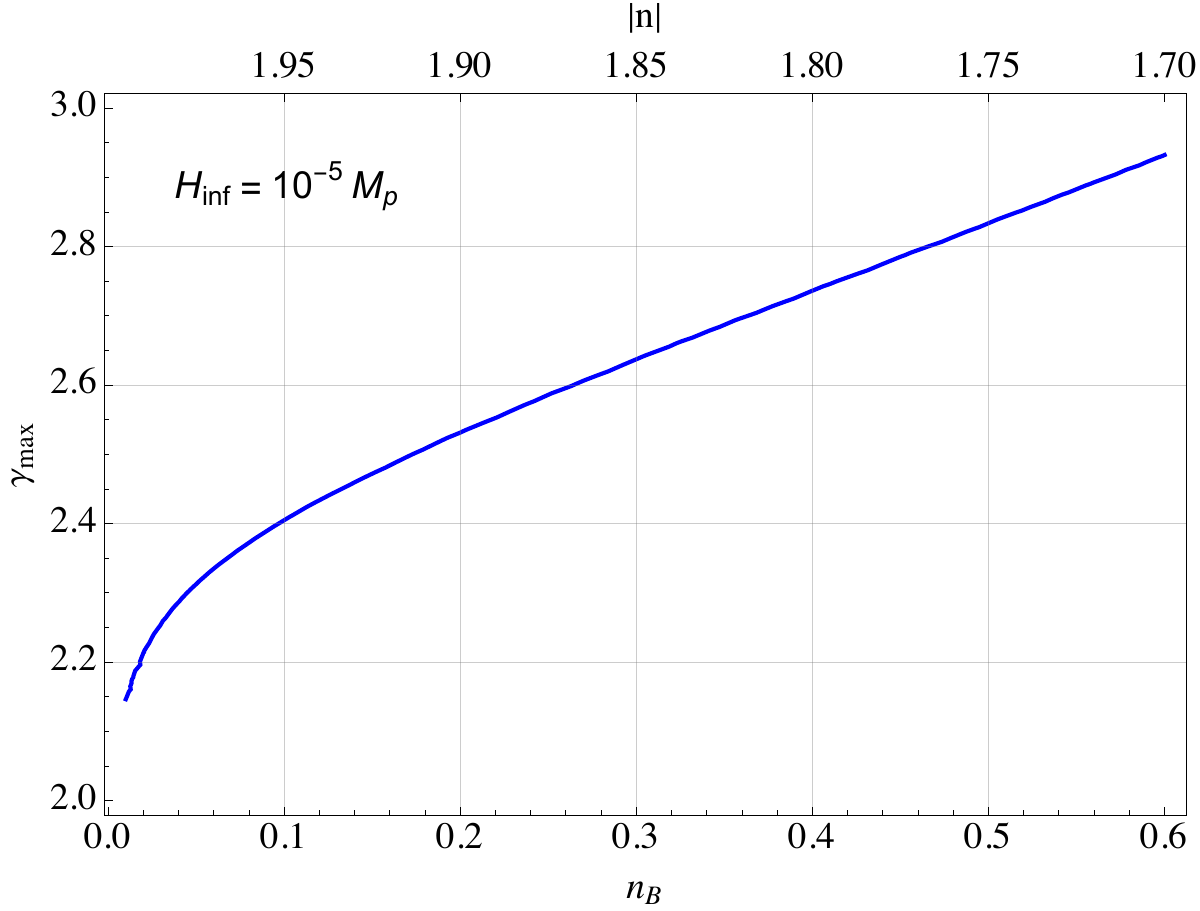}
\caption{In this figure, we plot the maximum allowed value of the coupling parameter, \(\gamma_{\rm max}\), as a function of the magnetic spectral index \(\nb\). The curve represents the boundary above which the generated electromagnetic energy density exceeds the inflaton energy density, i.e., $\rhoem>\rhop$.}
\label{fig:mag_b}
\end{figure}

To avoid the backreaction problem, we impose the condition that the total electromagnetic energy density remains smaller than the inflaton energy density, namely $\rhop(\ee)\geq\rhoem(\ee)$. We quantify the fractional EM energy density through the ratio $\delta_{\rm em}(\ee)=\rhoem(\ee)/\rhop(\ee)$, which can be expressed as
\begin{align}
    \delta_{\rm em}(\ee)\simeq \l(\frac{\HI}{\Mp}\r)^2\times \l\{
    \begin{matrix}
         \frac{n^2+1}{2}\l(\frac{\kc}{\ke}\r)^{\nb} & \nb<0\\
         \frac{\mcB(n,\gamma)}{3\nb}\l(1+\gamma^2+\frac{(1-2|n|\gamma^2)^2}{(1-2|n|)^2}\r) & \nb\geq 0
    \end{matrix}
   \r.
\end{align}

If the fractional EM energy density satisfies $\delta_{\rm em}(\eta)\simeq 1$, the electromagnetic energy density becomes comparable to the background energy density and consequently modifies the inflationary background dynamics. From the above expressions, it is evident that for $\nb<0$ (equivalently $|n|>2.0$), the total EM energy density depends only on the magnetogenesis parameter $n$ and the characteristic time $\eta_c$, or equivalently the characteristic wavenumber $\kc$. In contrast, for $\nb\geq0$, the generated EM energy density depends not only on the magnetogenesis parameter $n$ but also on the helicity parameter $\gamma$.

For an inflationary energy scale $\HI \simeq 10^{-5}\,\Mp$, in order to avoid significant backreaction, we find the constraint $\mcB(n,\gamma) \leq {10^{10}}/{(1+\gamma^2)}$.
In Fig.~\ref{fig:mag_b}, we show the maximum allowed value of the coupling parameter $\gamma$, denoted as $\gamma_{\rm max}$, as a function of the magnetic spectral index $\nb$, assuming $\HI \simeq 10^{-5}\,\Mp$. This constraint ensures that the magnetic field energy density does not exceed the background energy density, i.e., $\delta_{\rm em} < 1$. 
As an example, if we consider $|n| = 1.9$ (corresponding to $\nb = 0.2$), we obtain $\gamma_{\rm max} \simeq 2.52$. Similarly, for $n = 1.85$ (corresponding to $\nb = 0.3$), the bound becomes $\gamma_{\rm max} \simeq 2.6$.    

Since the coupling exhibits non-trivial evolution, the spectral behavior of different wavenumbers depends on the characteristic mode $k=k_c$ associated with the time scale $\eta =\eta_c$, which determines when the coupling function becomes active. As mentioned earlier, large-scale modes $k < k_c$ that cross the horizon before the characteristic time, i.e., $\eta<\eta_c$, remain non-helical in nature. In contrast, modes that exit the horizon after $\eta >\eta_c$, acquire helicity for $\gamma\neq 0$. In Fig.\ref{fig:gauge_field}, we plot the evolution of the mode function $\mVk$ to examine its behavior over time, particularly in our scenario where the coupling becomes active only after a certain time, i.e., $\eta>\eta_c$. 
The normalization is chosen such that the mode with $k=1.0$ exits the horizon at $\eta=-1.0$. With respect to this normalization, we set the characteristic time to $\eta_*=-5.0$.

With this setup, the modes with $k=1.0$ and $k=20$ exit the horizon after the coupling function becomes active, whereas the mode with $k=0.01$ is already outside the horizon when the coupling is activated. As a result, we observe that for $k=0.01$, both helicity modes have identical amplitudes, indicating that these modes are non-helical in nature. In contrast, for the modes $k=1.0$ and $k=20$, which exit the horizon after the coupling becomes active, the positive and negative helicity modes exhibit different growth rates. This implies that modes which are inside the horizon at $\eta>\eta_*$ acquire helicity. Furthermore, their amplitudes depend on both the coupling parameter $n$ and the helicity parameter $\gamma$, as shown in Eq.\eqref{eq:mPbe_inf}.

Thus, this class of magnetogenesis models possesses a distinctive feature: large-scale modes remain non-helical, whereas smaller-scale modes become helical when $\gamma\neq 0$.

%\color{blue}
\paragraph{\it Post-inflationary evolution of the large-scale magnetic field:}
As we address the strong coupling and backreaction problems, we restrict our analysis to the range $-2.1<n<0$. From Eqs.\eqref{eq:mpb_kc}, we find that the induced magnetic energy spectrum exhibits a scaling behavior $\mPbi(k)\propto k^4$, whereas the electric energy spectrum scales as $\mPei(k)\propto k^2$. In addition, both spectra contain an enhancement factor that depends on the characteristic time $\eta_c$, or equivalently on the associated wavenumber $\kc$.

Due to the strongly blue-tilted nature of the magnetic spectrum, it is not possible to generate sufficiently large magnetic fields on cosmological scales. Even if we choose $\eta_c$ such that all CMB modes are excited after horizon crossing, in the case of instantaneous reheating, the resulting magnetic field strength is only $B_0(1~\Mpc)\sim 10^{-28}$ Gauss. Therefore, additional mechanisms are required to enhance the large-scale magnetic field in order to match present-day observations.

In this work, we consider the mechanism proposed in \cite{Maiti:2025cbi, Kobayashi:2014sga, Haque:2020bip}, where the presence of a strong, large-scale electric field allows for the transfer of electric energy into magnetic energy in the absence of electrical conductivity during reheating. 
For our purpose in the present paper, we assume $V(\phi)\propto \phi^4$ with an effective reheating equation of state $w_\phi=1/3$. It is in this phase we assume conductivity to be vanishingly small, and the Faraday effect is effective. After the completion of reheating universe becomes standard radiation dominated.

%After reheating, assuming the Universe is filled with a relativistic plasma with effectively infinite conductivity, the electric field rapidly decays, while the magnetic field evolves adiabatically until the present epoch.
The present-day magnetic field $B_0$ can then be estimated as~\cite{Maiti:2025cbi, Kobayashi:2014sga, Haque:2020bip}
%\begin{eqnarray}
%a_0^4 \mathcal{P}^0_B(k,\ere) \simeq \frac{\mPei(k,\ee) k^2}{\are^2 H_{\rm re}^2} \implies B_0 \simeq \frac {\sqrt{\mPei(k,\ee)} }{{a_0^2}} \l(\frac{k}{\are\Hre}\r).
%\end{eqnarray}
\begin{eqnarray}
a_0^4 \mathcal{P}^0_B(k,\ere) \simeq \frac{\mPei(k,\ee) k^2}{\are^2 H_{\rm re}^2} \implies B_0 \simeq \frac {\ke^2\sqrt{(n^2+k^2\eta_c^2)} }{{2\pi}} \l(\frac{\kc}{\ke}\r)^{(\nb/2-1)}\l(\frac{k}{\ke}\r)\l(\frac{k}{\are\Hre}\r).
\end{eqnarray}
This expression clearly demonstrates the Faraday induction effect through the dependence on the parameters of the reheating phase, during which electrical energy is converted into magnetic energy. 
Furthermore, the present-day magnetic field strength depends sensitively on the duration of this inflaton-dominated phase through the characteristic time scale $\eta>\eta_c$. Here $\Hre=\l(\frac{\pi \gsre}{90}\r)^{1/2}\Tre^2\Mp^{-1}$ is the Hubble constant at the end of reheating, and $\gsre$ is the relativistic degree of freedom at the end of reheating. For large scale $k<k_c \simeq 1/\eta_c$ we therefore have $B_0 \propto (k_c/\ke)^{1-|n|}\Tre^{-2} k^2 \sim (\ke/k_c)^{1-\nb/2 }(\lambda\Tre)^{-2}$.
Therefore, lowering the reheating temperature enhances the present-day magnetic field strength due to a longer period of Faraday conversion.

%Furthermore, we note that the strength of the electric field on large scales ($k<\kc$) depends on the characteristic time $\eta_c$, which in turn determines the characteristic scale $\kc$.

%Although the Universe behaves as radiation-like during this phase, it is still dominated by the inflaton field, which is a neutral scalar field. In that scenario, when we defined the reheating temperature, we effectively defined the background temperature when the inflaton field decays and the radiation field is the dominating one. Although our background is expanding as $a^{-4}$ as during this phase the neutral inflaton field was the dominating one, we ignored the conductivity and due to the induction effect~\cite{Maiti:2025cbi}, the electric field continuously converted to the magnetic energy density up to the end of reheating. 
%After completion of the reheating, universe is as the SM field is the dominant one, and there is no further entropy production, our universe gets thermalised, and then conductivity is significantly high and decays the electric field part, so after reheating completion, there is no further conversion of energy from the electric field to the magnetic field.

\begin{figure}
    \centering
    \includegraphics[width=0.48\linewidth]{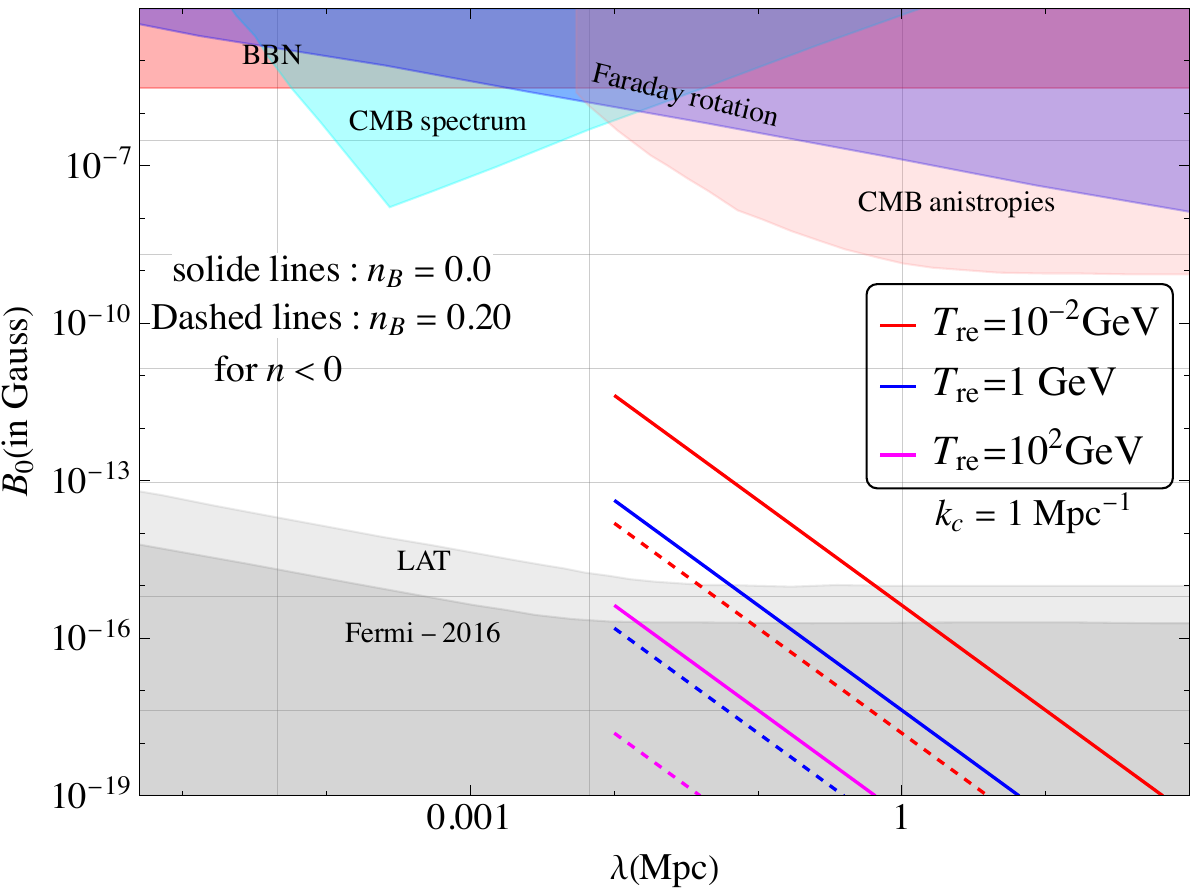}
    \includegraphics[width=0.48\linewidth]{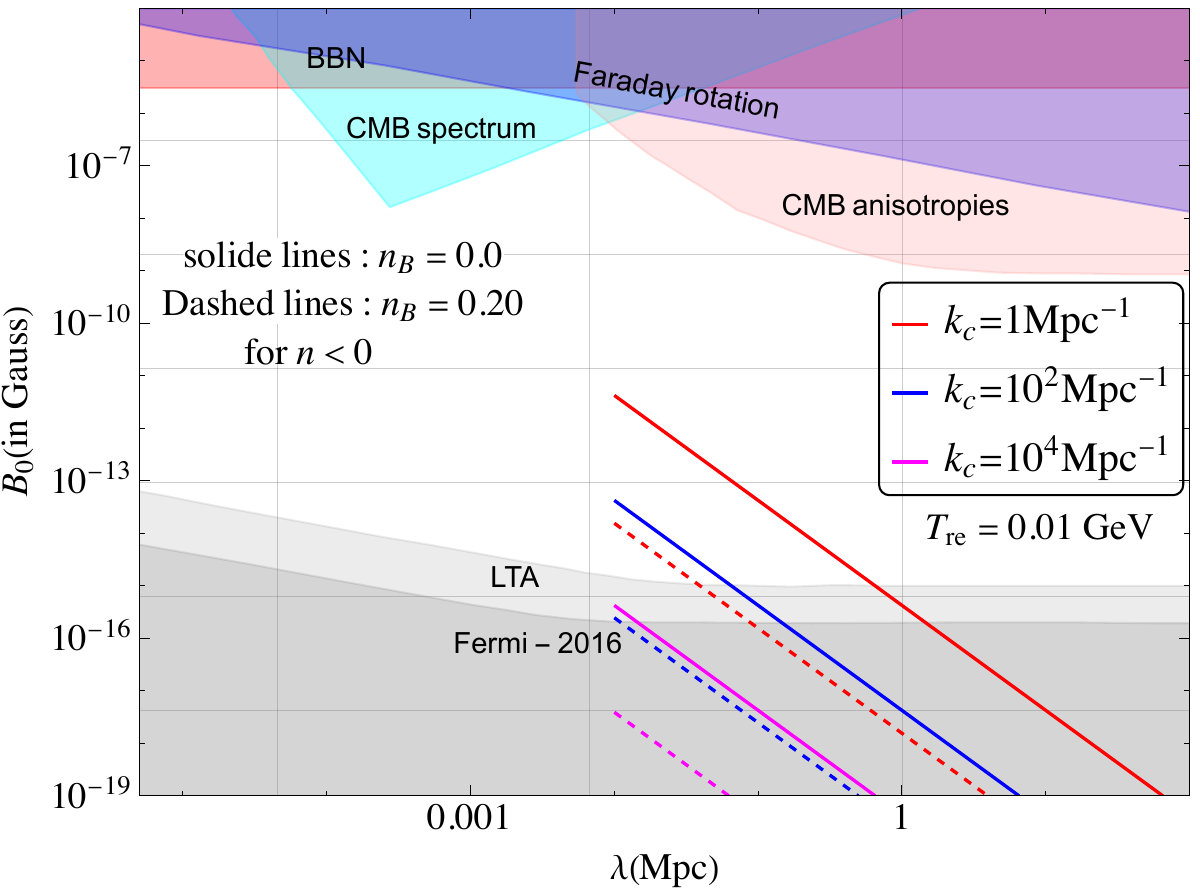}
    \caption{Here plotted the present-day magnetic field strength as a function of comoving present-day wavelength $\lambda$( in Mpc) for different scenarios. In the left panel, we consider three different reheating temperatures denoted by the three different colors, where we consider the characteristic scale $\kc=1\,\Mpc^{-1}$. In the right panel, we have plotted the same quantity, but now we consider three different values of the characteristic scale $\kc$ with a fixed reheating temperature $\Tre=0.01\,\GeV$. In both panels, solid lines are for $\nb=0.0$ and dashed lines are for $\nb=0.20$. Here, the shaded region indicates different observational constraints on the present-day magnetic field strength~\cite{Planck:2015zrl, Paoletti:2008ck,PhysRevLett.116.191302}.}
    \label{fig:B0_vs_kc}
\end{figure}
\begin{figure}
    \centering
    \includegraphics[width=0.48\linewidth]{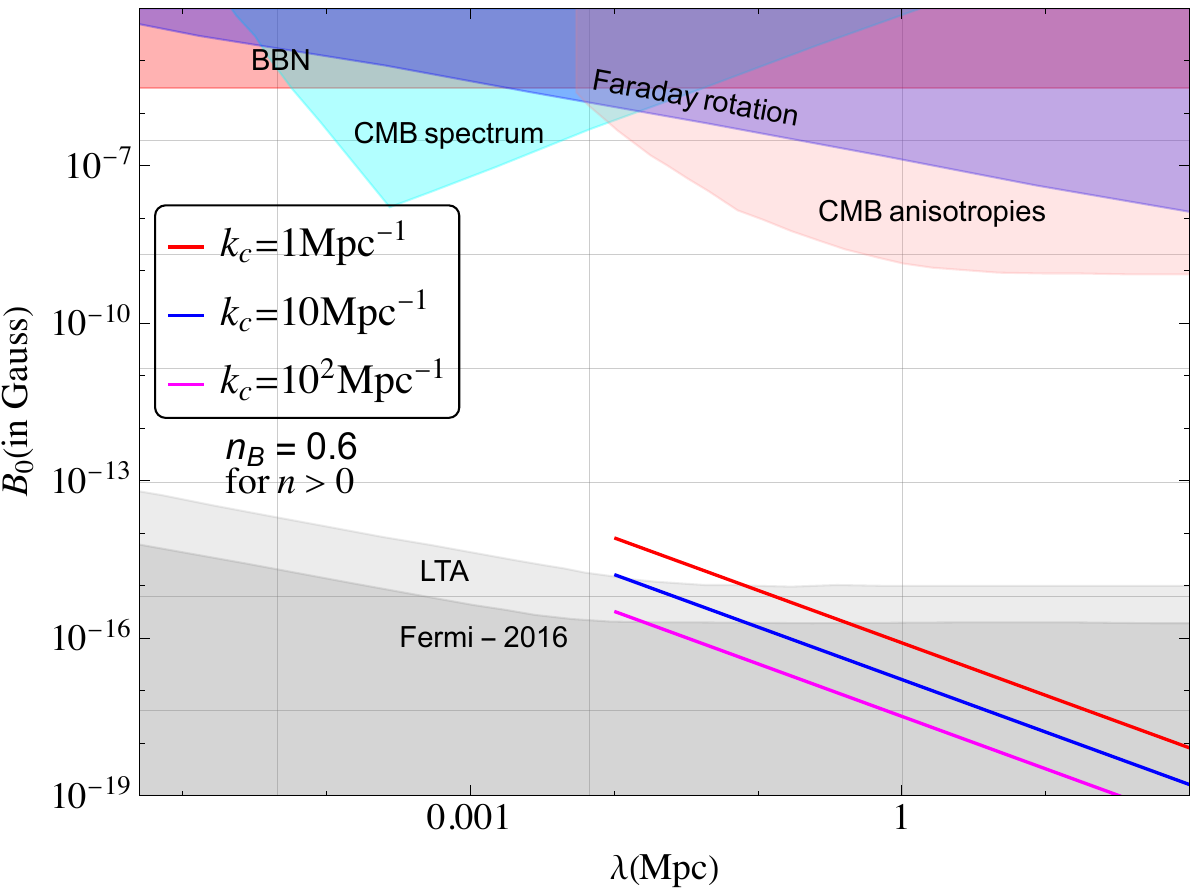}
    \includegraphics[width=0.48\linewidth]{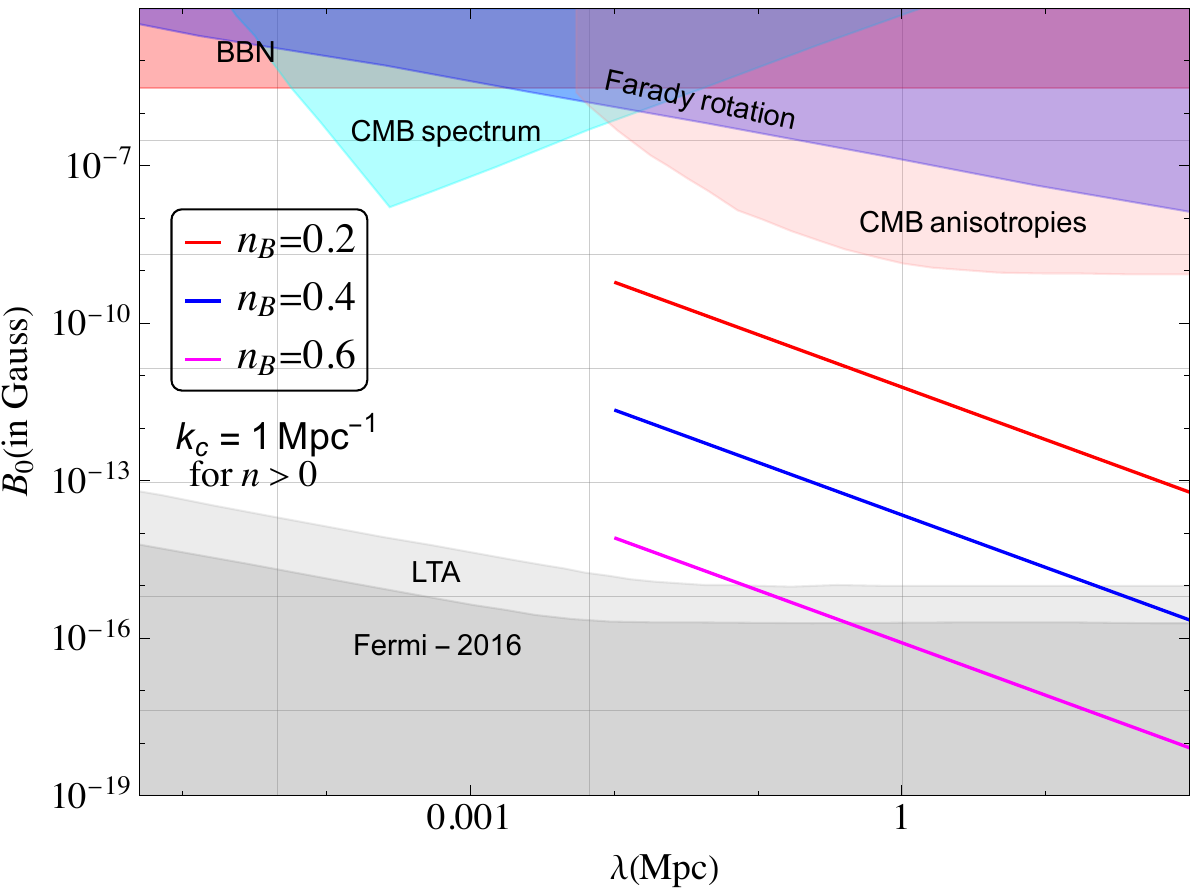}
    \caption{Here plotted the present-day magnetic field strength as a function of comoving present-day wavelength $\lambda$( in Mpc) for different scenarios. In the left panel, we consider three different reheating temperatures denoted by the three different colors, where we consider the characteristic scale $\kc=1\,\Mpc^{-1}$. In the right panel, we have plotted the same quantity, but now we consider three different values of the characteristic scale $\kc$ with a fixed reheating temperature $\Tre=0.01\,\GeV$. In both panels, solid lines are for $\nb=0.0$ and dashed lines are for $\nb=0.20$. Here, the shaded region indicates different observational constraints on the present-day magnetic field strength~\cite{Planck:2015zrl, Paoletti:2008ck,PhysRevLett.116.191302}.}
    \label{fig:B0_vs_kc_2}
\end{figure}
In Fig.\ref{fig:B0_vs_kc}, we plot the present-day magnetic field strength as a function of the present-day wavelength $\lambda(\Mpc)$. In the left panel, we consider three different reheating temperatures, $\Tre=10^{-2},\,1$, and $10^2\,\GeV$, represented by red, blue, and magenta, respectively. Here, we fix the characteristic scale to $\kc=1\,\Mpc^{-1}$, while the solid and dashed lines correspond to the magnetogenesis parameters $\nb=0.0$ and $\nb=0.2$, respectively. We recall that the magnetogenesis parameter is defined as $\nb=4-2|n|$.
%and throughout our analysis we restrict ourselves to the regime where the model does not suffer from strong coupling or backreaction problems.
The colored shaded regions indicate different observational bounds on the present-day magnetic field strength. 

In the right panel, we plot the same quantity for a fixed reheating temperature $\Tre=10^{-2}\,\GeV$, while considering three different characteristic scales corresponding to $\kc=1,10^2$, and $10^4\,\Mpc^{-1}$ represented by red, blue, and magenta, respectively. As shown in Eq.\eqref{eq:mpb_kc}, the overall enhancement depends explicitly on $\kc$. We find that the magnetic field strength decreases as $\kc$ increases. Physically, a larger value of $\kc$ corresponds to a shorter duration over which the coupling remains active, thereby significantly reducing the overall amplification of the gauge field. As a result, obtaining an observable magnetic field strength imposes a nontrivial constraint on the characteristic time, or equivalently on the characteristic scale, as illustrated in both panels.

\paragraph{Present-day magnetic field in the strong coupling regime:}

We have seen that, for the $n<0$ scenario, the electric field becomes dominant while the magnetic spectrum is strongly blue tilted. Consequently, explaining the observed large-scale magnetic field requires post-inflationary dynamics through which the electric field energy density is converted into magnetic energy density. Even after including this conversion mechanism, obtaining the observed present-day magnetic field strength restricts the analysis to the range $0\leq\nb<0.3$, where $\nb(=4-2|n|)$. However, our primary interest lies in scenarios where ultra-light PBHs can be generated from the magnetic field, which typically requires $\nb>0.5$ as will be discussed in detail in the next section. For $\nb<0.5$, the resulting PBHs form within mass windows where they evaporate during the radiation-dominated era. Such PBHs are strongly constrained by various observational bounds.

Therefore, for completeness, we also present the case $n>0$ for which one can simultaneously generate a large-scale magnetic field of sufficient strength and ultralight PBHs. 
For this case, the present-day magnetic field strength is given by
\begin{align}
    B_0(k)\simeq \frac{\ke^2}{2\pi}\sqrt{n^2+k^2\eta_c^2}\l(\frac{\kc}{\ke}\r)^{\nb/2-1}\l(\frac{k}{\ke}\r).
\end{align}
Note that the strength does not depend on the reheating parameters as opposed to the previous case. This can be attributed to the fact that the reheating equation of state is considered to be radiation-like. However, such behavior is no longer true for the reheating phase dominated by matter-like or stiff equation of state~\cite{Maiti:2025cbi}.
Nevertheless, for large scale $k<k_c \simeq 1/\eta_c$ we therefore have $B_0 \propto (k_c/\ke)^{1-|n|} k \sim (\ke/k_c)^{1-\nb/2 }(\lambda)^{-1}$. Therefore, large scale magnetic field has distinct scaling behavior for two different regime of $n$.

%4$ inflationary potential, the post-inflationary Universe evolves effectively as radiation, with the energy density redshifting as $a^{-4}$. As a result, the present-day magnetic field strength in this scenario becomes independent of the reheating temperature. 

In Fig.\ref{fig:B0_vs_kc_2}, we plot the present-day magnetic field strength $B_0(k)$ as a function of the present-day observable wavelength $\lambda(\Mpc)$. In the left panel, we fix $\nb=0.6$ and consider three different characteristic wavenumbers, $\kc=1,\,10$, and $10^2\,\Mpc^{-1}$. We find that, despite the strongly blue-tilted nature of the spectrum, there exists a region of parameter space where the present-day magnetic field strength remains consistent with current observational bounds.

In the right panel of Fig.\ref{fig:B0_vs_kc_2}, we show the same quantity for three different values of $n=0.2,\,0.4$, and $0.6$, while fixing the characteristic scale to $\kc=1\,\Mpc^{-1}$. We clearly observe that a broad region of parameter space allows the generation of sufficiently strong magnetic fields compatible with present-day observations.

Most importantly, all modes excited after the characteristic time $\eta_c$ share the same spectral behavior, with both the magnetic and electric spectra scaling as $\mathcal{P}_{\rm B/E}(k)\propto k^{\nb}$. Consequently, the curvature power spectrum induced by the electromagnetic field remains unaffected by the sign of the coupling parameter $n$. Therefore, the PBH formation mechanism is insensitive to the sign of $n$.

To this end, we emphasize once again that although the $n>0$ magnetogenesis scenario suffers from the strong coupling problem, it can simultaneously generate ultra-light PBHs and large-scale magnetic fields without requiring any post-inflationary energy conversion mechanism. In contrast, the $n<0$ scenario is free from both strong coupling and backreaction problems; however, it is considerably more difficult to simultaneously produce ultra-light PBHs and sufficiently strong large-scale magnetic fields within the same framework.

Overall, these results demonstrate that there exists a viable region of parameter space in which sufficiently strong large-scale magnetic fields can be generated while remaining consistent with present-day observational constraints. Moreover, depending on the choice of the coupling parameter, the same framework can also lead to the production of ultra-light PBHs with a broad mass distribution. Since the remaining calculations are largely insensitive to the sign of the magnetogenesis coupling parameter $n$, we express all relevant quantities in terms of the magnetic spectral index $\nb$, rather than $n$, for simplicity and convenience.

\color{black}               
\section{Induced Curvature Power Spectrum}
 In standard inflationary scenarios, the dominant contribution to the curvature perturbation arises from quantum fluctuations of the inflaton field. However, in the presence of an electromagnetic (EM) field generated during inflation, this field can also source curvature perturbations, and its contribution must be properly accounted for in the total curvature perturbation.

 The comoving curvature perturbation \(\zeta\) is defined as~\cite{Bassett:2005xm, Wands:2000dp, PhysRevD.94.043523, Baumann:2009ds, Malik:2008im}
\begin{align}
    \zeta = -\Psi - H \frac{\delta\rho}{\dot{\bar{\rho}}}, \label{eq:zeta_def}
\end{align}
where \(\Psi\) is the Bardeen potential, and \(\delta\rho\) is the total energy density perturbation. This expression is gauge-invariant. For adiabatic fluctuations and in the absence of anisotropic stress at leading order, one typically has \(\Psi \simeq \Phi\), and on super-horizon scales, \(\zeta\) remains conserved over time.

However, in the presence of an EM field, \(\zeta\) may not be conserved on super-horizon scales, as it can continue to be sourced by the EM field even after horizon exit. There are two distinct channels through which the EM field can contribute to the curvature perturbation: (i) it can directly modify the Bardeen potential \(\Psi\), and (ii) it can contribute via the energy density perturbation in the last term of Eq.~\eqref{eq:zeta_def}.

Since we are interested in curvature perturbations generated during inflation—when the relevant modes are already outside the horizon—we now turn to a quantitative analysis to determine which of these contributions dominantly affects the total induced curvature perturbation.

The perturbed Einstein constraint (in Newtonian gauge) gives~\cite{ Brandenberger:2003vk, Malik:2008im, Weinberg:2008zzc}
\begin{align}
    3\mH(\mH\Phi+\Psi')-\nabla^2\Psi=- \frac{a^2\delta\rho}{2\Mp^2} 
\end{align}
We can drop $-\nabla^2\Psi=k^2\Psi<<1$ as $k<<aH$ on the super-horizon scales. As we are working in the inflationary era. In this approximation, during inflation in super-horizon scales, we get
\begin{align}
   \zetai_{\Psi}= \Psi\simeq -\frac{a^2\delta\rho}{6\mH^2\Mp^2} =-\frac{1}{2}\frac{\delta\rho}{\rho} \label{eq:Psi_super_horizon}
\end{align}
Where the contribution came from the 2nd term of Eq.\eqref{eq:zeta_def} is
\begin{align}
    \zetai_{\delta\rho}\simeq-H\frac{\delta\rho}{\drho}\simeq-\frac{1}{2\epsilon}\frac{\delta\rho}{\rho} \label{eq:zeta_i_delta_rho}
\end{align}
where we used $\drho=-3H\dbphi^2$~\cite{Baumann:2009ds}. We recall that $\dbphi$ is the velocity of the inflation field. Now, if we compare Eq.\eqref{eq:Psi_super_horizon} with \eqref{eq:zeta_i_delta_rho}, then we can find that $\zetai_{\Psi}/\zetai_{\delta\rho}=\epsilon$. As for slow-roll inflationary scenarios, $\epsilon<<1$, so we can conclude that most of the contribution on the super-horizon scales due to EM fields originates from the last term of Eq. \eqref{eq:zeta_def}. 

In the following subsections, we are going to discuss the induced curvature perturbation due to the inflaton fluctuations $\delta\phi$ as well as from the inflationary EM fields.

\subsection{Curvature Power spectrum Due to Inflaton field Fluctuations}\label{inflation_curv}
In general scenarios, the dominant contribution to the curvature perturbations arises from the quantum fluctuation of the inflation field. In the constant density hypersurface, the comoving curvature perturbation due to inflaton fluctuations is~\cite{mukhbranden, Weinberg:2008zzc}
\begin{align}
    \zetav=-\Psi-H\frac{\delta\rho}{\drho}=-(1+\epsilon)\mR
\end{align}
where in the slow-roll inflation era, the solution of the metric fluctuation $\Psi$ on super-horizon scales is proportional to the velocity of the inflaton field, and it can be written as $\Psi\simeq\epsilon\mR$. Here we define $\mR=H\frac{\delta\phi}{\dphi}$ as the comoving curvature perturbation defined in spatially flat gauge, i.e,. $\Psi=0$. In slow-roll phase $\epsilon<<1$, so we get the well know relation between $\zeta$ and $\mR$ as $\zetav=-\mR$, \cite{mukhbranden, Weinberg:2008zzc}. Now utilizing the solution of the inflaton fluctuation $\delta\phik$ defined in Eq.\eqref{eq:sol_delta_phik}, we can write the induced curvature perturbation due to the inflaton field is ~\cite{mukhbranden, Weinberg:2008zzc}
\begin{align}
    \mPcv(k)\simeq \frac{1}{8\pi^2\epsilon}\l(\frac{\HI}{\Mp}\r)^2
\end{align}
Which is well know solution of the comoving curvature perturbation for the single field inflation models. Where $\epsilon = -\dot{H}/H^2$ is the first slow-roll parameter. This nearly scale-invariant spectrum provides an excellent match to observations of the CMB anisotropies.

\subsection{Curvature Power spectrum Due to IMFs}
In our magnetogenesis scenario, we introduce an additional coupling that breaks the conformal symmetry of the gauge field, allowing the electromagnetic field to be significantly amplified during inflation. For this particular choice of coupling, both the magnetic and electric fields exhibit the same spectral behavior, with their amplitudes differing only through the coupling parameter $\gamma$. 

For a nonvanishing helicity parameter, i.e., $\gamma\neq0$ (and not extremely small), the ratio between the electric and magnetic field spectra is approximately given by ${\mPe(k)}/{\mPb(k)}\simeq \gamma^2$, which is valid primarily for those modes that exit the horizon after the characteristic time $\eta_c$, namely for $k>\kc$.

The curvature power spectrum induced by the magnetic field, evaluated at the end of inflation, is given by~\cite{Fujita:2013qxa, PhysRevD.94.043523}
\begin{align}\label{eq:p_zeta_s_mag}
    \mPcs(k,\ee)=\frac{1}{8\epsilon^2\rho^2}\times \int \frac{dq}{q}\int_{-1}^1d\gamma\, F_2(\mu,\gamma)\frac{\mPb(q,\ee)\mPb(|\vk-\vq|,\ee)}{\l| 1-\vq/\vk\r|^3}.
\end{align}
where $F_2(\mu,\,\gamma)=1+(\hat{q}\cdot\widehat{\vk-\vq})^2$ and $\gamma=\hat{q}\cdot\hat{k}=\cos(\theta)$, where $\theta$ is the angle between two vector $\vk$ and $\vq$. Now replacing the values of $\mPb(k,\ee)$ from the above Eq.~\eqref{eq:mPbe_inf}, we get that the total induced comoving curvature perturbations due to the EM fields generated during inflation are
\begin{align}\label{eq:mPc_ind}
    \mPcs(k)=\frac{(1+\epsilon)^2(1+\gamma^2)^2 \mcB^2(\nb,\gamma)}{72\epsilon^2}\l(\frac{\HI}{\Mp}\r)^4\l(\frac{k}{\ke}\r)^{2\nb}\times\mFnb^{(1)}(k)
\end{align}
where $\mFnb^{(1)}(k)$ is defined as
\begin{align}
    \mFnb^{(1)}(k)\simeq \frac{4}{3}\l\{ \frac{1}{\nb}\l(1-\l(\frac{\kpv}{k}\r)^{\nb}\r)+\frac{1}{5-2\nb}\l(1-\l(\frac{k}{\ke}\r)^{5-2\nb}\r)\r\}
\end{align}
\begin{figure}[t]
\centering
\includegraphics[scale=0.43]{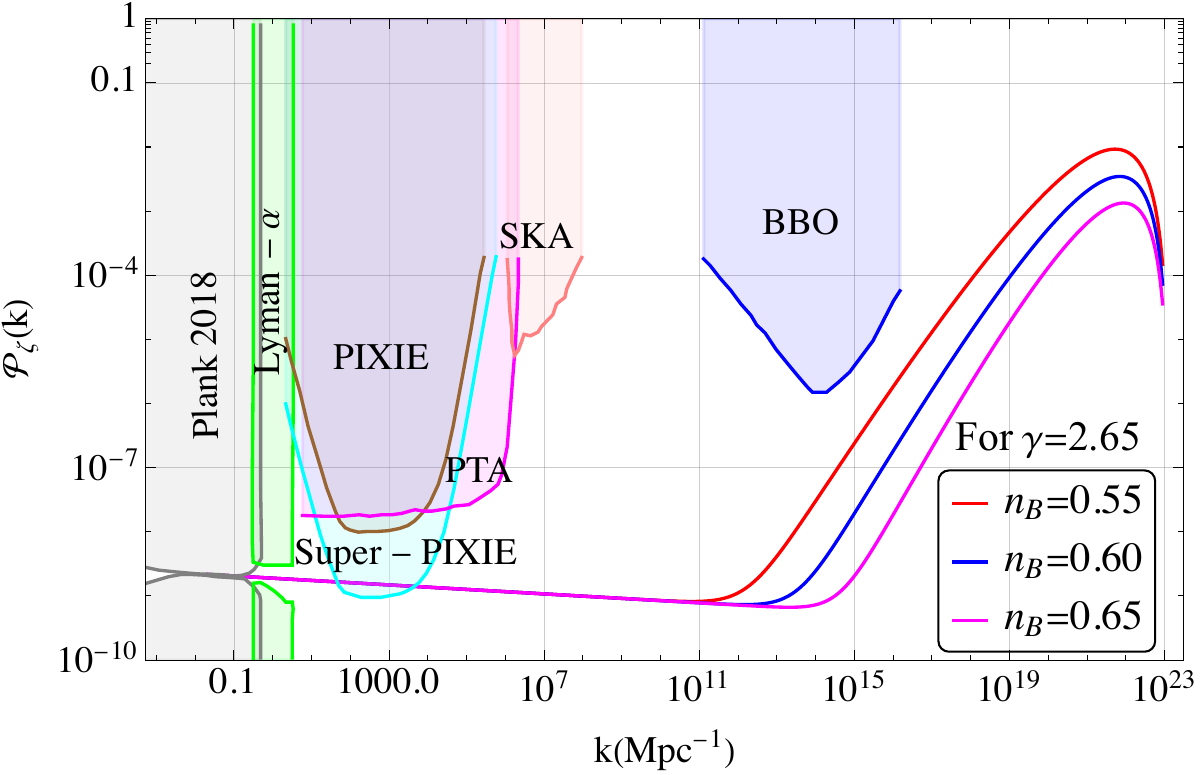}
\includegraphics[scale=0.43]{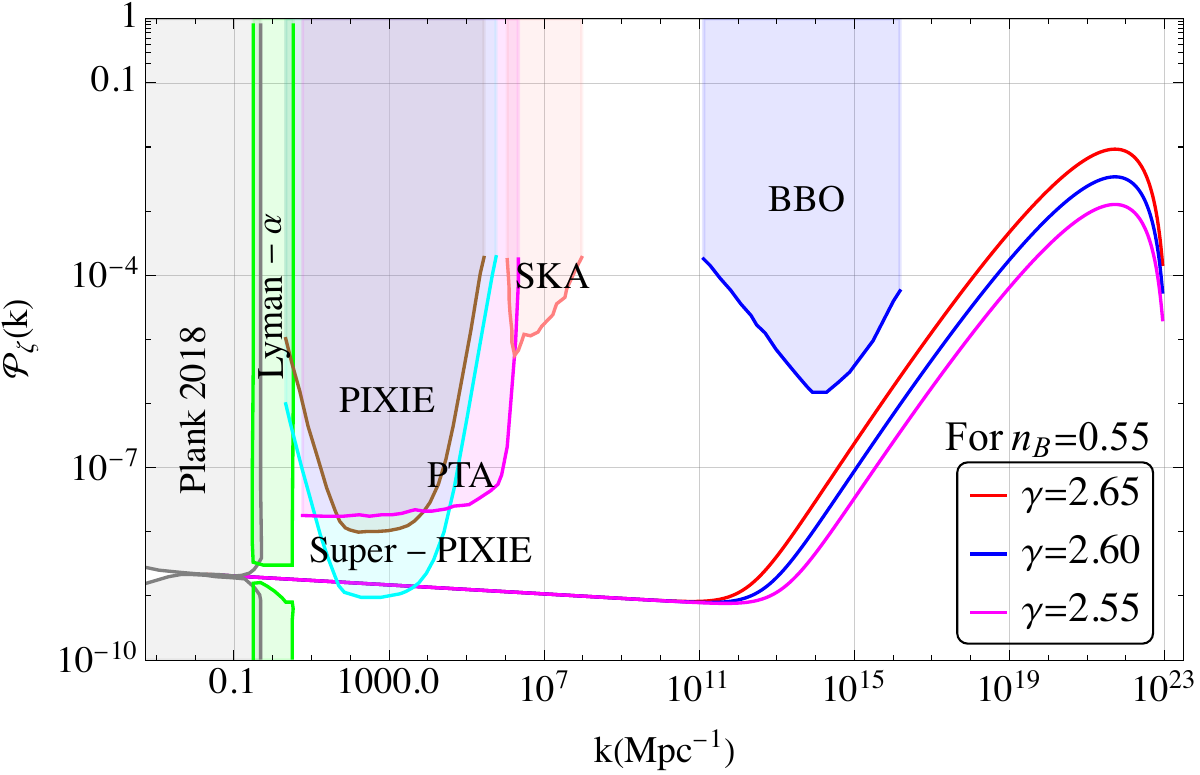}
\caption{ In this figure, we have plotted the curvature power spectrum as a function of comoving wavenumber for two different scenarios. In the left panel, we have plotted it for a fixed helicity parameter $\gamma=2.65$ where three different colors indicate three different values of the magnetic field spectral index $\nb=0.55,\,0.60$ and $0.65$. In the right panel, we have plotted the same things, but now we have varied the helicity parameter $\gamma$, which is denoted by three different colored lines for a fixed magnetic spectral index $\nb=0.55$. We also have to add the current and future observational bound those are denoted by the different color shaded regions.
}
\label{fig:curvature_k}
\end{figure}
To arrive at Eq. \eqref{eq:mPc_ind}, we replace $\rho=3\HI^2\Mp^2$, which is the background inflaton energy density at the end of inflation. Now, from the above Eq.\eqref{eq:mPc_ind}, we have found that the curvature power spectrum induced by the EM field during inflation depends on the initial amplitude of the EM field produced during inflation, as well as it depends on the slow-roll parameter $\epsilon$. On a large scale, the value of the slow-roll parameter is almost constant, so the spectrum behaves as $k^{2\nb}$, where $\nb$ is the magnetic spectral index related to the coupling parameter $n$. Now, for these helical magnetic field scenarios, the magnetic field amplitude is exponentially dependent on the helicity parameter $\gamma$, so there is a situation where the magnetic field at a larger scale is enough such that the secondary contribution to the curvature is dominated over the primary one.

To illustrate the spectral behavior of the induced curvature perturbations, Fig.~\ref{fig:curvature_k} shows the curvature power spectrum $\mPc(k)$ as a function of the comoving wavenumber $k~(\mathrm{Mpc}^{-1})$ for representative values of the magnetogenesis parameters $\gamma$ and $\nb$.

In the left panel of Fig.~\ref{fig:curvature_k}, we fix the helicity parameter to $\gamma = 2.65$ and vary the magnetic spectral index $\nb$. The three curves correspond to different choices of $\nb$, as indicated by the color coding. The shaded regions represent current and projected observational constraints on the curvature power spectrum. For a fixed value of $\gamma$, increasing the magnetic spectral index leads to a gradual suppression of the peak amplitude of $\mPc(k)$. This behavior follows from the relation $\nb = 4 - 2|n|$, which implies that a larger $\nb$ corresponds to a smaller absolute value of the coupling parameter $n$. Since the magnetic field amplitude depends exponentially on $n$ (see Eq.~\eqref{eq:mPbe_inf}), increasing $\nb$ effectively reduces the overall magnetic field strength and hence the induced curvature perturbations. We also observe a mild shift of the peak of the curvature power spectrum toward higher wavenumbers as $\nb$ increases. This reflects the increasingly blue-tilted nature of the magnetic field spectrum, for which most of the magnetic energy density is concentrated near the end-of-inflation scale $k \simeq \ke$.

In the right panel of Fig.~\ref{fig:curvature_k}, we fix the magnetic spectral index to $\nb = 0.55$ and vary the helicity parameter $\gamma$. The three curves correspond to different values of $\gamma$. In this case, the amplitude of the induced curvature perturbations increases exponentially with $\gamma$, consistent with the exponential dependence of the magnetic energy density on the helicity parameter (see Eq.~\eqref{eq:mPbe_inf}). Since the magnetic spectral index is fixed, the spectral shape of the magnetic field—and therefore the position of the peak in $\mPc(k)$—remains unchanged, while only the overall amplitude is enhanced.

In both panels, the curvature power spectrum on large scales exhibits a slightly red-tilted behavior. This is because, on these scales, the dominant contribution to $\mPcv(k)$ arises from inflaton fluctuations, which exceed the secondary contributions sourced by the magnetic field. Consequently, the spectrum follows the standard scaling $\mPcv(k) \propto k^{\ns - 1}$, where $\ns$ is the scalar spectral index. For definiteness, we adopt the central value $\ns \simeq 0.9649$ motivated by $\alpha$-attractor models, although our results are insensitive to the precise choice of $\ns$.

\section{ Ultra-light PBH Production}
PBHs are one of the prime candidates for dark matter. Apart from being DM, their small mass counterpart can efficiently evaporate, and yield distinct physical effect in the early universe cosmology. For example, it can help us explain the bayrogenesis problem~\cite{Majumdar:1995yr, Baumann:2007yr, Hook:2014mla, Smyth:2021lkn, Datta:2020bht, Boudon:2020qpo, Morrison:2018xla, PhysRevD.59.041301, Bernal:2022pue, Schmitz:2023pfy, Borah:2024lml, Barman:2022pdo, RiajulHaque:2023cqe, Hamada:2016jnq, DeLuca:2022bjs, DeLuca:2021oer}, lead to successful reheating ~\cite{ Hawking:1975vcx,DonPage:1976, Baumann:2007yr, RiajulHaque:2023cqe}. 
The conventional mechanism of forming PBHs is 
to have large curvature perturbation beyond a critical threshold. Such a large perturbation can be generated from inflation,     %can lead to the formation of PBHs.
post-inflationary pre-reheating  
~\cite{Khlopov:1980mg, Bullock:1996at, Garcia-Bellido:1996mdl, Yokoyama:1998pt, Kawasaki:1997ju, Garcia-Bellido:2017mdw, PhysRevD.103.083510, Bhaumik:2020dor, Solbi:2021wbo, Figueroa:2021zah, Frolovsky:2022qpg, Lin:2012gs,Cheng:2016qzb,Cheng:2018yyr}, bubble collisions in first-order phase transitions~\cite{PhysRevD.109.123030, PhysRevD.26.2681, Rubin:2001yw, Ai:2024cka} or the collapse of topological defects like cosmic strings~\cite{Hawking:1987bn, Polnarev:1988dh, PhysRevD.45.3447, Balaji:2025tun, Balaji:2024rvo, Balaji:2022rsy}. Following our earlier papers, in this paper, we investigate the formation of ultralight PBHs from the inflationary magnetic field. 
We have discussed the possible magnetogenetic scenarios that induce large
curvature perturbations 
at small scales.

The conventional and widely studied formation mechanism involves the direct collapse of the overdense regions during the radiation-dominated era. If a region with the density contrast exceeds the threshold value $\delta>\delta_c\simeq0.4$,~\cite{Escriva:2019phb, PhysRevD.105.124055} then the pressure forces are unable to counter gravity, and the region collapses to form a PBH  
of mass $\Mpbh\simeq\frac{4}{3}\pi\gc\rho(\th)H_{\th}^{-3}\simeq 4\pi\gc\Mp^2H^{-1}_{\th}$, where we have used the background energy density during horizon crossing, $\rho(\th)=3\Mp^2H^2_{\th}$. We consider the post-inflationary reheating period to be dominated by inflation fluid with equation of state $\wre\simeq \langle w_\phi\rangle=1/3$, such that $\rho \propto a^{-4}$. Here, angle brackets denote the average value of the effective equation of state. Where $H_{\th}$ is the Hubble constant during the horizon crossing, i.e., at $t=\th$. Here, $\gc\sim0.2$ is the efficiency factor that accounts for the dynamics of collapse.

The abundance of the PBHs is exponentially sensitive to the tail of the probability distribution of the curvature perturbations. The fraction of the density that collapses to PBHs, often referred to as the initial PBH mass fraction, i.e., $\beta(\rM)=\rhopbh(\th)/\rhot(\th)$, is defined as~\cite{Khlopov:1980mg, Bullock:1996at, Garcia-Bellido:1996mdl, Yokoyama:1998pt, Kawasaki:1997ju, Garcia-Bellido:2017mdw, PhysRevD.103.083510, Bhaumik:2020dor, Solbi:2021wbo, Figueroa:2021zah, Frolovsky:2022qpg, Motohashi:2017kbs, Byrnes:2018txb, Ballesteros:2018wlw, Raveendran:2022dtb, Ragavendra:2020sop, 
Braglia:2020eai, Karam:2022nym}
\begin{align}
    \beta(\rM)=\int_{\delta_c}^{\infty}\d \delta\,P(\delta)
\end{align}
Here $\rm P(\delta)$ denotes the probability distribution of the density contrast $\delta$. The parameter $\delta_c$ represents the critical threshold above which an overdense region can undergo gravitational collapse and form a PBH. In the presence of magnetically induced density perturbations, 
the total pressure and energy density
are given by $p_{\rm tot}=p_{\rm rad}+p_B$ and  $ 
\rho_{\rm tot}=\rho_{\rm rad}+\rho_B$,
where $r_B=\rho_B/\rho_{\rm rad}$. The corresponding effective equation of state can then be written as $w_{\rm eff}={p_{\rm tot}}/{\rho_{\rm tot}}=c_s^2+\alpha_r r_B$,
where the coefficient $\alpha_r$ encapsulates the dynamical evolution of the magnetic field on the relevant length scales. In Ref.~\cite{Kushwaha:2024zhd}, the authors analytically estimated $\alpha_r\simeq 2$; however, this value is sensitive to nonlinear magnetohydrodynamic evolution.

In our scenario, we always consider a blue-tilted magnetic spectrum, such that most of the energy of the electromagnetic field is stored at $k\sim\ke$. For PBH formation, the ratio between the electromagnetic energy density and the background energy density should be around $\delta\rhob/\rho_c\simeq \mathcal{O}(0.01)$.
As PBHs are mainly formed during the early radiation-dominated era, the critical density threshold in this phase is governed by the effective EoS, $\delta_c\simeq w_{\rm eff}$. As discussed above, if we include the effect of magnetic anisotropic stress up to linear order, we can approximately write the critical value of the density contrast as $\delta_c\simeq \sqrt{c_s^2+\alpha_r r_B}$.
Taking this as an upper bound and adopting $\alpha_r=2$, the corresponding threshold value for magnetic-field-induced collapse is $\delta_c\simeq 0.533$, Although a precise determination of $\delta_c$ would require solving the full set of magnetohydrodynamic equations while incorporating all relevant effects in the collapse mechanism, for simplicity we adopt a fiducial threshold value $\delta_c\simeq 0.55$.

The initial abundance of PBHs is highly sensitive to the value of $\delta_c$. For a Gaussian distribution, the initial PBH mass fraction $\beta$ depends exponentially on the threshold value of the density contrast $\delta_c$, implying that even a small change in $\delta_c$ can significantly modify the final PBH spectrum. For the dependence of the final spectrum on $\delta_c$ in the case of a non-Gaussian distribution, see Ref.~\cite{Maiti:2025ijr}.

Here $\rm P(\delta)$ is the probability distribution of the density contrast of the magnetic field. We recall that the density contrast is defined as $\delta=\delta\rhob/\rho_{\rm rad}\propto (B^2+E^2)$, and hence $\delta$, due to its quadratic dependence on electromagnetic field, is intrinsically non-Gaussian in nature~\cite{Saga_2020}. To capture the full non-Gaussian nature of the distribution function, one can use the MCMC analysis as discussed in \cite{Saga:2020ics}, but in our computation, we consider a generalized hyperbolic (GH) distribution function, which can be parameterized as
\begin{align}
    \Png(x)=C\,\exp\l[-\frac{3}{2}\sqrt{1+\frac{x^2}{b^2}}\r]
\end{align}
where $C=2.13\sigma^{-1}$ and $b=0.84\sigma$~\cite{Kushwaha:2024zhd,Maiti:2025ijr}.
Where $\sigma$ is the variance of the smoothed density contrast at mass scale $\Mpbh$, computed at the time of horizon re-entry of that scale. The variance of the smoothed density contrast on scale $R=1/k$ is
\begin{align}
    \sigma^2(\rm{M})=\frac{16}{81}\int \frac{\d q}{q}(qR)^4W^2(qR)\mPc(q)
\end{align}
where we commonly used window function $W(qR)$ as a Gaussian, i.e., $W(qR)=\exp(-q^2R^2/2)$ and utilizing this function we get
\begin{align}\label{eq:sigma}
    \sigma^2(\rm{M})=\frac{16}{81}\int_{0}^{\infty}\frac{\d q}{q}(qR)^4e^{-q^2R^2}\mPc(q)
\end{align}
Our goal is to compute the total number of PBHs or the total energy density of the produced PBHs as a function of their initial mass $\Mpbh$. To do so, we need to relate their mass $\Mpbh$ to the smoothing scale $R$ that we can introduce through the window function. In the scenario where a scale with wavenumber `$k$' re-enters the Hubble radius, we can express the mass of the PBHs formed to be $\Mpbh=\gc \rm{M}_{H}$, with efficiency factor $\gc\sim 0.2$. Since no other scale is present, it seems reasonable to set $k=R^{-1}$ and use the fact that $k=aH$, where the perturbation with wavenumber `$k$' re-enters the Hubble radius, to obtain the relation between $R$ and $\Mpbh$. It can be easily shown that `$R$' and `$\Mpbh$' are related as follows~\cite{Ragavendra:2020vud,Ragavendra:2020sop} 
\begin{align}
    R=\frac{1}{k}=\l(\frac{2}{\gc}\r)^{1/2}\l(\frac{\gsk}{\gseq}\r)^{1/12}\l(\frac{1}{\keq}\r) \l(\frac{\Mpbh}{\Meq}\r)^{1/2}
\end{align}
where $\keq$ denotes the wave number that re-enters the Hubble radius at the time of matter-radiation equality, and the quantity $\Meq$ represents the mass within the Hubble radius at equality. In quantities $\gsk$ and $\gseq$ denote the effective number of relativistic degrees of freedom at the times of PBH formation and matter radiation equality, respectively. One find that $\Meq=5.83\times 10^{50}~\text{gm}$, and using this result, the above relation between `$R$' and `$\Mpbh$' can be expressed in terms of the solar $\Msolar $ as follows~\cite{Ragavendra:2020vud,Ragavendra:2020sop}
\begin{align}\label{eq:R}
    \l(\frac{R}{\Mpc}\r)=\l(\frac{\Mpc^{-1}}{k}\r)=4.72\times 10^{-7}\l(\frac{\gc}{0.2}\r)^{-1/2}\l(\frac{\gsk}{\gseq}\r)^{1/12}\l(\frac{\Mpbh}{\Msolar}\r)^{1/2}
\end{align}
Once we know the inflationary power spectrum, we can make use of Eq.\eqref{eq:R} in Eq.\eqref{eq:sigma} to compute the variance $\sigma^2(\Mpbh)$. Using this, we then obtain the initial fractional energy density of PBHs of mass $\Mpbh$ through Eq.~\eqref{eq:sigma}. As is well known, the initial PBHs abundance is highly sensitive to both the variance $\sigma$ and the value of the critical density contrast $\delta_c$.

\subsection{Computing the critical initial PBH mass fraction $\beta_c$:}

In this scenario, the enhancement of curvature perturbations occurs on small scales, leading to the formation of low-mass PBHs that can evaporate before the onset of BBN. Since the amplitude of the curvature power spectrum depends on the initial profile and strength of the electromagnetic fields, the resulting PBH abundance can be sufficiently large for PBHs to temporarily dominate the cosmic energy density before complete evaporation.

The mass loss rate of a PBH due to Hawking evaporation is given by~\cite{PhysRevD.13.198, Das:2025vts}
\begin{align}
    \frac{\d M}{\d t}=-\epsilon_{\rm M}(M)\frac{\Mp^4}{M^2}\,,
\end{align}
where $\epsilon_{\rm M}(M)=\frac{27}{4}\frac{\gst(\TBH)\pi}{480}$. Here $\gst(\TBH)=106.75$ is the relativistic degree of freedom. Integrating this equation from the initial mass $\Mpbh$ to complete evaporation yields the PBH lifetime,~\cite{Carr:2020gox}
\begin{align}\label{eq:tev}
    \tau=\frac{\Mpbh^3}{3\epsilon_{\rm M}\Mp^4}
    \simeq 407\l(\frac{\Mpbh}{10^{10}\,\gm}\r)^3\sec\,.
\end{align}

Since PBHs form with masses related to the horizon mass, their formation time $(\tf)$ can also be expressed in terms of $\Mpbh$. During radiation domination, one finds 
\begin{align}\label{eq:tf}
    \tf=\frac{\Mpbh}{2\pi\gc \Mp^2}
    \simeq 5.11\times10^{-39}\l(\frac{0.2}{\gc}\r)\l(\frac{\Mpbh}{1\,\gm}\r)\sec\,,
\end{align}
where $\gc$ denotes the collapse efficiency factor.

 For illustration, a PBH with initial mass $\Mpbh=1\,\gm$ forms at
$\tf\simeq 5.11\times10^{-39}\sec$ and evaporates at
$\tev=\tf+\tau\simeq 4.07\times10^{-29}\sec$.
Similarly, a PBH with mass $5\times10^{14}\,\gm$ has a lifetime comparable to the age of the Universe, while a PBH with mass $10^{9}\,\gm$ evaporates well before the onset of BBN. 

During radiation domination, the scale factor evolves as $a(t)\propto t^{1/2}$. Since the radiation background redshifts faster than the PBH energy density, the PBH fractional abundance grows with time as $\beta(t)=\beta(\tf)\l({a(t)}/{a(\tf)}\r)$.
Therefore, depending on the initial abundance $\beta(\tf)$, PBHs may dominate the total energy density before complete evaporation. Once PBH domination is reached, both the background and PBH energy densities scale as $a^{-3}$, so the fractional abundance remains constant for a monochromatic PBH population.

For each PBH mass, there exists a critical initial fraction $\beta_c$ such that PBH domination occurs before evaporation whenever $\beta(\tf)>\beta_c$. Evaluating the abundance at the evaporation time $\tev$, we obtain
\begin{align}
    \beta(\tev)=\beta(\tf)\l(\frac{a(\tev)}{a(\tf)}\r)
    =\beta(\tf)\l(\frac{\tev}{\tf}\r)^{1/2}.
\end{align}
The minimal condition for a PBH-dominated era is $\beta(\tev)\geq1$. Using the expressions for $\tf$ and $\tev$, this yields
\begin{align}
    \beta_c(\tf)\simeq 3.54\times10^{-6}\l(\frac{1\,\gm}{\Mpbh}\r).
\end{align}
Hence, for a $1\,\gm$ PBH, domination occurs before evaporation provided
$\beta(\tf)\geq3.54\times10^{-6}$.

\begin{figure}[t]
\centering
\includegraphics[scale=0.29]{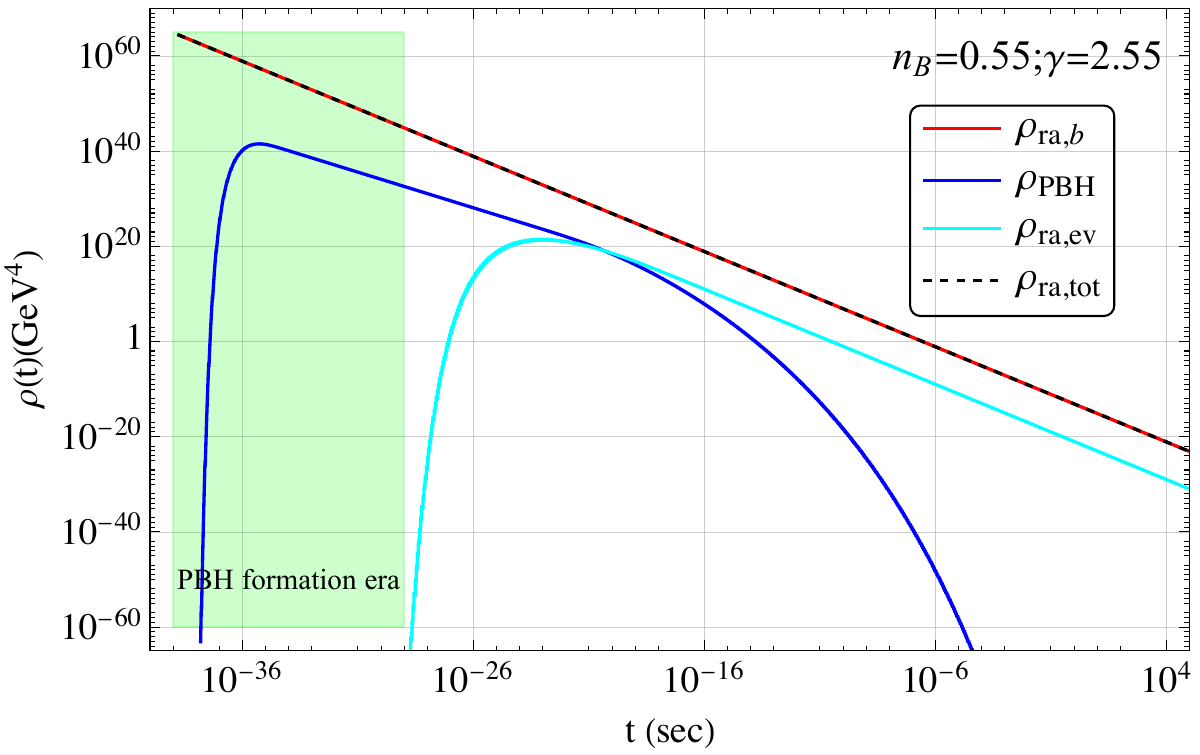}
\includegraphics[scale=0.29]{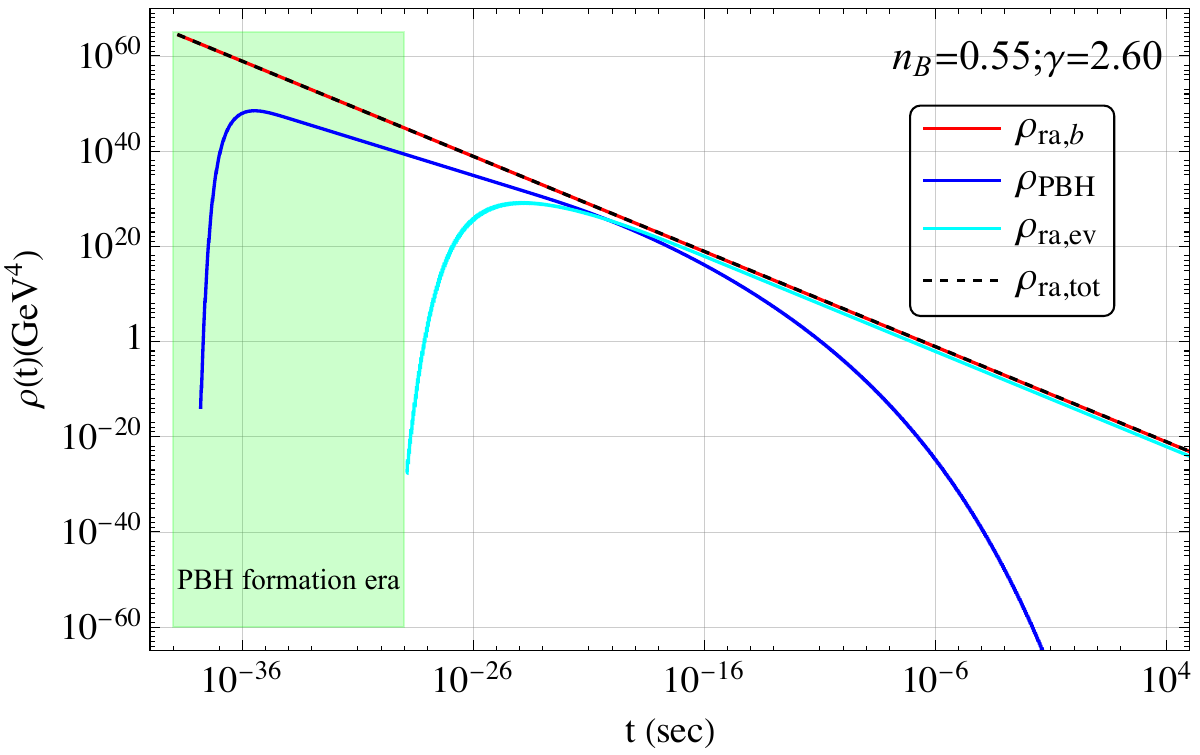}
\includegraphics[scale=0.29]{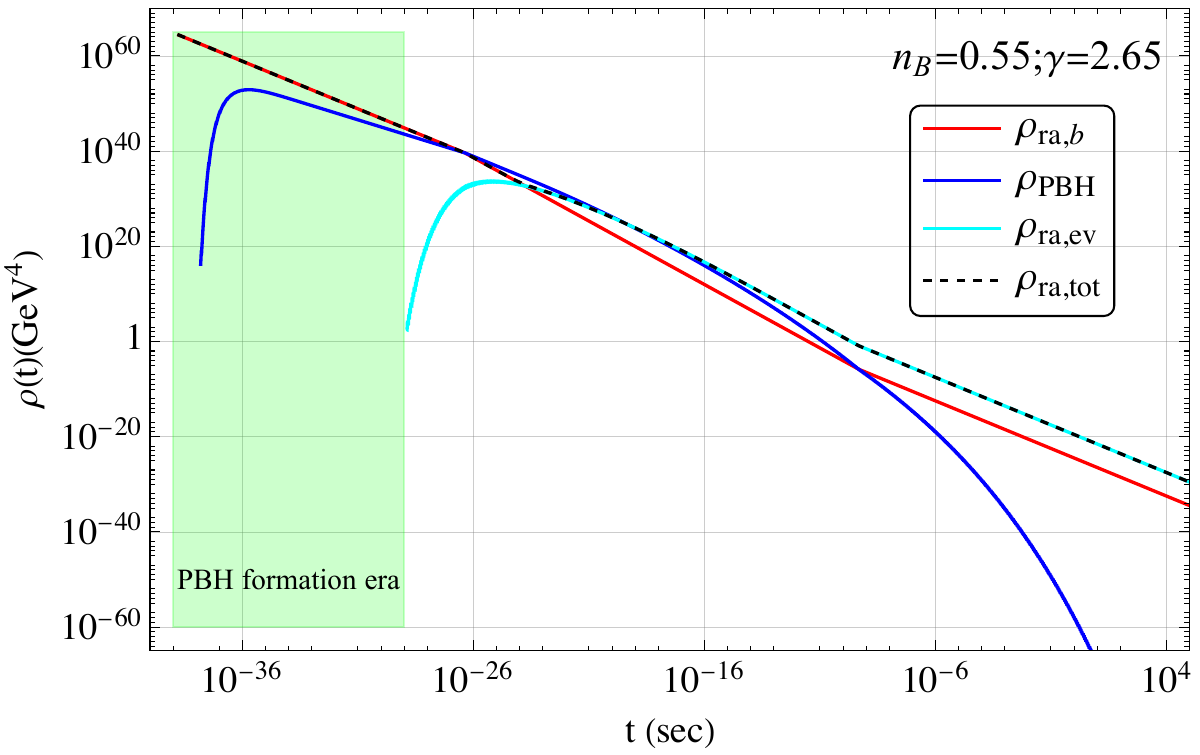}
\caption{In this figure, we have plotted the evolution of the energy density as a function of cosmic time $t$(in sec) for the different components. Here, the red line of each panle are indicating the background radiation energy density originating from the inflaton field. The blue line indicates the evolution of the PBH energy density, and the cyan line indicates the contribution of the radiation energy density came from the PBH evaporation. The black dashed lines indicate the total energy density of the radiation field. Here, the variable green shaded region main indicate the time scales where the PBH production is mainly effective. The different panels indicate three different values of helicity parameter $\gamma=2.55$(left), $\gamma=2.60$( middle), and $\gamma=2.65$(right). For each plot, we have fixed the magnetic spectral index $\nb=0.55$.}
\label{fig:rho_vs_t_g}
\end{figure}

As an illustrative example, for $\nb=0.55$ and $\gamma=2.65$, we find several PBH masses satisfying $\beta(M)>\beta_c(M)$. In particular, for $\Mpbh=100\,\gm$, the initial fractional energy density is $\beta(\Mpbh)\simeq5.8\times10^{-7}$, whereas the critical value required to obtain a PBH-dominated era is $\beta_c(\Mpbh)\simeq3.54\times10^{-8}$. This implies a PBH-dominated phase prior to evaporation, consistent with Fig.~\ref{fig:rho_vs_t_g}. By contrast, for $\nb=0.60,\,\gamma=2.65$ and $\nb=0.65,\,\gamma=2.65$, no significant mass range satisfies $\beta(M)>\beta_c(M)$, and therefore no PBH-dominated epoch is expected before BBN.
%%%%%%%%%%%%%%%%%%%%%%%%%%%%%%%%%%%%%%%%%%%%%%%%%%%%%%%%%
\begin{figure}[t]
\centering
\includegraphics[scale=0.43]{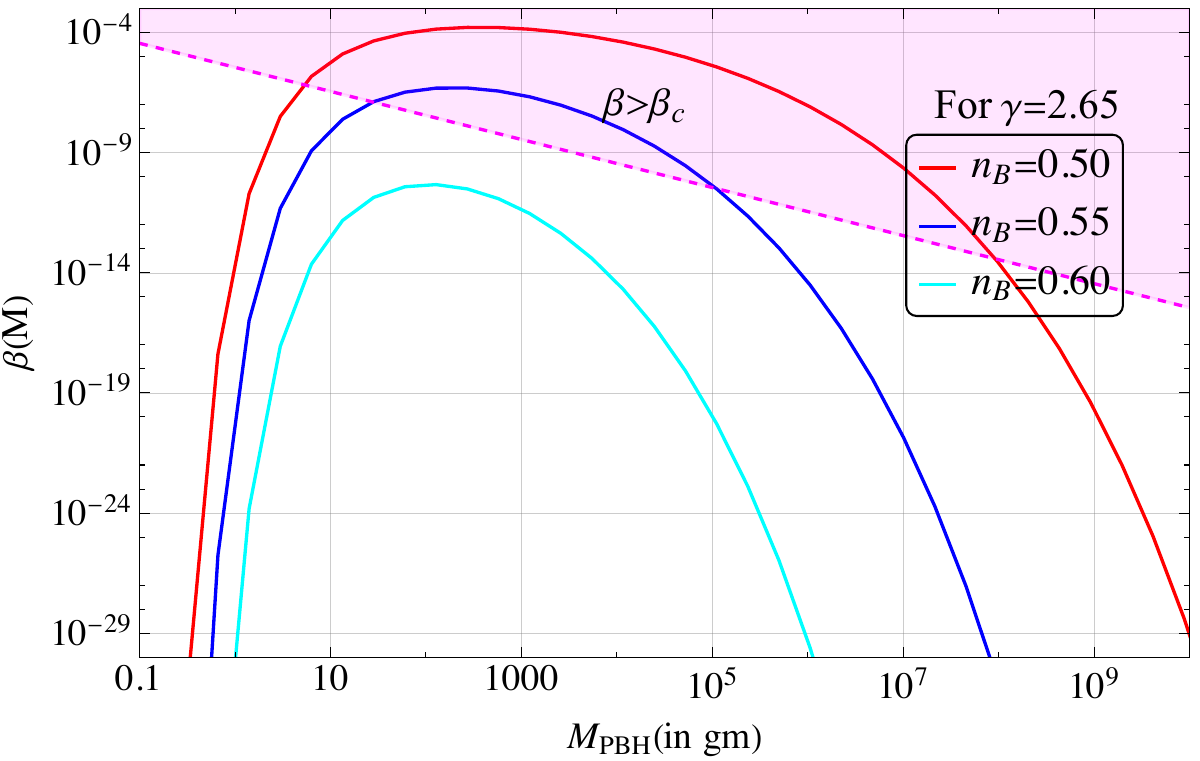}
\includegraphics[scale=0.43]{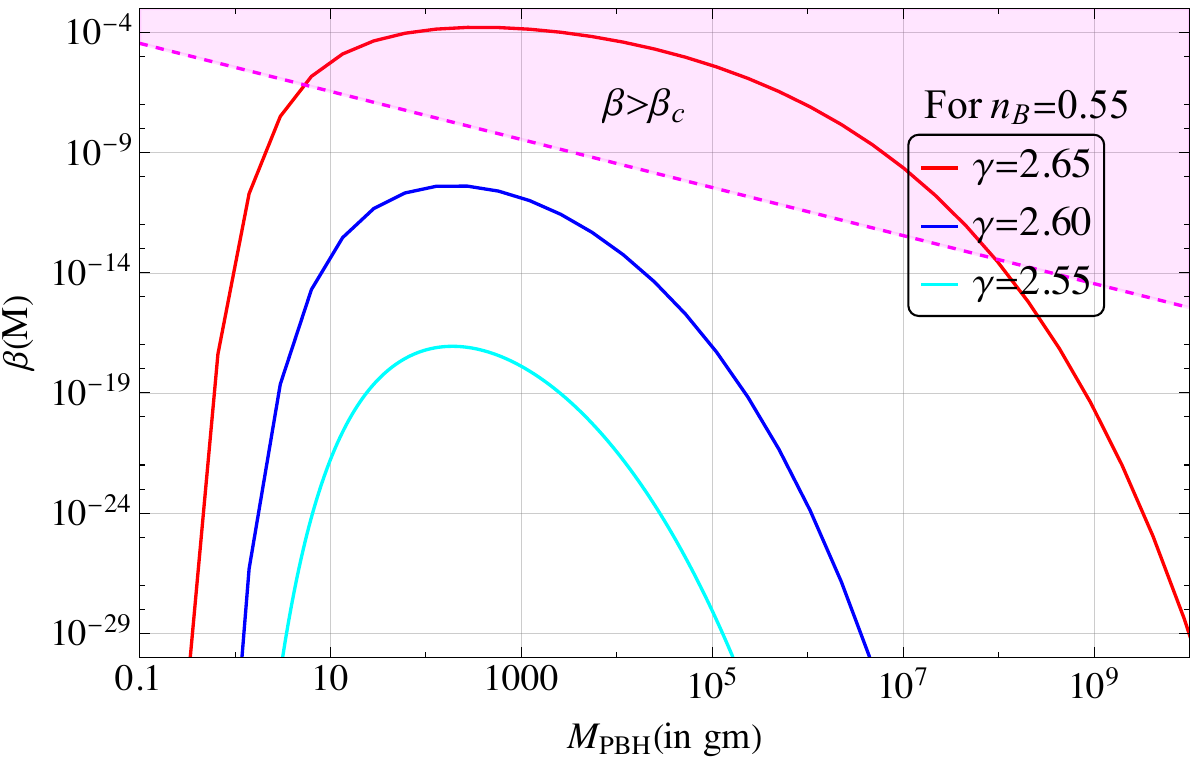}
\caption{n these figures, we present the initial fractional energy density of primordial black holes (PBHs) for a given mass under two distinct scenarios. In the left panel, the helicity parameter is fixed at $\gamma = 2.65$, while the three colored curves correspond to different values of the magnetic spectral index, $\nb = 0.50$, $0.55$, and $0.60$. In the right panel, the magnetic spectral index is held fixed, and the helicity parameter $\gamma$ is varied instead. In both panels, the magenta dashed line denotes the critical initial fractional energy density, $\beta_c(M)$, above which PBHs of a given mass dominate the energy density of the Universe before completely evaporating.
}
\label{fig:beta_M}
\end{figure}
%%%%%%%%%%%%%%%%%%%%%%%%%%%%%%%%%%%%%%%%%%%%%%%%%%%%%%%%%

In Fig.\ref{fig:beta_M}, we show the initial fractional energy density of the produced primordial black holes (PBHs), defined as $\beta \equiv \rhopbh/\rho_{\rm total}$, as a function of the PBH mass $M$ (in grams). The left panel corresponds to a fixed helicity parameter $\gamma = 2.65$, where the three curves represent different values of the magnetic spectral index: $\nb = 0.50$ (red), $0.55$ (blue), and $0.60$ (cyan). For a fixed value of $\gamma$, we find that increasing the magnetic spectral index leads to a suppression of the PBH energy density at the time of formation. This behavior can be traced back to the reduction of the total electromagnetic energy density. As shown in Eq.\eqref{eq:mPbe_inf}, the overall amplitude of the electric and magnetic fields depends exponentially on the coupling parameter $n$. An increase in $\nb$ effectively corresponds to a smaller value of $n$, thereby decreasing the total energy density stored in the generated electromagnetic fields.

We further observe that the peak of the PBH mass spectrum shifts toward smaller masses as $\nb$ increases. This trend can be understood from the fact that larger values of $\nb$ lead to a more blue-tilted magnetic power spectrum, with a larger fraction of the energy density concentrated around $k \simeq \ke$. As a result, the production of lower-mass PBHs is enhanced. The magenta dashed line indicates the critical initial fractional energy density $\beta_c(M)$ for a given PBH mass. We find that, for $\nb = 0.50$ and $0.55$, a broad range of PBH masses exceeds this critical threshold, implying that these PBHs dominate the energy density of the Universe before completely evaporating.

In the right panel of Fig.~\ref{fig:beta_M}, we plot the initial fractional energy density of PBHs, $\beta(M)$, as a function of the PBH mass $M$ (in grams) for a fixed value of the magnetic spectral index $\nb=0.55$. We consider three different values of the helicity parameter, $\gamma=2.55$, $2.60$, and $2.65$. Since the magnetic spectral index $\nb$ is held fixed, the spectral shapes of the inflationary electric and magnetic fields remain identical for all values of $\gamma$, with the only modification arising in the overall amplitude. Consequently, the peak position of the $\beta(M)$ spectrum remains unchanged.

As the amplitudes of the generated magnetic and electric fields depend exponentially on the helicity parameter $\gamma$ (see Eq.~\eqref{eq:mPbe_inf}), larger values of $\gamma$ lead to an enhancement of the total electromagnetic energy density. This, in turn, increases the overall PBH energy density at formation. Moreover, for higher values of $\gamma$, a wider range of overdensities satisfies the collapse condition $\delta>\delta_c$, resulting in a broader PBH mass spectrum.

The most important outcome is that this PBH production mechanism naturally generates a broad mass spectrum, spanning a wide range of mass windows.

\subsection{PBH dominated Universe}
In our scenario, the magnetic field induced curvature spectrum is so constructed that PBHs are formed immediately after inflation. As discussed earlier, depending on the values of the magnetic spectral index \( \nb \) and the helicity parameter \( \gamma \), a wide range of PBH masses can be generated (see Fig.~\ref{fig:beta_M}). For certain regions of parameter space, the initial PBH fraction exceeds the critical value \( \beta_c \), implying the formation of a PBH-dominated Universe. In the following subsection, we compute the subsequent evolution of the PBH energy density and the radiation background.
%%%%%%%%%%%%%%%%%%%%%%%%%%%%%%%%%%%%%%%%%%%%%%%%%%%
%%%%%%%%%%%%%%%%%%%%%%%%%%%%%%%%%%%%%%%%%%%%%%%%%%%
\begin{figure}[h]
\centering
\includegraphics[scale=0.43]{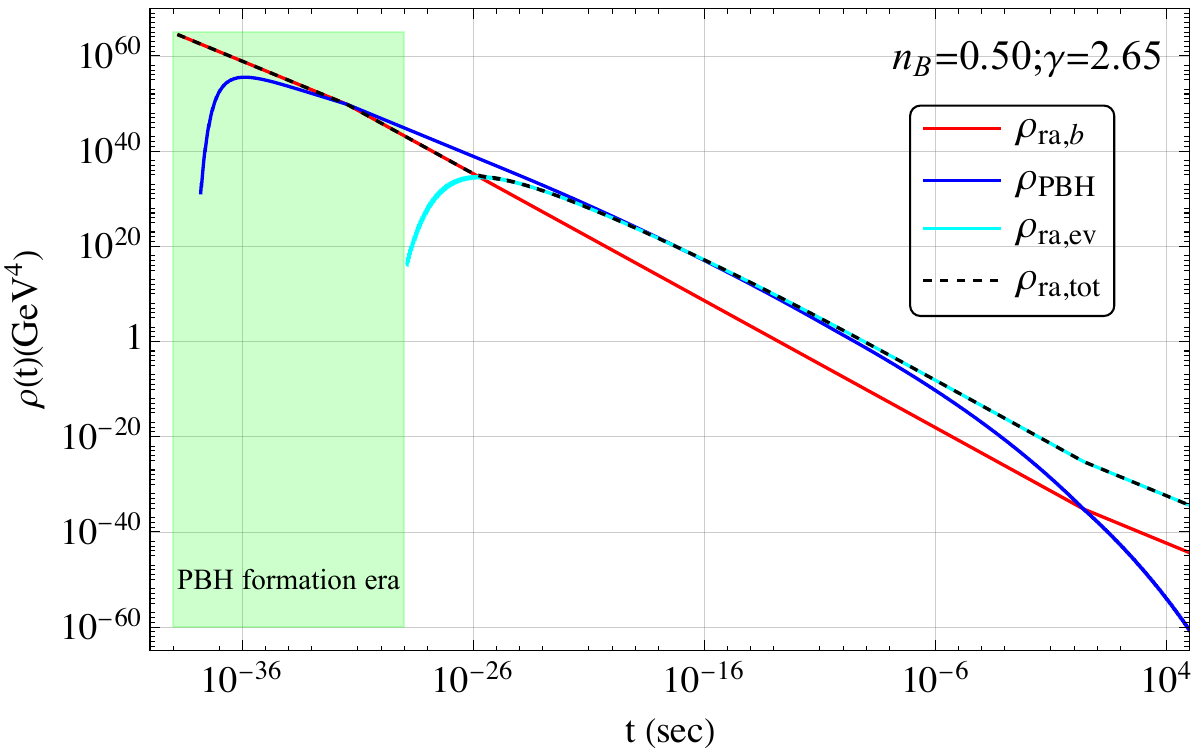}
\includegraphics[scale=0.43]{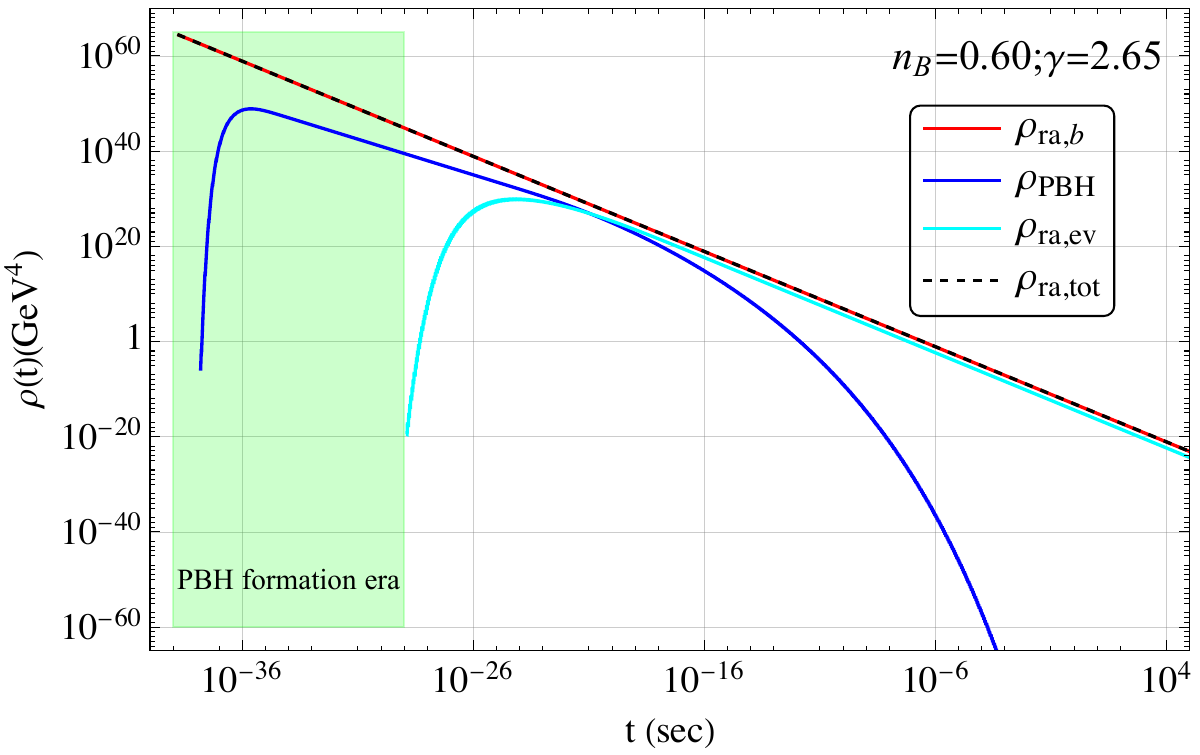}
\caption{In this figure, we show the evolution of the energy density as a function of cosmic time \( t \) (in seconds) for two different scenarios. The left panel corresponds to \( \nb = 0.50 \) with \( \gamma = 2.65 \), while the right panel shows the case \( \nb = 0.60 \) with \( \gamma = 2.65 \). In both panels, different colored curves represent the contributions from various energy-density components, with the black solid line denoting the total radiation energy density. The green shaded region indicates the epoch of maximal PBH production. In our scenario, however, the majority of PBHs are formed at the onset of the early radiation-dominated era. }
\label{fig:rho_vs_t_nb}
\end{figure}

\subsubsection{Computing the Evolution of the Radiation and PBH Energy Density}
 
Since PBHs are formed immediately after inflation, one can estimate the lowest possible PBH mass using $\Mpbh\simeq 4\pi\gc\Mp^2H^{-1}_{\tf}$. For a Hubble parameter $\HI\simeq 10^{13}\GeV$, the minimum PBH mass is $\Mpbh^{\rm min}\simeq 2.57\,\gm$. These ultra-light PBHs evaporate rapidly and thereby generate a thermal bath. 

If we require that such PBHs evaporate before the onset of BBN, or that their fractional abundance remains sufficiently small so that it does not effect the BBN process, one finds that PBHs with masses $\Mpbh^{\rm bbn}\simeq 10^9\,\gm$ evaporate before BBN. Using Eq.~\eqref{eq:tf}, the formation time of a $10^9\,$gm PBH is approximately $\th(10^9~\gm)\simeq 5.1\times 10^{-29}\,\sec$,
while the lifetime of the smallest-mass PBH is $\tau_{\rm ev}(\Mpbh=2.57\,\gm)\simeq 7\times 10^{-27}\,\sec$.
From this comparison, it is evident that the contribution from PBH evaporation becomes relevant only for $t\geq t_{\rm ev}(\Mpbh=2.57\,\gm)$.

Because our scenario predicts a broad PBH mass spectrum, determining the evolution of the PBH and radiation energy densities requires solving the Boltzmann equations, including Hawking evaporation. However, the PBH spectrum is highly non-linear and varies continuously with the magnetogenesis parameters, making a full numerical treatment technically involved. Since such an analysis lies beyond the main scope of this work, we defer it to future studies.
Instead, we employ a semi-analytic approach to estimate the background evolution. The continuous PBH mass distribution is decomposed into a set of discrete masses, denoted by $M_i$, and the total PBH contribution is obtained by summing over the full mass range.

We assume that a PBH of mass $M_i$ evaporates instantaneously at the end of its lifetime $\tau_{\rm ev}(M_i)$. Thus, a PBH formed at $\tf(M_i)$ completely evaporates at 
\[
t_{\rm ev}(M_i)=\tf(M_i)+\tau_{\rm ev}(M_i).
\]
Although approximate, this prescription captures the essential physics. Since PBHs are continuously formed over time, and subsequently evolve for $t>t_{\rm ev}(\Mpbh^{\rm min})$, the total PBHs energy density at time $t$ can be written as
\begin{align}
    \rhopbh(t)=\sum_i \beta(M_i)\rho_c(\tf(M_i))
    \left(\frac{a(\tf(M_i))}{a(t)}\right)^3
    \Theta\!\left(t_{\rm ev}(M_i)-t\right)
    \Theta\!\left(t-\tf(M_i)\right).
\end{align}
The first Heavisine function removes PBHs that have evaporated before time $t$, while the second one ensures that only PBHs formed before $t$ contribute.

The radiation energy density receives two contributions: the background radiation present due to the decay of the inflaton field, and the radiation coming from the PBH-evaporation. Hence, at any time $t$,
\begin{align}
    \rho_{\rm ra}(t)
    =\rho_c(t_e)\left(\frac{a(t_e)}{a(t)}\right)^4
    +\rho_{\rm ra}^{\rm ev}(t),
\end{align}
where $t_e$ denotes the cosmic time at the end of inflation, and $\rho_c(t_e)=3\HI^2\Mp^2$ 
is the background energy density at that epoch. The quantity $\rho_{\rm ra}^{\rm ev}(t)$ represents the radiation contribution from PBH evaporation.

Since the evaporation time of any PBH is much larger than its formation time, i.e.\ $t_{\rm ev}(M_i)\gg \tf(M_i)$, and since all PBHs are formed during the early radiation-dominated era (except in regions where a transient PBH-dominated era may appear depending on the parameters $\gamma$ and $\nb$), we adopt the instantaneous-evaporation approximation for each mass bin of the broad PBH spectrum.

The energy density associated with PBHs of mass $M_i$ at their formation time is
\begin{align}
    \rhopbh^{(i)}(\tf(M_i))
    =\beta(M_i)\rho_c(\tf(M_i)),
\end{align}
where $\rho_c(\tf)=3H^2(\tf)\Mp^2$ is the background energy density at PBH formation. After formation, PBHs behave as non-relativistic matter, and their energy density redshifts as $a^{-3}$ until evaporation. Thus, the energy density of PBHs with mass $M_i$ at the time of evaporation is
\begin{align}
    E_i\equiv\rhopbh(t_{\rm ev}(M_i))
    =\rhopbh(\tf(M_i))
    \left(\frac{a(\tf(M_i))}{a(t_{\rm ev}(M_i))}\right)^3 .
\end{align}

After complete evaporation, the released energy redshifts as radiation. Therefore, for times $t>t_{\rm ev}(M_i)$, the radiation energy density contributed by PBHs of mass $M_i$ is
\begin{align}
    \rho_{\rm ra,i}^{\rm ev}(t\geq\tev)
    =E_i\left(\frac{a(t_{\rm ev}(M_i))}{a(t)}\right)^4.
\end{align}
Since we know the discrete PBH mass bins and their corresponding initial fractional energy densities $\beta(M_i)$, the total radiation energy density from PBH evaporation is obtained by summing over all mass bins:
\begin{align}
    \rho_{\rm ra}^{\rm ev}(t)
    =\sum_i
    E_i\left(\frac{a(t_{\rm ev}(M_i))}{a(t)}\right)^4
    \Theta\!\left(t-t_{\rm ev}(M_i)\right).
\end{align}
Here, the Heaviside function ensures that only PBHs that have evaporated before time $t$ contribute to the radiation energy density.

In Fig.~\ref{fig:rho_vs_t_g}, we present the evolution of the radiation and PBH energy densities as functions of cosmic time \( t \) (in seconds) for three representative values of the helicity parameter:
\( \gamma = 2.55 \) (left panel), \( \gamma = 2.60 \) (middle panel), and \( \gamma = 2.65 \) (right panel),
while keeping the magnetic spectral index fixed at \( \nb = 0.55 \).

Different colored curves denote the evolution of the various energy-density components. The red curve represents the background radiation energy density originating from inflaton decay, the blue curve corresponds to the PBH energy density, and the cyan curve shows the radiation produced via PBH evaporation. In all panels, the black dashed curve indicates the total radiation energy density of the Universe.

We find that for the parameter sets \( \nb = 0.55 \) with \( \gamma = 2.55 \) (left panel) and
\( \nb = 0.55 \) with \( \gamma = 2.60 \) (middle panel), PBHs are produced efficiently; however, their initial fractional energy density is insufficient to trigger a PBH-dominated era. In the left panel, both the PBH energy density and the radiation from PBH evaporation remain subdominant compared to the background radiation sourced by inflaton decay. In contrast, in the middle panel, corresponding to \( \gamma = 2.60 \), the PBH energy density and the radiation generated by PBH evaporation grow at later times and become comparable to the background radiation component.

On the other hand, for \( \nb = 0.55 \) and \( \gamma = 2.65 \) (right panel), there exist epochs during which the PBH energy density dominates over the background radiation, leading to a PBH-dominated Universe. We also observe that once the PBH-dominated phase begins at
\( t_{\rm PBH} \simeq t_{\rm dom} \), the slope of the radiation energy-density evolution changes. This behavior reflects the transition in the cosmic expansion rate: during PBH (matter-like) domination, the scale factor evolves as \( a(t) \propto t^{2/3} \), whereas during radiation domination it follows \( a(t) \propto t^{1/2} \).

In Fig.~\ref{fig:rho_vs_t_nb}, we show the time evolution of the various energy-density components for a fixed value of the helicity parameter \( \gamma = 2.65 \). The left panel corresponds to \( \nb = 0.50 \), while the right panel shows the case \( \nb = 0.60 \). As discussed earlier, the green shaded region denotes the epoch during which PBHs are efficiently produced.

In both panels, the red curve represents the background radiation energy density originating from inflaton decay, the blue curve denotes the PBH energy density, and the cyan curve corresponds to the radiation produced via PBH evaporation. We observe that for \( \nb = 0.50 \), compared to \( \nb = 0.60 \), the PBH-dominated phase persists for a significantly longer duration.
This behavior can be understood from the dependence of the magnetic-field strength on the parameter \( n \), as given in Eq.~\eqref{eq:mPbe_inf}. Recall that \( n \) is the parameter associated with the coupling function \( I \) (see Eq.~\eqref{eq:coupling_function}), and the magnetic spectral index is related to \( n \) via
\( \nb = 4 - 2|n| \).
Therefore, a smaller value of \( \nb \) corresponds to a larger value of \( |n| \). 
 Since the magnetic-field strength depends exponentially via the dependence of $\cosh{(n\pi\gamma)}$ on \( n \) (see Eq.\ref{eq:fn}), lowering the magnetic spectral index for a fixed coupling function effectively enhances the magnetic-field energy density.

Because the density contrast is proportional to the magnetic-field energy density, this enhancement leads to a more efficient production of PBHs. As a result, for smaller values of the magnetic spectral index and a fixed helicity parameter, the resulting PBH energy density is sufficient to trigger a PBH-dominated era.

From this analysis, we conclude that achieving a long-lived PBH-dominated phase requires either a smaller magnetic spectral index for a fixed helicity parameter or a larger helicity parameter for a fixed value of the magnetic spectral index.

\section{Gravitational waves}
In our scenario, we have seen that a tilted magnetic field generated during inflation can also generate significant curvature perturbations in a slow-roll inflationary background. For a certain value of the magnetic field strength, the induced curvature power spectrum can be larger than the threshold value of the curvature power spectrum $\mPc(k>>k_*)\mathcal{O}(10^{-2})$, which can easily form PBHs at the early stage of our universe (just after the end of inflation). Moreover, depending on the magnetic field strength, controlled by the magnetic field parameters $\nb$ and $\gamma$, the initial fractional energy density of the PBHs is sufficient to easily lead to a PBH-dominated era. 

Thus, there are different ways of generating gravitational waves (GWs). In this paper, we focus mainly on two different sources of GWs; the magnetic field induced GWs, which encodes the signature of magnetognesis mechanism itself ~\cite{Sorbo:2011rz, Caprini:2014mja, Ito:2016fqp, Sharma:2019jtb, Okano:2020uyr, Maiti:2025cbi, Maiti:2025rkn, Maiti:2024nhv, Bhaumik:2025kuj}, and PBH evaporation.

\subsection{Primary GWs}
During inflation, the quantum fluctuations of the gravitational field itself are stretched to cosmological scales, giving rise to a stochastic background of primordial gravitational waves. In the absence of any kind of source terms, the EoM for the tensor fluctuations $h_{ij}(\vx,\eta)$ is simply governed by ~\cite{Starobinsky:1979ty, Grishchuk:1974ny, Guzzetti:2016mkm, Haque:2021dha}
\begin{align}
    h_{ij}''(\vx,\eta)+2\mH h_{ij}'(\vx,\eta)-\nabla^2 h_{ij}(\vx,\eta)=0,
\end{align}
where $h_{ij}$ is the transverse stressless tensor, i.e., $\partial^i h_{ij}=h^i_i=0$. Now we can decompose the $h_{ij}(\vx,\eta)$ in term of the Fourier modes, say, $\hkl(\eta)$ as follows~\cite{Maiti:2024nhv}
\begin{align}
    h_{ij}(\vx,\eta)=\sum_{\lambda\pm}\int \frac{d^3\vk}{(2\pi)^3}e^{\lambda}_{ij}(\hat{k})\hkl(\eta)e^{i\vk\cdot\vx},
\end{align}
where $e^{\lambda}_{ij}(\vk)$ is the polarization tensor corresponding to the mode $\vk$ and $\lambda$ is the polarization index.

For a quais-de sitter spacetime, the solution of the tensor power spectrum
\begin{align}
    h_{\vk}^{\lambda}(\eta) = \frac{\sqrt{-\eta}}{\Mp\, a(\eta)} e^{i(\nu+1/2)\pi/2}\mathrm{H}^{(1)}_{\nu}(-k\eta)\,,
\end{align}
here $\nu=\epsilon+3/2$, we recall that $\epsilon$ is the slow-roll parameter~\cite{PhysRevD.50.7222,Di:2017ndc, Balaji:2022dbi, Figueroa:2021zah, PhysRevD.28.679}.
For the vacuum solution, both polarization modes have been excited equally. We can define the tensor-power spectrum associated with the quantum fluctuations as $\mPt(k,\eta)=\frac{k^3}{2\pi^2}\sum_{\lambda=\pm}\langle h^{\lambda}_{\vk}(\eta)h^{\lambda *}_{\vk}(\eta)\rangle$ as~\cite{MUKHANOV1992203, PhysRevD.89.123503, Baumann:2009ds}
\begin{align}
    \mPtv(k,\eta)\simeq \frac{8}{\Mp^2}\l(\frac{\HI}{2\pi}\r)^2\l(\frac{k}{aH}\r)^{-2\epsilon}
\end{align}
In de-sitter inflationary background, as $\epsilon=0$, the tensor power spectrum at super-horizon scales is simply scale invariant in nature,i.e., $\mPtv(k<\ke)\propto\HI^2$, which is a well known results~\cite{Maiti:2024nhv,Maiti:2025cbi,Maiti:2025ijr,Chakraborty:2024rgl,Haque:2021dha,Hoory:2025qgm}

After the end of inflation, the tensor modes evolve through different phases, and depending on the evolution history, the nature of the tensor power spectrum will be modified. If we simply introduced a non-trivial reheating era, described by the equation of state $\wre$, then during the radiation-dominated era, the produced tensor power spectrum due to the vacuum fluctuations is written as~\cite{Maiti:2025awl, Maiti:2025cbi, Maiti:2024nhv, Chakraborty:2024rgl}
\begin{align}
    \mPt^{\rm vac}(\eta>\ere)\simeq 
    \l\{
    \begin{matrix}
        \mPt^{\rm vac}(k,\ee) & k\leq \kre\\
        (k/\kre)^{2l(\wre)+1}\mPt^{\rm vac}(k,\ee) & \kre<k<\ke
    \end{matrix}
    \r.
\end{align}
where $l(\wre)=3(\wre-1)/2(1+3\wre)$. Here it has been clear that the non-trivial reheating phase has a significant modification to the tensor power spectrum for those modes which are inside the horizon at the end of reheating, i.e., $k>\kre$ (for details see~\cite{Haque:2021dha, Maiti:2025cbi, Maiti:2024nhv}).

\subsection{SGWs from IMFs}
The magnetic field can source gravitational waves (GWs) at different stages of the cosmic evolution. During inflation, magnetic fields can be generated with significant amplitude, and hence  contribute to GW production. 
After inflation, we consider an inflaton behaving like a radiation, and it further decays into relativistic particles to reproduce hot universe. 
During this reheating phase PBHs form, and depending on its initial abundance, they may eventually dominate the energy density of the universe, leading to a matter like PBH-dominated phase.

Therefore, after inflation, there are two distinct stages during which magnetic fields can efficiently source GWs: the early radiation-dominated (eRD) era and the PBH-dominated era. However, as shown in \cite{Maiti:2024nhv, Maiti:2025cbi}, during a matter-like phase the fractional energy density of the magnetic field evolves as $\delta_{\rm B}=\rhob/\rho_c\propto a^{-1}$. Consequently, the effective anisotropic stress sourced by the magnetic field is significantly diluted, suppressing its ability to generate substantial secondary gravitational waves (SGWs) in later radiation-dominated phases.

As magnetic fields act as a source of tensor perturbations, we begin with the well-known equation of motion (EoM) for the metric perturbation $h_{ij}(\vx,\eta)$ in position space~\cite{Sorbo:2011rz, Caprini:2014mja, Ito:2016fqp, Sharma:2019jtb, Okano:2020uyr, Maiti:2025cbi, Maiti:2025rkn, Maiti:2024nhv, Bhaumik:2025kuj}.
\begin{align}
    h_{ij}''(\vx,\eta)+2\mH h_{ij}'(\vx,\eta)-\nabla^2 h_{ij}(\vx,\eta)=\frac{2}{\Mp^2}P^{lm}_{ij}T^{\rm EM}_{lm}(\vx,\eta)
\end{align}
where $\mH=a'/a$ is the comoving Hubble parameter, $P^{lm}_{ij}=P^l_iP^m_j-P_{ij}P^{lm}/2$, is the transverse tressless projector with $P_{ij}=\delta_{ij}-\partial_i\partial_j/\Delta$ and $T^{\rm EM}_{lm}(\vx,\eta)$ is the source term due to EM field.  Now in Fourier space, the EoM for the tensor fluctuation is governed by
\begin{align}\label{eq:hk_fourier}
    {\hkl}''(\eta)+2\frac{a'}{a}{\hkl}'(\eta)+k^2\hkl=\mathcal{S}^{\lambda}_{\vk}(\eta)
\end{align}
where the source term due to the EM field is defined as~\cite{Sorbo:2011rz, Caprini:2014mja, Ito:2016fqp, Sharma:2019jtb, Okano:2020uyr, Maiti:2025cbi, Maiti:2025rkn, Maiti:2024nhv, Bhaumik:2025kuj}
\begin{align}
    \mathcal{S}_{\vk}^{\lambda}(\eta)=-\frac{2}{\Mp^2}e^{ij}_\lambda(\hat{k})\int \frac{d^3\vk}{(2\pi)^3}[E_i(\vq,\eta)E_j(\vk-\vq,\eta) +B_i(\vq,\eta)B_j(\vk-\vq,\eta)]
\end{align}
As our source is helical in nature, it is best to work with circular polarization basis~\cite{Sorbo:2011rz, Caprini:2014mja, Sharma:2019jtb}. Now the tensor power spectrum due to the helical EM field is defined as~\cite{ Maiti:2025cbi, Maiti:2025rkn, Maiti:2024nhv}
\begin{align}\label{eq:pt_def}
    \mPts(k,\eta) &=\frac{2}{\Mp^4}\int_0^{\infty} \frac{dq}{q}\int_{-1}^1\d \mu \frac{F_1(\mu\beta,\lambda)}{[1+(q/k)^2-2\mu(q/k)]^{3/2}}\nn\\
    &\times \l[ \int_{\eta_i}^{\eta}\d \eta\, a^2(\eta_1)\mGk(\eta,\eta_1)(\mPb^{1/2}(q,\eta_1)\mPb^{1/2}(|\vk-\vq|,\eta)+\mPe^{1/2}(q,\eta_1)\mPe^{1/2}(|\vk-\vq|,\eta)) \r]^2
\end{align}
For the helical magnetic field, we have defined $F_1(\mu,\beta,\lambda)=(1+\lambda\mu)^2(1+\lambda\beta)^2$, with $\mu=\hat{k}\cdot\hat{q}$ and $\beta=\widehat{\vk-\vq}\cdot\hat{k}$. Here $\eta_i$ is the starting time, when the EM field is produced, and $\eta$ is the conformal time when we compute the tensor power spectrum.  
There are different phases where the EM field can source the tensor fluctuations and produce a significant amount of the GWs. 

\paragraph{\underline{Production during inflation}\\}
To compute the tensor power spectrum during inflation, we employ the Green's function method. The functional form of the Green's function is given by
\begin{align}
    \mGk(x,x_1)=\frac{1}{W(x_1)}\l[ h_1(x)h_2(x_1)-h_1(x_1)h_2(x) \r]\Theta(x-x_1)
\end{align}
where $h_1$ and $h_2$ are two independent solutions of the homogeneous tensor perturbation equation defined in Eq.\eqref{eq:hk_fourier}. Here, $\Theta$ denotes the Heaviside step function, and $W(x_1)$ is the Wronskian, defined as
\begin{align}
    W(x_1)=h_1(x_1)\partial_{x_1}h_2(x_1)-\partial_{x_1}h_1(x_1)h_2(x_1).
\end{align}

During inflation, including slow-roll corrections, the scale factor evolves as $a(\eta)=(-\HI \eta)^{-1-\epsilon}$, where we assume that the variation of the slow-roll parameter is negligible over time. Substituting this form into Eq.\eqref{eq:hk_fourier}, the two independent solutions of the tensor perturbation equation are given by
\begin{align}
    h_1(x_1)=x_1^{\frac{3}{2}+\epsilon}\bJ_{\frac{3}{2}+\epsilon}(x_1);\,\,\,\,
    h_2(x_1)=x_1^{\frac{3}{2}+\epsilon}\bY_{\frac{3}{2}+\epsilon}(x_1).
\end{align}
Here, $\bJ(x)$ and $\bY(x)$ denote the Bessel functions of the first and second kind, respectively.
Using these solutions, the Green's function becomes
\begin{align}
    \mGk(x,x_1)=\frac{\pi}{2}x^{\frac{3}{2}+\epsilon}x_1^{-\frac{1}{2}-\epsilon}\l[ \bJ_{\frac{3}{2}+\epsilon}(x)\bY_{\frac{3}{2}+\epsilon}(x_1) -\bJ_{\frac{3}{2}+\epsilon}(x_1)\bY_{\frac{3}{2}+\epsilon}(x)\r].
\end{align}

We are interested in the regime where all modes are outside the horizon at the end of inflation. Moreover, the gauge field is excited only after horizon exit. In this limit, the Green's function simplifies to $\mGk(x,x_1)\simeq -\frac{x_1}{3+\epsilon}$.

On the other hand, for a pure de Sitter background with $\epsilon=0$, the Green's function during inflation is well known~\cite{Maiti:2024nhv, PhysRevD.85.023534,Teuscher:2025xke,Teuscher:2025jhq} and is given by
\begin{align}\label{eq:mgk_inf}
\mGk^{\rm inf}(\eta,\eta_1)=\frac{\Theta(\eta-\eta_1)}{k^3\eta_1^2} \l[(1+k^2\eta_1^2)\sin(k(\eta-\eta_1))+k(\eta_1-\eta)\cos(k(\eta-\eta_1))\r].
\end{align}

Using this expression, the tensor power spectrum induced by the electromagnetic (EM) field at the end of inflation can be written as~\cite{Teuscher:2025xke}
\begin{align}
    \mPti(k,\ee)\simeq 2 \frac{\HI^4\mcB^2}{\Mp^4}\l(\frac{k}{a_*\HI}\r)^{-4\epsilon} \l( \int_1^{\xe} d x_1 \frac{\mGki(\xe,x_1)}{x_1^{2(1+\epsilon)}}x_1^{\nb}\r)^2 \int_{\umin}^{\umax} \frac{du}{u}\int_{-1}^1\d \mu \frac{F_1(u,\mu)}{[1+u^2-2\mu u]^{(3-\nb)/2}}.
\end{align}

After performing the above integrals, the tensor power spectrum induced by the EM field at the end of inflation can be approximated as~\cite{PhysRevD.85.023534, Teuscher:2025xke, Teuscher:2025jhq}
\begin{align}
    \mPti(k,\ee)\simeq \frac{2(1+\gamma^2)\HI^4}{\Mp^4}\mI_{\rm inf}^2\l(\frac{k}{\kpv}\r)^{-4\epsilon}{\Fn}(k)\mcB^2(\nb,\gamma),
\end{align}
where $\mI_{\rm inf}^2\simeq \l\{ 1-(k/\ke)^{\nb-2\epsilon}\r\}/3\nb$~\cite{Maiti:2025cbi}$,$ and
\begin{align}
    \Fn(k,\nb)\simeq \frac{8}{3\nb}\l[ 1-\l(\frac{\kpv}{k}\r)^{\nb} \r] + \frac{56}{15(2\nb-3)}\l[ \l(\frac{\ke}{k}\r)^{2\nb-3}-1\r].
\end{align}
We recall that $\mcB(\nb,\gamma)$ has already been defined in Eq.~\eqref{eq:fn}. For an exact de Sitter background with $\epsilon=0$, the induced tensor power spectrum reduces to $\mPti(k,\ee)\simeq (\HI/\Mp)^4$, corresponding to a scale-independent spectrum for a blue-tilted magnetic field. If the magnetic spectrum is red tilted, the induced tensor spectrum becomes scale dependent, $\mPti(k,\ee)\propto k^{-4\epsilon+\nb}$, where $\nb<0$. The overall amplitude is also controlled by $\mcB(\nb,\gamma)$, which depends exponentially on the coupling parameter $n(=2-\nb/2)$ and the helicity parameter $\gamma$. Therefore, varying these parameters significantly affects the tensor signal, as discussed later. We also note that, for a blue-tilted magnetic spectrum with slow-roll corrections, the induced and primary tensor spectra acquire different spectral tilts: the sourced contribution scales as $k^{-4\epsilon}$, whereas the vacuum tensor spectrum scales as $k^{-2\epsilon}$~\cite{Teuscher:2025xke,Teuscher:2025jhq}.

\paragraph{\underline{Production of the tensor power spectrum during early-radiation domination}\\}
After inflation, there is no further production of the magnetic field, as the electromagnetic field restores its conformal invariance. So the magnetic field evolves adiabatically and its energy dilution is $a^{-4}$. In the radiation-dominated background, the Green's function associated with the tensor fluctuation is~\cite{Maiti:2024nhv}
\begin{align}\label{eq:greens_erd}
    \mGkerd(\eerd,\eta_1)=\Theta(\eerd-\eta_1)\frac{\eta_1}{k\,\eerd}\sin(k(\eta_1-\eerd)).
\end{align}
Here $\eerd$ is the conformal time  at the end of radiation domination. As the magnetic energy density evolves adiabatically, it comoving spectral energy density can be defined as
\begin{align}
\tmPb(k)=\ke^4\,\mcB(\nb,\gamma)\l(\frac{k}{\ke}\r)^{\nb} .
\end{align}
We recall that $\ke$ is the comoving wavenumber that left the horizon at the end of inflation. The tensor power spectrum associated with the magnetic field during the eRD era can be written as
\begin{align}
    \mPt^{\rm sec}(k,\eerd)=\frac{2\HI^4}{\Mp^4}\mFnb(k)\,\mcB^2(\nb,\gamma)\l(\frac{k}{\ke}\r)^{2\nb}\overline{\mI_{\rm eRD}^2(k,\eerd,\ee)}
\end{align}
where overline denotes the oscillation average of the quantity and $\mI_{\rm eRD}(k,\eerd,\ee)$ is defined as
\begin{align}
    \mI_{\rm eRD}(k,\eerd,\ee)=\int_{\xe}^{\xerd}\d x_1\,x_1^{-2}\tmGkerd(\xerd,x_1)
\end{align}
Here $\tilde{\mathcal{G}}_k(x,x_1)=k\mGk(x,x_1)$, which is a dimensionless quantity.

\paragraph{\underline{Production of the tensor power spectrum during PBH-domination}\\}
After the production of the PBHs, their energy density evolves as $a^{-3}$ and once the there energy density dominates our universe, then our universe behaves matter-like fluid. In the matter-dominated background, the Green's function assumes
\begin{align}
    \mGkmd(\ere,\eta_2)=\Theta(\ere-\eta_2)\frac{k \eta_2^3}{\ere}\l\{ j_1(k\eta_2)y(k\ere)-j_1(k\ere)y(k\eta_2)\r\}
\end{align}
Similarly, we can write the tensor-power spectrum in the following fashion
\begin{align}
    \mPtsmd(k,\ere)=\frac{2\HI^4}{\Mp^4}\mFnb(k) \mcB^2(\nb,\gamma)\l(\frac{k}{\ke}\r)^{2\nb}\overline{\mIemd^2(k,\ere,\eerd)}
\end{align}
where we defined $\mIemd(k,\ere,\eerd)$ as
\begin{align}
    \mIemd(k,\ere,\eerd)=\xerd^2\times\int_{\xerd}^{\xre}\d x_2\, x_2^{-4}\,\tmGkmd(\xre,x_2)
\end{align}

\paragraph{\underline{Production of the tensor power spectrum during late-radiation domination}\\}
After the PBH-dominated era ends, the universe transitions into the standard radiation-dominated epoch, during which the magnetic field component continues to exist. The anisotropies generated by inflationary magnetic fields can still act as a source for stochastic gravitational waves (SGWs) up to the epoch of neutrino decoupling, which occurs at $T_\nu \simeq 1\,\text{MeV}$. Once neutrinos decouple from the thermal plasma, they efficiently erase magnetic field anisotropies, thereby suppressing any further production of tensor fluctuations sourced by the magnetic field~\cite{PhysRevD.70.043011, KAGRA:2021kbb}.

The functional form of the Green’s function in the radiation-dominated era has already been given in Eq.~\eqref{eq:greens_erd}, allowing us to identify $\tmGkerd(x,x_1) = \tmGkrd(x,x_1)$.
Using this relation, the tensor power spectrum sourced by the magnetic field during the late radiation-dominated era can be expressed as
\begin{align}
    \mPtsrd(k,\eta_\nu)=2\l(\frac{\HI}{\Mp}\r)^4\mcB^2(\nb,\gamma)\mFnb(k)\l(\frac{k}{\ke}\r)^{2\nb}\l(\frac{\xerd}{\xre}\r)^4\overline{\mI_{\rm RD}(k,\eta_\nu,\ere)}^2
\end{align}
Similarly, $\mI_{\rm RD}(k,\eta_\nu,\ere)$ is define as
\begin{align}
    \mI_{\rm RD}(k,\eta_\nu,\ere)=\int_{\xre}^{x_\nu}\d x_3\,x_3^{-2}\tmGkrd(\xre,x_3)
\end{align}
As we have seen, there is an extra damping factor associated with the tensor power spectrum, i.e., $(\xerd/\xre)^4$. This additional suppression arises due to the relative dilution of the magnetic field compared with the background during the PBH-dominated era. For a matter-like background, the background energy density dilutes as $\rho \propto a^{-3}$, whereas the magnetic field energy density dilutes as $\rho_{\rm B} \propto a^{-4}$. Since the source dilutes faster than the background, by the end of the PBH-dominated era the anisotropies induced by the magnetic field are significantly reduced. Consequently, the secondary production of tensor fluctuations in the radiation-dominated era, compared with the previous PBH-dominated era, becomes suppressed. Therefore, in the presence of a PBH-dominated phase, we can safely neglect the contribution arising from the magnetic field during the subsequent radiation-dominated era.

\paragraph{\underline{The present-day fractional energy density in gravitational waves}\\}
The gravitational waves weekly interact with ordinary matter as well, and their self-interaction is negligible. In general scenarios, for the sub-hubble scale, the GWs propagate freely through space after their production. Due to the expansion of our universe, the GW energy density decreases similarly to radiation, i.e, $\rho_{\rm gw}\propto a^{-4}$. So, during the radiation-dominated era, the relative energy density of the produced GW is conserved. So, the fractional energy density of the GW energy density  per logarithmic frequency interval, normalized by the critical energy density today, i.e., $\rho_{\rm c}=3H_0^2\Mp^2$, is defined as~\cite{Haque:2021dha, Maiti:2025cbi, Maiti:2025rkn, Maiti:2024nhv}
\begin{align}
    \Omega_{\rm gw}h^2(f)=\frac{\Omega_{\rm ra}h^2}{12}\l(\frac{g_{*,0}}{g_{*,\rm eq}}\r)^{1/3}\mPt^{\lambda}(f)
\end{align}
where $\mPt^{\lambda}(f)$, is the total produced tensor power spectrum defined at the time of the matter-radiation equality and {$f(=2\pi/k)$ is the comoving frequency, we observed today. Here $g_{*,\rm eq}\simeq g_{*,0}\simeq 3.36$  are the relativistic degrees of freedom at the time of matter-radiation equality and the present day, respectively. Her $\Omega_{\rm ra}h^2\simeq 4.3\times 10^{-5}$ is the fractional energy density of the radiation component at present-day.

\paragraph{\underline{GWs in the presence of PBH-dominated era}:\\}
After inflation, the universe first enters into the early-radiation-dominated era (e-RD), followed by a PBH-dominated era (e-MD), and finally transitions into the standard radiation-dominated era (RD). Due to this intermediate matter-dominated phase, there is an additional suppression of modes that re-enter the horizon during e-MD. This suppression arises from the relative dilution of the GW energy density compared to the background.  

The primordial GWs (PGWs) produced from vacuum fluctuations during inflation can then be written as~\cite{Teuscher:2025xke, Teuscher:2025jhq}
\begin{align}
    \ogwp(f)\simeq\frac{\omegara}{6\pi^2}\l(\frac{\HI}{\Mp}\r)^2\l(\frac{f}{f_{*}}\r)^{-2\epsilon}\times
    \l\{
    \begin{matrix}
        1 & f_*<f<\fre\\
        (f/\fre)^{-2} & \fre<f<\fpbh\\
        (\fpbh/\fre)^{-2} & \fpbh<f<\fe
    \end{matrix}
    \r.
\end{align}
where $\fpbh$ and $\fe$ correspond to the comoving frequencies at the PBH-dominated epoch and at the end of inflation, respectively. Here, $f_{\rm re}$ denotes the comoving frequency at the end of the PBH-dominated era, marking the onset of the standard radiation-dominated (RD) phase.

As discussed earlier, modes that remain outside the horizon at the end of reheating retain an approximately scale-invariant spectrum, since $\epsilon\sim 10^{-4}$. In contrast, modes that re-enter the horizon during the PBH-dominated era acquire an additional scale dependence due to the effective matter-like evolution of the background. In this case, the gravitational wave spectrum scales as $\ogwh(f)\propto f^{-2(\epsilon+1)}$.

For modes re-entering during the early radiation-dominated era, the spectrum follows $\ogwh(f>\fpbh)\propto f^{-2\epsilon}$, as expected in standard scenarios. However, the presence of an intermediate PBH-dominated phase leads to an overall suppression of the amplitude in this frequency range.

In the absence of a PBH-dominated era after inflation, the primordial gravitational wave spectrum would simply scale as $\ogwh(f)\propto f^{-2\epsilon}$ across all relevant frequencies.

%%%%%%%%%%%%%%%%%%%%%%%%%%%%%%%%%%%%%%%%%%%%%%%%%%%
\begin{figure}[t]
\centering
\includegraphics[scale=0.43]{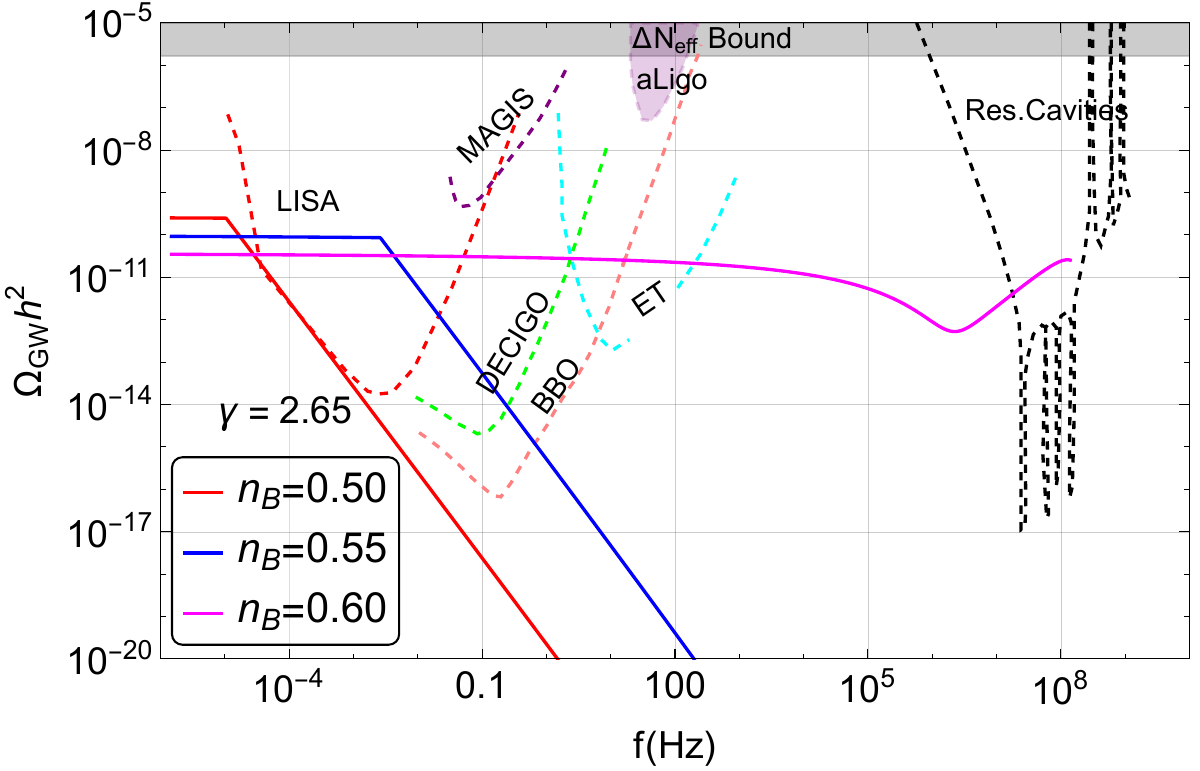}
\includegraphics[scale=0.43]{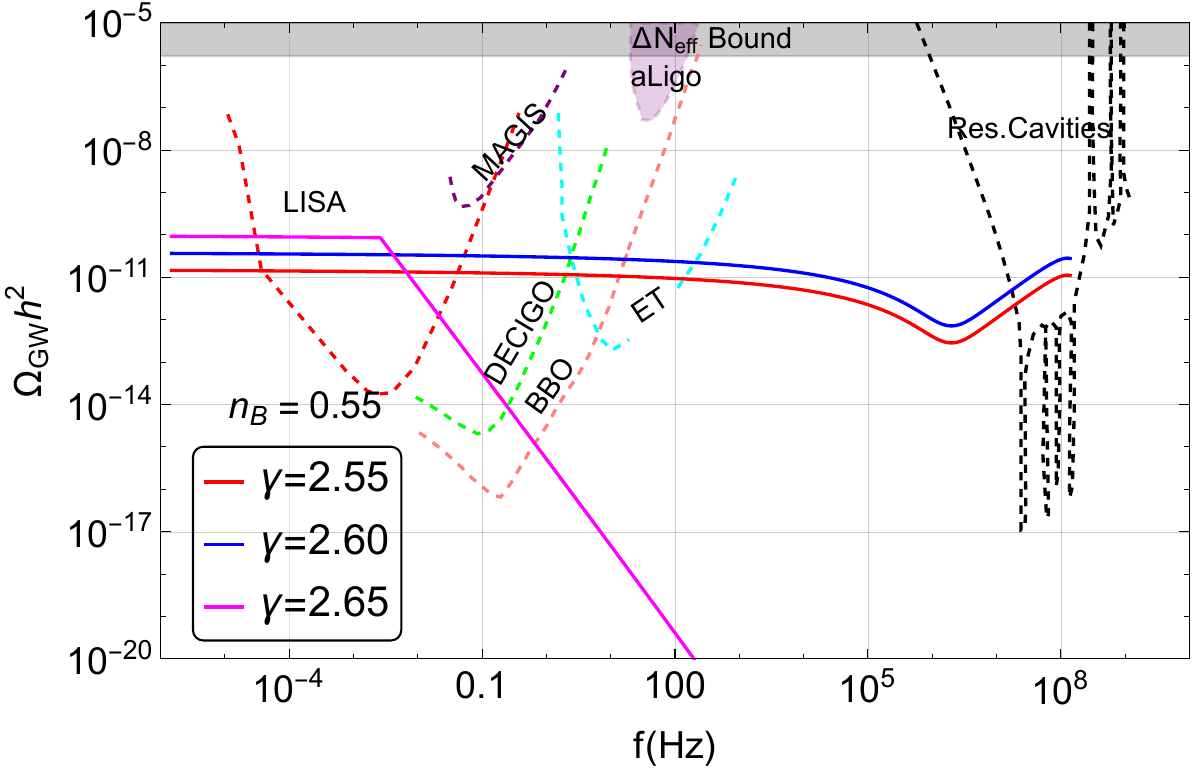}
\caption{In the above figure, we have plotted the total SED of the present-day GWs 
$\ogwh(f)$ as a function of the present-day observable frequency 
$f \, (\text{Hz})$ for different sets of paramerters.
In the left panel, we have plotted for a fixed value of helicity parameter $\gamma$ where three different colors indicate three different values of magnetic spectral index $\nb=0.50,\,0.55$ and $0.60$. In the right panel, we have plotted for a fixed value of the magnetic spectral index $\nb=0.55$ where we consider three different values of helicity parameter $\gamma=2.65,\,2.60,$ and $2.55$.
}
\label{fig:ogws}
\end{figure}
In our scenario, we assume that immediately after inflation the inflaton field behaves effectively as radiation, during which primordial black holes (PBHs) can form. The initial energy density of the PBHs is determined by the magnetogenesis parameters $n$ and $\gamma$. If the initial fractional PBH energy density exceeds a critical threshold, namely $\beta>\beta_c$, the Universe subsequently enters a PBH-dominated phase. Consequently, the duration of the PBH-dominated era is governed by the initial magnetogenesis parameters.

The present-day gravitational wave spectrum induced by the electromagnetic field in the presence of a PBH-dominated era can then be expressed as
\begin{align}
    \Omega^{\rm sec}_{\rm gw}h^2\simeq \frac{\Omega_{\rm ra}h^2}{6}\l(\frac{\HI}{\Mp}\r)^4\l(\frac{f}{f_*}\r)^{-4\epsilon}\Fn(k)\mcB^2(\nb,\gamma)\times
    \l\{ 
    \begin{matrix}
        1 & f_*<f<\fre\\
        (f/\fre)^{-2} & \fre<f\leq \fpbh\\
        (\fpbh/\fre)^{-2}& \fpbh<f<\fe
    \end{matrix}
    \r.
\end{align}
The presence of a PBH-dominated phase significantly modifies the spectral behavior of the present-day GW spectrum. Modes that re-enter the horizon during the early radiation-dominated era, just after inflation, typically follow the scaling $\ogwh(f)\propto f^{-4\epsilon}$. However, due to the subsequent PBH-dominated phase, their amplitude is suppressed by a factor $(\fpbh/\fre)^{-2}$.

On the other hand, modes that re-enter during the PBH-dominated era inherit the effects of the matter-like background evolution, leading to a modified spectral scaling $\ogwh(f)\propto f^{-2(1+2\epsilon)}$. This behavior differs from that of the primary GW spectrum, which scales as $\ogwp h^2(f)\propto f^{-2(\epsilon+1)}$.

Finally, modes that remain outside the horizon until the end of reheating (i.e., the end of the PBH-dominated phase and the onset of the standard radiation-dominated era) retain their inflationary scaling, following $\ogwh(f<\fre)\propto f^{-4\epsilon}$.

\paragraph{\underline{GWs in absence of PBH dominated era:}\\}
There exists an alternative scenario, as shown in Fig.~\ref{fig:rho_vs_t_g}, where ultra-light PBHs with extended mass spectrum are produced. However, their initial fractional energy density is insufficient to trigger a PBH-dominated era. This situation arises, for instance, when considering an inflationary potential with $p=2$, corresponding to $V(\phi)\propto \phi^4$, for which the effective equation of state after inflation is $\wre=1/3$. In the absence of a PBH-dominated phase, the universe continues to evolve as a radiation dominated after inflation.

In contrast to scenarios involving a non-instantaneous reheating phase with an intermediate PBH-dominated era, where the GW spectrum exhibits non-trivial features over a specific frequency window, the present case leads to a simpler yet distinct spectral behavior. In particular, at high frequencies, the secondary GW spectrum receives additional contributions from late-time production during the radiation-dominated era.

Over most of the frequency range, the tensor modes generated during inflation dominate, leading to a spectral scaling $\ogws(f\ll\fe)\propto f^{-4\epsilon}$. However, in a narrow frequency window around $f\sim \fe$, the secondary tensor production sourced by the magnetic field during the radiation-dominated era becomes dominant over the inflationary contribution. In this regime, the GW spectrum exhibits a blue-tilted behavior, $\ogws(f\sim\fre)\propto f^{2\nb}$, as clearly illustrated in Fig.~\ref{fig:ogws}.

In Fig.~\ref{fig:ogws}, we present the combined present-day GW spectrum for different magnetogenesis and reheating scenarios. In the left panel, we fix the helicity parameter $\gamma=2.65$ and consider three values of the magnetic spectral index, $\nb=0.50,\,0.55,$ and $0.60$. For $\nb=0.50$ and $\nb=0.55$, modes re-entering during the radiation-dominated era ($f<\fre$) exhibit a nearly scale-invariant spectrum, $\ogwh(f<\fre)\propto f^{-4\epsilon}$, due to the smallness of $\epsilon\sim 10^{-4}$. Modes that re-enter during the PBH-dominated era instead develop a strongly red-tilted spectrum, $\ogwh(\fre<f<\fpbh)\propto f^{-2(1+2\epsilon)}$. For modes re-entering during the early radiation-dominated phase ($f>\fpbh$), the spectral behavior follows $\ogwh(f>\fpbh)\propto f^{-4\epsilon}$; however, their amplitude is significantly suppressed by a factor $(\fpbh/\fre)^2$, rendering them far below the sensitivity of near-future GW detectors.

On the other hand, for $\nb=0.60$, although a substantial number of PBHs are produced, their initial energy density is insufficient to dominate the total energy budget before evaporation. As a result, no PBH-dominated era occurs, and the GW spectrum develops a characteristic feature near $f\sim\fe$. In this case, we observe a dip in the combined GW spectrum, arising from the interplay between two competing contributions: the inflationary production and the post-inflationary production of tensor modes. Specifically, the tensor power spectrum generated during inflation scales as $\mPts(f)\propto f^{-4\epsilon}$, whereas the contribution from the radiation-dominated era scales as $\mPts(f)\propto f^{2\nb}$. At high frequencies, the latter dominates, leading to the observed spectral transition.

Furthermore, decreasing the magnetic spectral index $\nb$ for a fixed helicity parameter $\gamma$ enhances the overall GW amplitude. This can be understood from the corresponding increase in the coupling parameter $n$, which controls the strength of the electromagnetic field. Since the electric and magnetic field amplitudes depend exponentially on $n$ (see Eq.~\eqref{eq:mPbe_inf}), a smaller $\nb$ results in a stronger GW signal.

In the right panel of Fig.~\ref{fig:ogws}, we show the present-day GW spectrum for a fixed magnetic spectral index $\nb=0.55$, while varying the helicity parameter $\gamma=2.55,\,2.60,$ and $2.65$. The overall spectral features remain similar. For $\gamma=2.65$, a PBH-dominated era is realized (as shown in Fig.~\ref{fig:rho_vs_t_g}), leading to a red-tilted and suppressed high-frequency spectrum. In contrast, for $\gamma=2.55$ and $\gamma=2.60$, although PBHs are formed, their energy density is insufficient to induce a PBH-dominated phase (see Fig.~\ref{fig:rho_vs_t_nb}). Nevertheless, the electromagnetic field remains sufficiently strong to generate an observable GW signal over a broad frequency range.

In these cases, due to a resonance-like enhancement at high frequencies near $f\sim\fe$, the GW production during the radiation-dominated era dominates over the inflationary contribution, resulting in a blue-tilted spectrum in this regime. Notably, this high-frequency enhancement lies within the sensitivity range of resonant cavity experiments. Finally, increasing the helicity parameter $\gamma$ leads to an overall amplification of the GW spectrum, as the magnetic field strength depends exponentially on $\gamma$, as evident from Eq.~\eqref{eq:mPbe_inf}.

\subsection{High frequency GW from evaporating PBHs}

Black-Hole emits quasi-thermal radiation via the well-known process of Hawking evaporation~\cite{Majumdar:1995yr, Baumann:2007yr, Hook:2014mla, Smyth:2021lkn, Datta:2020bht, Boudon:2020qpo, Morrison:2018xla, PhysRevD.59.041301, Bernal:2022pue, Schmitz:2023pfy, Borah:2024lml, Barman:2022pdo, RiajulHaque:2023cqe, Hamada:2016jnq, DeLuca:2022bjs, DeLuca:2021oer,Bhaumik:2024qzd}. 
They can emit all sorts of particles with mass around or below the associated black-hole temperature~\cite{Hawking:1974rv, Hawking:1975vcx, PhysRevD.13.198, PhysRevD.16.2402, PhysRevD.14.3260}. 
As they can emit massive particles, they can also emit gravitons. 
As such, black hole evaporation directly produces gravitational radiation~\cite{Ireland:2023avg,Domenech:2021wkk, PhysRevD.101.123533, Bhaumik:2022zdd}, 
and they can contribute to the stochastic background of gravitational waves. 

The spectrum of such emission depends on both the mass and the spin of the black holes. 
After the graviton is emitted, it will redshift from the formation time of the PBH to today. 
Typically, small mass PBHs are formed early, and they will evaporate early, so emitted gravitons experience more redshift, 
and they will produce comparatively larger wavelength GW spectra. 
As we are interested in the present-day gravitational wave energy density evaporating from the PBHs, 
we parametrize it by the spectral density parameter defined as $\Omega_{\rm gw}$:
\begin{align}\label{eq:gws_d}
    \Omega_{\rm gw} = \frac{1}{\rho_{\rm crit}} \frac{\d \rho_{\rm gw}}{\d \ln f},
\end{align}
with $\rho_{\rm crit} = 3H_0^2 \Mp^2$ being the critical energy density today. 

For computational purposes, we define a new quantity $Q_{\rm gw}(t,\omega) = \frac{\d N_{\rm grav}}{\d t \, \d \omega}$,
which represents the instantaneous graviton flux emitted by PBHs, 
where $t$ is the cosmic time and $\omega$ is the wavenumber associated with the emitted graviton. 

Using this, the corresponding instantaneous power of the emitted graviton can be written as $\frac{\d E_{\rm grav}}{\d t \, \d\omega} 
    = \frac{\omega}{2\pi} Q_{\rm gw}(t,\omega)$,
and hence the instantaneous energy density emitted in the form of gravitational waves due to evaporation is~\cite{Ireland:2023avg}
\begin{align}
    \frac{\d \rho_{\rm gw}}{\d t \, \d \omega} 
    = \npbh(t)\,\frac{\omega}{2\pi} Q_{\rm gw}(t,\omega),
\end{align}
where $\npbh(t)$ is the PBH number density at time $t$. 

To obtain the total energy density emitted in the form of gravitational waves, we integrate the above quantity over the PBH lifetime, i.e., from the formation time 
$\th \simeq \Mpbh/\Mpl^2$ to the evaporation time $t_{\rm ev}$, 
the latter being defined in Eq.~\eqref{eq:tev}. 
For a single-mass PBH distribution, the energy density per logarithmic frequency interval at the evaporation time is given by~\cite{Ireland:2023avg}
\begin{align}
    \frac{\d \rho^*_{\rm gw}}{\d \ln \omega^*} 
    &= \npbhi(t_{\rm h})
    \left(\frac{a_i}{a_*}\right)^3 
    \frac{\omega_*^2}{2\pi}
    \int_{t_{\rm h}}^{t_{\rm ev}} \d t \;
    \frac{a_*}{a(t)} \;
    Q_{\rm gw}\!\left(t, \, \omega_* \frac{a_*}{a(t)}\right),
\end{align}
where the symbol `$*$' denotes quantities evaluated at the end of the PBH evaporation. 
Since the gravitational wave (GW) energy density redshifts as $\rho_{\rm gw}\propto a^{-4}$, to define the present-day GW strength, we must consider the dilution factor as well as the redshifting of the frequency. 
The present-day frequency $\omega_0$ and the emitting frequency $\omega_*$ are related by  $\omega_0 = a_* \, \omega_*$.

Now, to define the spectral energy density (SED) of GWs due to PBH evaporation,  
we normalize by the critical density (as defined in Eq.~\eqref{eq:gws_d}), and obtain the present-day SED as  
\begin{align}
    \ogwevp(\omega_0)
    &= \frac{4 \, \npbhi \, a_i^3}{3 \, \Mp^2 \, H_0^2} \, \omega_0^2 
    \int_{\th}^{\tev} \frac{dt}{a(t)} \, \Qgw\!\left(t, \frac{\omega_0}{a(t)}\right) .
\end{align}

To make this expression more transparent, one can rewrite $\npbhi$ in terms of the initial fractional energy density $\beta_i$ and the PBH mass $\Mpbh$.  
The expression for the SED then becomes  
\begin{align}
    \ogwevp(\omega_0)
    &= \frac{4 \, \Mp^4 \, a_i^3}{32\pi \, H_0^2 \, \Mpbh^3} \, \beta_i(\th) \, \omega_0^2
    \int_{\th}^{\tev} \frac{dt}{a(t)} \, \Qgw\!\left(t, \frac{\omega_0}{a(t)}\right) .
\end{align}

As discussed above, these scenarios typically lead to a \emph{broad PBH mass spectrum}, with PBHs of different masses forming at different cosmic times. Consequently, the final gravitational-wave (GW) spectrum cannot be computed directly using \textit{BlackHawk}, which is designed for a single PBH mass formed at a fixed time. For simplicity, we therefore assume non-rotating (Schwarzschild) PBHs, setting the angular momentum to zero. In this case, the peak frequency obtained from analytical estimates agrees well with that extracted from \textit{BlackHawk} simulations~\cite{Ireland:2023avg}. The main difference between the two approaches lies in the overall amplitude, which differs by approximately one order of magnitude, while the spectral shape remains identical for frequencies $f<f_{\rm peak}$. However, for $f>f_{\rm peak}$, the numerical \textit{BlackHawk} spectra exhibit a red-tilted behavior over a wide frequency range, rather than the sharp high-frequency cutoff predicted by analytical approximations.

Motivated by this, we adopt a \emph{semi-analytical approach} in this work. Specifically, we compute the GW spectrum for discrete PBH masses spanning the full mass range of interest and then sum the individual contributions over all frequencies. This procedure preserves the overall spectral shape, with deviations appearing only at the high-frequency end when compared with full numerical results. A detailed numerical analysis using \textit{BlackHawk} for the complete PBH mass spectrum will be presented in a forthcoming work.

The present-day spectral energy density of GWs emitted via PBH evaporation can be written as
\begin{align}
    \ogwevp \simeq 
    \frac{\beta(M_{\rm PBH})}{H_0^2\, M_{\rm PBH}}\, \omega_0^4\, I(\omega_0),
\end{align}
where $H_0 = 100\,h\,{\rm km\,s^{-1}\,Mpc^{-1}}$ is the present Hubble parameter with $h\simeq 0.67\text{--}0.7$, and $I(\omega_0)$ is defined as~\cite{Ireland:2023avg}
\begin{align}
    I(\omega_0) \simeq 
    \int_{\tf}^{t_{\rm ev}} dt \,
    \frac{(a_i/a)^3}{\exp\!\left(\omega_0/(a T_{\rm BH})\right)-1}.
\end{align}

To evaluate this integral numerically, we introduce the dimensionless variable $x = t/\tf$, where $\tf$ denotes the PBH formation time. The above expression can then be rewritten as
\begin{align}
    \ogwevp(\omega_0) 
    &\simeq 
    \frac{\beta(M_{\rm PBH})\, \tf}{H_0^2\, M_{\rm PBH}}\, \omega_0^4
    \int_{1}^{x_{\rm ev}} dx \,
    \frac{(a_i/a)^3}{\exp\!\left[\alpha (a_i/a)\right]-1}
    =
    \frac{3\,\beta(M_{\rm PBH})}{2\pi\,\gamma_c\,\rho_{\rm crit,0}}\, \omega_0^4
    \int_{1}^{x_{\rm ev}} dx \,
    \frac{(a_i/a)^3}{\exp\!\left[\alpha (a_i/a)\right]-1},
\end{align}
where $\rho_{\rm crit,0}$ denotes the present critical energy density. In scenarios where PBHs dominate the energy density of the Universe for an extended period, the cosmic background effectively behaves as a matter-dominated fluid. The corresponding modification of the expansion history is automatically incorporated in our calculation through the appropriate choice of the scale-factor evolution and integration limits.
\begin{figure}[t]
\centering
\includegraphics[scale=0.43]{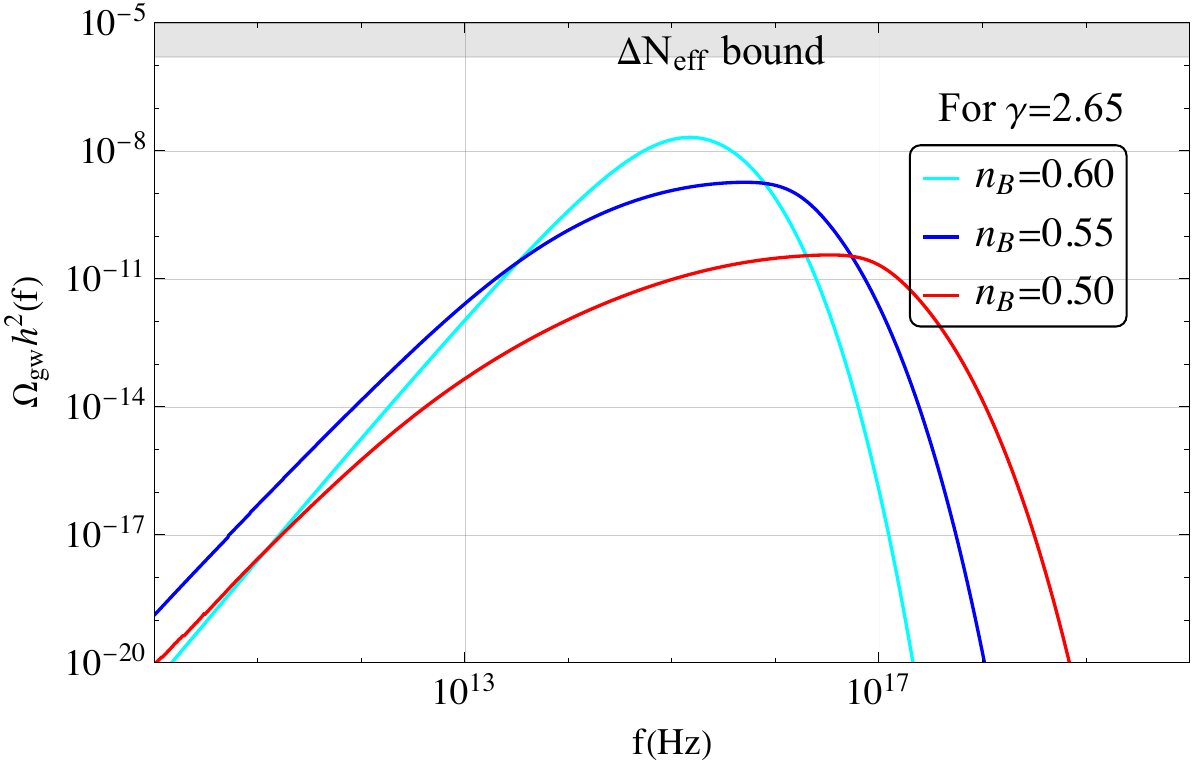}
\includegraphics[scale=0.43]{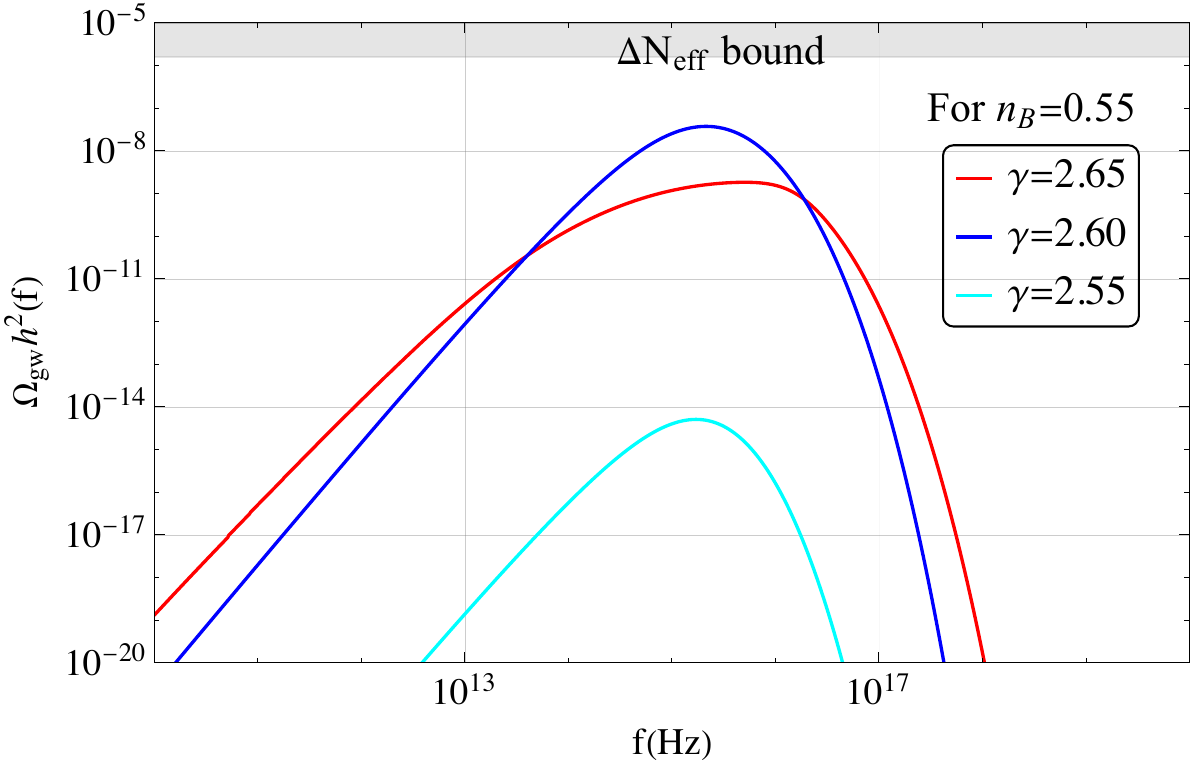}
\caption{Here we present the spectra of gravitational waves sourced by the evaporation of PBHs for a representative set of initial parameters $\gamma$ and $\nb$. 
In the left panel, the three different colors correspond to distinct values of the magnetic spectral index $\nb$, while keeping the coupling parameter fixed at $\gamma = 2.65$. 
In the right panel, the three colors correspond to different values of the coupling parameter $\gamma$, with the magnetic spectral index fixed at $\nb = 0.55$.}
\label{fig:ogws_eva}
\end{figure}

In Fig.~\ref{fig:ogws_eva}, we present the GW spectra from PBH evaporation for two representative scenarios. In the left panel, we vary the magnetic spectral index $\nb$ while fixing $\gamma=2.65$. We find that the peak amplitude is maximized for $\nb=0.60$, reaching $\Omega_{\rm GW}^{\rm peak}\simeq 2.1\times 10^{-8}$. In this case, the spectrum is significantly broader than that obtained from a monochromatic PBH mass function~\cite{Ireland:2023avg}, reflecting the underlying broad PBH mass distribution. For smaller values of $\nb$, the GW spectrum becomes even broader; however, the present-day amplitude is mildly suppressed due to additional dilution effects associated with PBH domination. This suppression arises because, for a broad mass spectrum, no single PBH mass dominates the energy density throughout the entire evolution, as also indicated in Fig.~\ref{fig:beta_M}.

In the right panel, we show the GW spectra for three different values of $\gamma$ at a fixed magnetic spectral index $\nb=0.55$. The maximum amplitude is obtained for $\gamma=2.60$. Increasing $\gamma$ leads to a broader spectrum with a reduced peak amplitude, as for $\gamma=2.65$ the Universe undergoes a prolonged PBH-dominated phase. Conversely, for $\gamma=2.55$, the resulting GW amplitude is strongly suppressed, even though no PBH-dominated epoch occurs. This suppression originates from the reduced initial PBH energy fraction for this parameter choice, as shown in Fig.~\ref{fig:beta_M}.

In conclusion, both the amplitude and the spectral shape of the GW background from PBH evaporation are highly sensitive to the PBH mass spectrum and the background cosmological evolution, particularly the presence and duration of any PBH-dominated era.

\section{Conclusions}

In this work, we have presented an alternative mechanism for the formation of ultra-light primordial black holes (PBHs) sourced by inflationary magnetogenesis, thereby establishing a direct and natural connection between PBH formation and the generation of primordial magnetic fields. We considered a class of inflationary models in which the gauge field is kinetically coupled to the inflaton through the interaction term $\gamma I^2(\eta) F_{\mu\nu}\tilde{F}^{\mu\nu}$. This coupling generically leads to the production of helical magnetic fields, with the spectral properties determined by the time dependence of the coupling function $I(\eta)$. For nonvanishing values of the helicity parameter $\gamma$, the magnetic field energy density grows exponentially. In particular, for blue-tilted spectra ($\nb>0$), the energy density is dominated by modes near the end of inflation, $k \sim k_{\rm e}$.

Such amplified magnetic fields act as efficient sources of both stochastic gravitational waves (SGWs) and curvature perturbations. We have shown that, once the magnetogenesis parameters $(\nb,\gamma)$ exceed a critical threshold, the resulting small-scale curvature power spectrum satisfies $\mPc(k\gg k*) \gtrsim \mathcal{O}(10^{-2})$, which is sufficient to trigger PBH formation. This enhancement corresponds to a fractional electromagnetic energy density $\delta\rho_{\rm em}/\rho_c \sim \mathcal{O}(0.01)$. The PBHs formed in this scenario typically emerge immediately after inflation with extremely small initial masses and therefore evaporate well before the onset of big bang nucleosynthesis (BBN). Owing to the broad enhancement of the curvature power spectrum, PBH formation occurs over a wide range of scales, leading naturally to a broad PBH mass spectrum. The subsequent Hawking evaporation of these PBHs, with mass-dependent lifetimes, generically gives rise to an extended PBH-dominated era in the early Universe.

We have further investigated the associated gravitational-wave signatures arising from two distinct sources: (i) the anisotropic stresses of the primordial magnetic fields, and (ii) the direct emission of gravitational waves during PBH evaporation. The gravitational-wave spectrum sourced by magnetic anisotropies exhibits rich spectral features that depend sensitively on the magnetogenesis parameters $(\nb,\gamma)$. When the PBH energy density becomes dominant, additional spectral breaks appear, accompanied by a suppression of the high-frequency gravitational-wave amplitude. In contrast, the gravitational waves produced by PBH evaporation form a broad and nearly scale-invariant spectrum with an amplitude $\Omega_{\rm gw} h^2 \simeq 2.5\times 10^{-9}$. The position of the spectral peak is directly related to the PBH mass and, consequently, to the underlying magnetogenesis parameters.

In summary, we have demonstrated that helical  inflationary magnetogenesis model can simultaneously generate primordial magnetic fields at high frequency and source the formation of ultra-light PBHs. This framework naturally produces a broad PBH mass spectrum and can lead to a prolonged PBH-dominated epoch between the end of inflation and BBN, with its duration controlled by the magnetogenesis parameters $\nb$ and $\gamma$. Moreover, the scenario predicts two distinct and potentially observable gravitational-wave signals, providing a novel observational window into the physics of inflationary magnetogenesis and PBH formation. Beyond offering an alternative pathway to ultra-light PBH production, the evaporation of such PBHs may have far-reaching implications for particle cosmology, including possible connections to baryogenesis and leptogenesis. Our results, therefore, provide a unified and self-consistent framework linking inflationary magnetogenesis, primordial black holes, and gravitational-wave phenomenology.

\appendix

\section{Computing the Radiation Energy Density from PBH Evaporation}

To compute the contribution of primordial black hole (PBH) evaporation to the total radiation energy density, we adopt the simplifying assumption that PBHs evaporate instantaneously at the end of their lifetime, $\tau_{\rm ev}$. Thus, a PBH formed at time $\tf$ completely evaporates at
\begin{align}
    t_{\rm ev} = \tf + \tau_{\rm ev}.
\end{align}

Consider a PBH of mass $M_i$ formed at time $t = t_i$. The formation time can be approximately related to its mass by
\begin{align}
    \tf \simeq \frac{M_i}{4\pi \gc x_c\, M_{\rm Pl}^2},
\end{align}
where $\gamma$ is the standard collapse parameter, and $\alpha$ is a constant determined by the background equation of state during PBH formation:
\begin{align}
x_c = 
\begin{cases}
    1/2, & \text{for PBH formation during a radiation-dominated background}, \\
    1,   & \text{for PBH formation during a matter-like background}.
\end{cases}
\end{align}
The mass of the PBHs are typically equal to the horizon mass at the time of horizon re-entry of the corresponding mode, where the horizon mass at the formation time is expressed as $M_{H}=\frac{4\pi}{3}\rho(\th)H^{-3}(\th)$,
where $\th$ is the cosmic time when the PBHs are formed and $\rho(\th)$ is the total background energy density during the formation time. Now, if we consider the efficiency factor accounting for the collapse dynamics, then we can write the PBH mass as 
\begin{align}\label{eq:pbh_mass}
    \Mpbh\simeq 4\pi\gc\Mp^2H^{-1}(\th)
\end{align}
We recall that $\Mp=1/\sqrt{8\pi G}$ is the reduced Planck mass. Here, $H(\th)$ is the Hubble constant during the formation of the PBHs.

Now, if we put the value of the $\Mp\simeq 2.4\times 10^{18}\,\GeV$ in the above equation, we can write the formation time in sec as
\begin{align}\label{eq:tf_mpbh}
    \tf\simeq 5.11\times 10^{-39}\l(\frac{0.2}{\gc}\r)\l(\frac{0.5}{x_c}\r)\l(\frac{M_i}{1\,\gm}\r)\,\sec
\end{align}

The lifetime of a PBH evaporating into Standard Model particles can be approximated as~\cite{Carr:2009jm, Carr:2020gox}
\begin{align}
    \tau_{\rm ev} \simeq 407 \left( \frac{M_i}{10^{10}\,\mathrm{g}} \right)\, \mathrm{sec}.
\end{align}

Using the above expressions, the evaporation time of a PBH with mass $M_i$ is
\begin{align}
    t_{\rm ev,f} = \tf + \tau_{\rm ev} 
    \simeq \tf + 407 \left( \frac{M_i}{10^{10}\,\mathrm{g}} \right)\, \mathrm{sec}.
\end{align}
Since the PBH lifetime is much larger than its formation time ($\tau_{\rm ev} \gg t_i$), we can safely approximate $t_{\rm ev,i} \simeq \tau_{\rm ev}$.

The energy density associated with PBHs of mass $M_i$ at formation time $t_i$ is
\begin{align}
    \rho_{\rm PBH}^{(i)}(t_i) = \beta_i\, \rho_{\rm ra}(t_i),
\end{align}
where $\rho_{\rm ra}(t_i) = 3H_i^2 M_{\rm Pl}^2$ is the background radiation energy density at formation.  
Since PBHs behave as non-relativistic matter, their energy density redshifts as $a^{-3}$. Therefore, the PBH energy density evaluated at the evaporation time $t_{\rm ev,f}$ is
\begin{align}
    E_i \equiv \rho_{\rm PBH}^{(i)}(t_{\rm ev,f}(M_i))
    = \beta_i\, \rho_{\rm ra}(\tf(M_i))
      \left( \frac{a(\tf(M_i))}{a(t_{\rm ev,f}(M_i))} \right)^3 .
\end{align}

After evaporation, the emitted radiation redshifts as $a^{-4}$. Thus, for any later time $t > t_{\rm ev,i}$, the radiation energy density contributed by this PBH mass component is
\begin{align}\label{eq:rho_ev_pbh}
    \rho_{\rm ra,i}^{\rm ev}(t) 
    = E_i \left( \frac{a(t_{\rm ev,i})}{a(t)} \right)^4 .
\end{align}

To obtain the total radiation energy density sourced by PBH evaporation, we sum over all PBH mass components that have already evaporated by time $t$. This yields
\begin{align}
    \rho_{\rm ra}^{\rm ev}(t)
    = \sum_i 
      E_i \left( \frac{a(t_{\rm ev,f})}{a(t)} \right)^4
      \Theta(t - t_{\rm ev,f}(M_i)),
\end{align}
where the Heaviside step function $\Theta(t - t_{\rm ev,i})$ ensures that only PBHs that have evaporated before time $t$ contribute to the radiation energy density.

\section{Computing the evolution of the radiation and PBH energy density}
In our scenarios, PBHs are produced due to the induced curvature perturbation when the overdensity regions re-enter the horizon after inflation. Now as the PBHs are formed after inflation, one can easily estimate the lowest mass a PBH can form from the above Eq.\eqref{eq:pbh_mass}. For the Hubble parameter $\HI\simeq 10^{13}\GeV$, the lowest possible PBH mass is $\Mpbh^{\rm min}\simeq 2.57\, \gm$.Now these ultra-light PBHs are evaporating and producing the thermal bath. If we demand that these PBHs should evaporate before the BBN takes place, or even if they evaporate, the fractional energy density should be very small, such that it should not affect the BBN process. From the computation, one can easily find the mass which can evaporate before the BBN is around $\Mpbh^{\rm bbn}\simeq 10^9$ gm. Now Eq.\eqref{eq:tf_mpbh}, one can easily find that the formation time of the $10^9$ gm PBH is around $\tf(10^9~\gm)\simeq 5.1\times 10^{-29}\,\sec$, whereas the life time of the smallest mass PBH is $\tau_{\rm ev}(\Mpbh=2.57\,\gm)\simeq 1.04\times 10^{-7}\,\sec $. From this comparison, it is clear that the contributions from the PBH evaporation are only significant after $t\geq t_{\rm ev}(\Mpbh=2.57\,\gm)$.

Now in our scenario, we have a broad PBH mass spectrum, so to track the evolution of the PBH energy density and the radiation energy density, we have to solve the Blotzmann equation, take the evaporation effect of the PBH, but as the mass spectrum of the PBH are not linear function and its shape is continuously changed when we change the magnetogenesis parameters, so it is very difficult to solve it. As it is not only the main goal to exactly solve the Boltzmann equation for the broad spectrum, we keep it for our future publications, so here we adopt a semianalytic formula to compute the evolution of the PBH energy density and radiation energy density. Here we assume that PBH with mass $M_i$ evaporates instantaneously at the end of its lifetime $\tau_{\rm ev}(M_i)$, so if the PBH formed at $\tf(M_i)$ then it completley evaporate at $t_{\rm ev}(M_i)=t_i(M_i)+\tau_{\rm ev}(M_i)$. Although it is an approximate solution, it will carry most of the physics. Now, over time, PBH are continuously produced, and it will continuously evolve after $t>t_{\rm ev}(\Mpbh^{\rm min})$, so we can write the total energy density associated with the PBH for time $t$ as
\begin{align}
    \rhopbh(t)=\sum_i \beta(M_i)\rho_c(\tf(M_i))\l(\frac{a(t_i(M_i)}{a(t)}\r)^3\Theta(t_{\rm ev}(M_i)-t)\Theta(t-\tf(M_i))
\end{align}
Here we add two Theta functions, where the 1st Theta function subtracts those PBH whichh are evaporate before the time $t$ and 2nd Theta function indicates only those PBH that are formed before the time $t$. So, due to the evaporation, one part of the radiation energy should add to the background radiation energy density. So the total radiation energy density at any time $t$ can be written as
\begin{align}
    \rho_{\rm ra}(t)=\rho_c(t_e)\l(\frac{a(t_e)}{a(t)}\r)^4+ \rho_{\rm ra}^{\rm ev}(t)
\end{align}
where $t_e$ is the cosmic time defined at the end of the inflation and $\rho_c(t_e)=3\HI^2\Mp^2=12\Mp^2/t_e^2$ is the background energy density at the end of inflation. Here $\rho_{\rm ra}^{\rm ev}(t)$ is the energy contribution to the radiation energy density that comes from the PBH evaporation, and it is defined in Eq.\eqref{eq:rho_ev_pbh}.

We have seen that the evaporation time of a PBH is far away from its formation time, i.e., $t_{\rm ev}(M_i)>>\tf(M_i)$. In our scenarios, all the PBHs are formed almost during the early-radiation dominated era (if there have PBH dominated era, which depends on the magnetogenesis parameters $\gamma$ and $\nb$). As we have a broad PBH mass spectrum and the shape depends on the magnetogenesis parameters, to compute the total evaporation energy density, we adopted a semi-analytical approach, where we assume a singular mass PBH is instantiniously evaporate at the end of its lifetime. Now the energy density associated with PBHs mass $M_i$ at its formation time $\tf(M_i)$ is
\begin{align}
    \rhopbh^{(1)}(\tf(M_i))=\beta(M_i)\rho_c(\tf(M_i))
\end{align}
where $\rho_c(\tf)=3H^2(\tf)\Mp^2$ is the background energy density when the PBH with mass $M_i$ is formed. After its formation, as PBHs are non-relativistic species, their energy density dilutes as matter-like fluid, i.e., $\rhopbh(t\leq \tev)\propto a^{-3}$ before it completely evaporates. So at the time of evaporation, the corresponding energy density associate with PBH mass $M_i$ is
\begin{align}
    E_i\equiv\rhopbh(\tev(M_i))=\rhopbh(\tf(M_i))\l(\frac{a(\tf(M_i)}{a(\tev(M_i))}\r)^3
\end{align}
After its completely evpoartion, its energy density goes like radiation, so the radiation energy density due to the PBH mass $M_i$ at time $t>\tev$ is $\rho_{\rm ra,i}^{\rm ev}(t>\tev)=E_i(a(\tev)/a(t))^4$. Now, as we know the discrete value of the PBH mass and its associated initial fractional energy density $\beta(M_i)$, to get the total radiation energy density, we have to sum over all the mass range. So the radaition energy density due to the PBH evaporation at time $t$ is 
\begin{align}
    \rho_{\rm ra}^{\rm ev}(t)=\sum_i E_i\l(\frac{a(\tev(M_i)}{a(t)}\r)^4\Theta(t-\tev(M_i))
\end{align}
Here we add an extra Theta function to ensure only the energy density that came from those PBHs that evaporate before the time $t$.

 %%%%%%%%%%%%%%%%%%%%%%%%%%%%%%%%%%%%%%%%%%%%%%%%%%%%%%%%%%%%%%%%%%%%%%%%%%
\bibliographystyle{apsrev4-1}
\bibliography{references}
\end{document}